\documentclass[twocolumn]{aastex63}
\usepackage{natbib}
\usepackage{amsmath}
\usepackage{verbatim}
\usepackage{supertabular}
\usepackage{longtable}
\usepackage[shortlabels]{enumitem}
\usepackage[T1]{fontenc}
\def\CC{{C\nolinebreak[4]\hspace{-.05em}\raisebox{.4ex}{\tiny\bf ++}}}

\newcommand{\file}[1]{\textbf{\detokenize{#1}}}  
\newcommand{\param}[1]{{\it{\detokenize{#1}}}}  
\newcommand{\parvalue}[1]{{`{\detokenize{#1}'}}}   

\newcommand{\pcmd}[1]{{\tt \detokenize{#1}}} 
\newcommand{\ccmd}[1]{{\tt \detokenize{#1}}} 
\newcommand{\recipe}[1]{{\tt \detokenize{procedure_#1}.xml}} 
\newcommand{\tsys}{$T_{\rm sys}$}
\newcommand{\trx}{$T_{\rm rx}$}
\newcommand{\median}{\texttt{Mdn}}
\newcommand{\checksources}{check sources}
\newcommand{\checksource}{check source}
\newcommand{\Checksource}{Check source}
\newcommand\titlelowercase[1]{\texorpdfstring{\lowercase{#1}}{#1}}

\def\bwForSensitivity{$\Delta\nu_{\rm sens}$}
\def\spwbw{$\Delta\nu_{\rm spw}$}
\def\spwchanwidth{$\Delta\nu_{\rm spw,chan}$}
\def\cubechanwidth{$\Delta\nu_{\rm cube,chan}$}
\def\repSpwBW{$\Delta\nu_{\rm repSpw}$}
\def\nbinFC{$nbin_{\rm findcont}$}
\def\nbinRepSpw{$nbin_{\rm repSpw}$}

\received{April 21, 2023}
\revised{June 3, 2023}
\accepted{June 9, 2023}
\graphicspath{{./}{figures/}}

\begin{document}

\title{The ALMA Interferometric Pipeline Heuristics}
\correspondingauthor{T. R. Hunter}
\email{thunter@nrao.edu}
\author[0000-0001-6492-0090]{Todd R. Hunter}
\affiliation{National Radio Astronomy Observatory, 520 Edgemont Rd, Charlottesville, VA 22903, USA}
\affiliation{Center for Astrophysics $\mid$ Harvard \& Smithsonian, Cambridge, MA 02138, USA}

\author[0000-0002-4663-6827]{Remy Indebetouw}
\affiliation{National Radio Astronomy Observatory, 520 Edgemont Rd, Charlottesville, VA 22903, USA}
\affiliation{Department of Astronomy, University of Virginia, P.O. Box 3818, Charlottesville, VA 22903, USA}

\author[0000-0002-6558-7653]{Crystal L. Brogan}
\affiliation{National Radio Astronomy Observatory, 520 Edgemont Rd, Charlottesville, VA 22903, USA}
\affiliation{Department of Astronomy, University of Virginia, P.O. Box 3818, Charlottesville, VA 22903, USA}

\author[0000-0001-9757-9706]{Kristin Berry}
\affiliation{National Radio Astronomy Observatory, 520 Edgemont Rd, Charlottesville, VA 22903, USA}

\author{Chin-Shin Chang}
\affiliation{Joint ALMA Observatory, Alonso de C\'ordova 3107, Vitacura, Santiago 763-0355, Chile}

\author[0000-0001-8123-0032]{Harold Francke}
\affiliation{Joint ALMA Observatory, Alonso de C\'ordova 3107, Vitacura, Santiago 763-0355, Chile}

\author[0000-0003-2692-8926]{Vincent C. Geers}
\affiliation{UK Astronomy Technology Centre, Royal Observatory Edinburgh, Blackford Hill, Edinburgh EH9 3HJ, UK} 

\author[0000-0002-3781-3537]{Laura G\'omez}
\affiliation{Joint ALMA Observatory, Alonso de C\'ordova 3107, Vitacura, Santiago 763-0355, Chile}

\author[0000-0002-2202-608X]{John E. Hibbard}
\affiliation{National Radio Astronomy Observatory, 520 Edgemont Rd, Charlottesville, VA 22903, USA}
\affiliation{Department of Astronomy, University of Virginia, P.O. Box 3818, Charlottesville, VA 22903, USA}

\author[0000-0001-9549-6421]{Elizabeth M. Humphreys}
\affiliation{Joint ALMA Observatory, Alonso de C\'ordova 3107, Vitacura, Santiago 763-0355, Chile}

\author[0000-0002-8990-1811]{Brian R. Kent}
\affiliation{National Radio Astronomy Observatory, 520 Edgemont Rd, Charlottesville, VA 22903, USA}

\author[0000-0002-3227-4917]{Amanda A. Kepley}
\affiliation{National Radio Astronomy Observatory, 520 Edgemont Rd, Charlottesville, VA 22903, USA}
\affiliation{Department of Astronomy, University of Virginia, P.O. Box 3818, Charlottesville, VA 22903, USA}

\author[0000-0002-1568-579X]{Devaky Kunneriath}
\affiliation{National Radio Astronomy Observatory, 520 Edgemont Rd, Charlottesville, VA 22903, USA}

\author[0000-0002-6667-7773]{Andrew Lipnicky}
\affiliation{National Radio Astronomy Observatory, 520 Edgemont Rd, Charlottesville, VA 22903, USA}

\author[0000-0002-8932-1219]{Ryan A. Loomis}
\affiliation{National Radio Astronomy Observatory, 520 Edgemont Rd, Charlottesville, VA 22903, USA}
\affiliation{Department of Astronomy, University of Virginia, P.O. Box 3818, Charlottesville, VA 22903, USA}

\author[0000-0002-8472-836X]{Brian S. Mason}
\affiliation{National Radio Astronomy Observatory, 520 Edgemont Rd, Charlottesville, VA 22903, USA}

\author{Joseph S. Masters}
\affiliation{National Radio Astronomy Observatory, 520 Edgemont Rd, Charlottesville, VA 22903, USA}

\author[0000-0002-7675-3565]{Luke T. Maud}
\affiliation{ESO Headquarters, Karl-Schwarzschild-Str.\ 2, D-85748 Garching, Germany}

\author[0000-0002-2315-2571]{Dirk Muders}
\affiliation{Max-Planck-Institut f\"ur Radioastronomie, Auf dem H\"ugel 69, D-53121, Bonn, Germany}

\author[0000-0003-1149-6294]{Jose Sabater}
\affiliation{UK Astronomy Technology Centre, Royal Observatory Edinburgh, Blackford Hill, Edinburgh EH9 3HJ, UK} 
\affiliation{SUPA, Institute for Astronomy, University of Edinburgh, Royal Observatory, Blackford Hill, Edinburgh, EH9 3HJ, UK}

\author{Kanako Sugimoto}
\affiliation{National Astronomical Observatory of Japan, National Institutes of Natural Sciences, 2-21-1 Osawa, Mitaka, Tokyo 181-8588, Japan}

\author{L\'aszl\'o Sz\H{u}cs}
\affiliation{National Radio Astronomy Observatory, 520 Edgemont Rd, Charlottesville, VA 22903, USA}
\affiliation{Leibniz Supercomputing Centre, Boltzmannstra{\ss}e 1, D-85748 Garching, Germany}

\author[0000-0002-5038-9267]{Eugene Vasiliev}
\affiliation{UK Astronomy Technology Centre, Royal Observatory Edinburgh, Blackford Hill, Edinburgh EH9 3HJ, UK}

\author{Liza Videla}
\affiliation{Joint ALMA Observatory, Alonso de C\'ordova 3107, Vitacura, Santiago 763-0355, Chile}

\author[0000-0003-4314-4947]{Eric Villard}
\affiliation{ESO Headquarters, Karl-Schwarzschild-Str.\ 2, D-85748 Garching, Germany}

\author{Stewart J. Williams}
\affiliation{UK Astronomy Technology Centre, Royal Observatory Edinburgh, Blackford Hill, Edinburgh EH9 3HJ, UK}

\author[0000-0001-7689-9305]{Rui Xue}
\affiliation{National Radio Astronomy Observatory, 520 Edgemont Rd, Charlottesville, VA 22903, USA}

\author[0000-0001-9163-0064]{Ilsang Yoon}
\affiliation{National Radio Astronomy Observatory, 520 Edgemont Rd, Charlottesville, VA 22903, USA}
\affiliation{Department of Astronomy, University of Virginia, P.O. Box 3818, Charlottesville, VA 22903, USA}

\shorttitle{ALMA Pipeline Heuristics}
\shortauthors{Hunter et al.}

\begin{abstract}

We describe the calibration and imaging heuristics developed and deployed in the Atacama Large Millimeter/sub-millimeter Array (ALMA) interferometric data processing pipeline, as of ALMA Cycle 9 operations.  The pipeline software framework is written in Python, with each data reduction stage layered on top of tasks and toolkit functions provided by the Common Astronomy Software Applications package. This framework supports a variety of tasks for observatory operations, including science data quality assurance, observing mode commissioning, and user reprocessing. It supports ALMA and VLA interferometric data along with ALMA and NRO\,45\,m single dish data, via different stages and heuristics. In addition to producing calibration tables, calibrated measurement sets, and cleaned images, the pipeline creates a WebLog which serves as the primary interface for verifying the quality assurance of the data by the observatory and for examining the contents of the data by the user. Following the adoption of the pipeline by ALMA Operations in 2014, the heuristics have been refined through annual prioritized development cycles, culminating in a new pipeline release aligned with the start of each ALMA Cycle of observations. Initial development focused on basic calibration and flagging heuristics (Cycles 2-3), followed by imaging heuristics (Cycles 4-5). Further refinement of the flagging and imaging heuristics, including the introduction of parallel processing, proceeded for Cycles 6-7. In the 2020 release, the algorithm to identify channels to use for continuum subtraction and imaging was substantially improved by the addition of a moment difference analysis.  A spectral renormalization stage was added for the 2021 release (Cycle 8) to correct high spectral resolution visibility data acquired on targets exhibiting strong celestial line emission in their autocorrelation spectra. The calibration heuristics used in the low signal to noise regime were improved for the 2022 release (Cycle 9). In the two most recent Cycles, 97\% of ALMA datasets were calibrated and imaged with the pipeline, ensuring long-term automated reproducibility of results.  We conclude with a brief description of plans for future additions, including a self-calibration stage, support for multi-configuration imaging, and complete calibration and imaging of full polarization data. 

\end{abstract}



\section{Introduction} \label{sec:intro}

Located on the 5000\,m plateau of Chajnantor in northern Chile, the Atacama Large Millimeter/submillimeter Array (ALMA) \citep{Wootten2009} enables new discoveries and insight on the origin of planets, stars, galaxies, and the universe by acquiring and delivering uniquely powerful and quality-assured data products for the international astronomical community. The ALMA Science Pipeline provides automated calibration and imaging of ALMA
interferometric and single-dish data. ALMA interferometric data refers to observations obtained with either the 50-antenna array of 12\,m antennas processed on the baseline correlator \citep[BLC,][]{Escoffier07} or the 12-antenna Atacama Compact Array (ACA) of 7\,m antennas processed on the ACA correlator \citep[ACAC,][]{Kamazaki12}.  ALMA single-dish data refers to observations obtained with the four 12\,m antennas of the ALMA Total Power Array (TPA).  The ACA and TPA \citep{Iguchi2009} are collectively known as the Morita Array \citep{Morita2008}.

From the beginning of the design stage of the ALMA observatory, the ALMA scientific specifications and requirements called for an automated pipeline that delivers ``standard data reduction...resulting in a properly calibrated image cube'' which ``shall require minimal input from the astronomer in most cases'' \citep{Wootten2005,Tarenghi2008}, thereby enabling all astronomers to use ALMA effectively.  Preliminary work on the pipeline framework \citep{Lightfoot2006,Lucas2002,Lucas2000} commenced many years in advance of the first science user observations in Cycle 0, which began on 2011\,September\,30 \citep{Lundgren2012}.  In retrospect, the development of the minimum viable set of heuristics for data flagging and calibration required a few years of experience with real ALMA data, including some project-wide refinement of the exact definitions of ALMA observing modes and the institution of strict adherence to the required scan intents included in each scheduled observation.   The calibration portion of the interferometric pipeline was first placed into operation by ALMA in 2014\,October, a few months after the start of Cycle\,2 observing (2014\,June\,03).  The metadata content of ALMA raw data reached a mature level by the start of Cycle\,3 (2015\,October\,01). The imaging portion of the pipeline was introduced in operations during Cycle\,4 with the first automated image products delivered to users in 2017. 

Further refinement of the pipeline heuristics has occurred on an annual prioritized development cycle, culminating in a new pipeline release aligned with the start of each Cycle of observations.  A list of the pipeline releases, including the corresponding CASA versions and downloadable tarballs, is hosted on the ALMA science portal\footnote{\url{https://almascience.org/processing/science-pipeline}}. 
Management of the pipeline branches is handled by the distributed version control system \ccmd{git} \citep{git}.  The historical and ongoing developments of the pipeline are tracked using the JIRA ticket system from Atlassian.  Major improvements include the introduction of advanced visibility flagging (\S\ref{correctedampflagtext}) and automask imaging for Cycle\,5, parallel processing to reduce the time required to create images for Cycle\,6, and the migration of the entire software stack 
from Python\,2 to Python\,3 in the second release for Cycle\,7 (2020\,October).   This release also contained a substantial improvement to the continuum finding algorithm via the addition of the moment difference analysis (\S\ref{momdiff}). During Cycle\,7 and 8, 96.5\% and 97.2\% of ALMA datasets were calibrated and imaged with the pipeline.   A spectral renormalization stage (\S\ref{renormalization}) was added for Cycle\,8 (2021\,October) to correct high spectral resolution ALMA
visibility data acquired on targets exhibiting strong celestial line emission in their autocorrelation
spectra. The calibration heuristics employed in the low signal to noise regime were improved for Cycle\,9 (2022\,October).

In this paper, we describe the heuristics of the ALMA interferometric data processing pipeline as of the 2022.2.0.64 release for ALMA Cycle 9 (2022 October) based on CASA 6.4.1.12 Python 3.6.  The heuristics encompass a collection of strategies, rules, and algorithms designed and refined over many years to effectively accommodate the broad range of ALMA data covering a factor of $>$10 in frequency and $>$1000 in angular resolution. An overview of the pipeline framework is provided in \S\ref{framework}. The heuristics for the calibration stages, visibility flagging stages, and imaging stages are described in \S\ref{calsection}, \S\ref{visibilityFlagging}, and \S\ref{imgsection}, respectively.  In \S\ref{findcont}, we describe the algorithm for identifying channels to use for continuum.   The heuristics for quality assessment (QA) scoring are described in \S\ref{QAscores}. The pipeline validation process is described in \S\ref{validation}. Finally, in \S\ref{future} we briefly describe future development plans including the addition of a self-calibration stage to the imaging pipeline, the completion of full polarization calibration, and the ability to perform combined imaging of calibrated data from multiple configurations. A glossary of acronyms is provided in Table~\ref{glossary} in Appendix~\ref{appendixA}. 



\section{Overview and nomenclature }
\label{framework}

\subsection{Framework}

\begin{figure*}  
\centering
\includegraphics[scale=0.77]{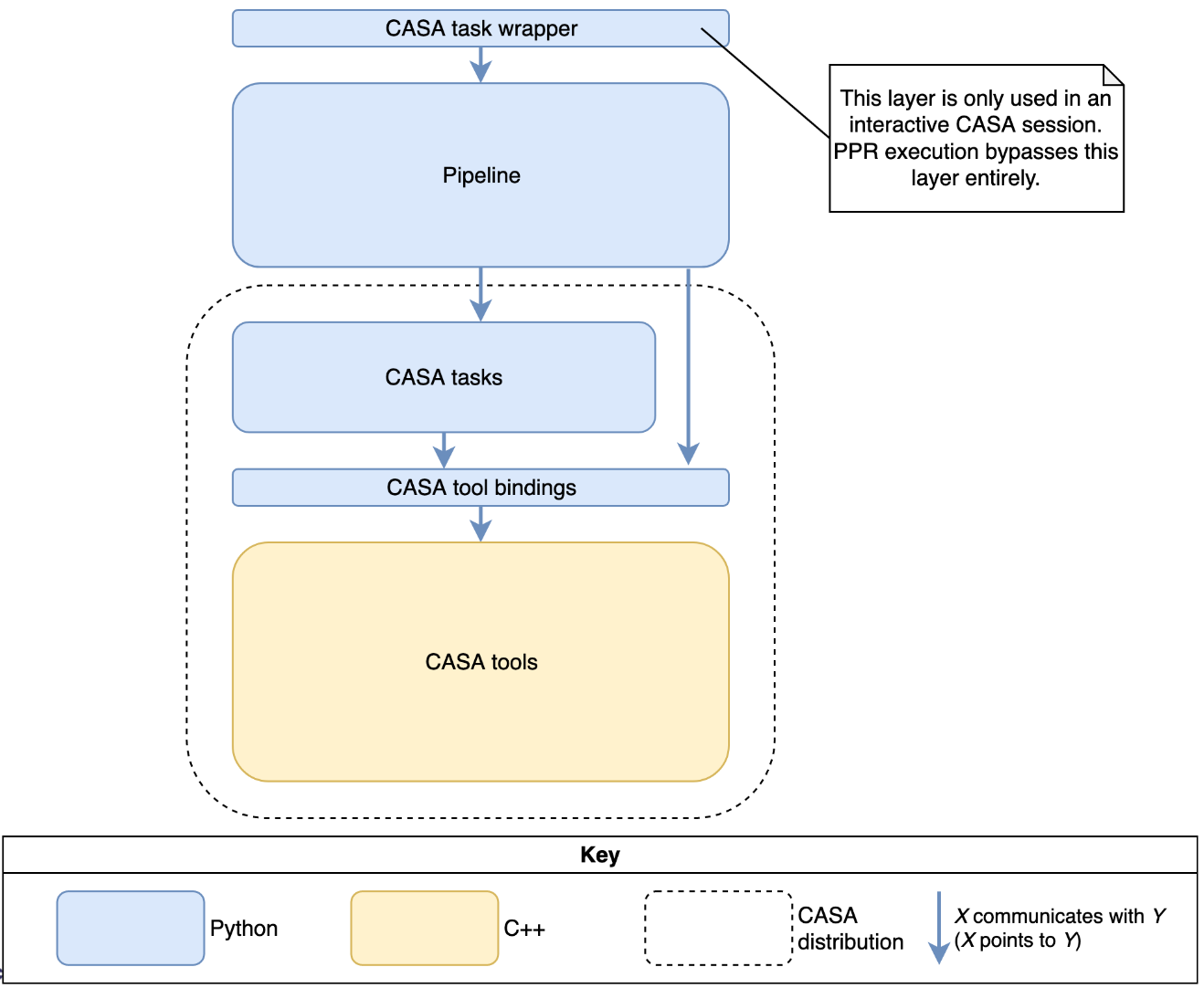}
\caption{Diagram showing the layering of the pipeline on top of CASA tasks and tools. In the 2023 release, the CASA task wrapper will no longer be used by the pipeline for interactive CASA sessions, in favor of a simpler, easier to maintain python interface.}
\label{fig:layering}
\end{figure*}

The ALMA pipeline software framework \citep{Davis15} consists of modular calibration and imaging tasks \citep{Geers19,Muders2014} written in Python \citep{python3} and built on top of the Common Astronomy Software Applications (CASA) data reduction package \citep{CASA2022}. 
CASA provides tasks and tools primarily written in \CC\/ and an IPython \citep{IPython} interface (see Figure~\ref{fig:layering}). A pipeline task may call other pipeline tasks or CASA tasks, and may utilize CASA tools and other Python modules from third party packages or packages written specifically for the pipeline. A pipeline stage is a pipeline task with a well-defined function, interface, WebLog page (\S\ref{Products}), and default behavior. During execution, stages share a common context object with all other stages.  This object maintains the pipeline state and is updated at the completion of each stage and written to disk upon pipeline completion.  Pipeline tasks also generate a results object, parts of which may be merged to the context for later use by other tasks.   The pipeline framework allows modifications to extend support for other similar observatories, as evidenced by the recently-deployed pipeline to process data from the 45\,m telescope of the Nobeyama Radio Observatory \citep[NRO\,45\,m,][]{Nakazato2022}.

An important component of the pipeline context is the pipeline calibration library (``CalLibrary''), which tracks the state of what calibration tables are to be (pre-)applied to each MeasurementSet (MS), and what calibrations have already been applied via the pipeline's ``applycal'' task. This object is distinct from CASA's ``callibrary'' mechanism, which provides a means of expressing calibration application instructions\footnote{\url{https://casadocs.readthedocs.io/en/stable/notebooks/cal_library_syntax.html}}.
Pipeline calibration tasks may register new calibration tables to the CalLibrary as so-called ``CalApplications'', where each CalApplication is composed of a ``CalFrom'' representing the calibration table and the CASA arguments that should be used when (pre-)applying that calibration table, and a ``CalTo'' representing the target data selection to which this calibration may be applied. The CalLibrary is implemented as an interval tree, where each CalApplication is registered with its target data selection represented as an interval. This allows for efficient retrieval of all CalApplications that should be applied when invoking the CASA task \ccmd{applycal} or pre-applied when invoking the CASA task \ccmd{gaincal}.

\subsection{Recipes}
\label{Recipes}
The pipeline stages are called in a specific order based on standard prescriptions known as recipes, which meet the processing requirements of ALMA's standard interferometric and single dish observing modes.  
The pipeline recipes are defined in text files written in the Extensible Markup Language (XML).  These XML files contain the sessions and ASDMs to be processed along with the sequence of pipeline stages each triggering the execution of a single pipeline command.  The interferometric observing modes are described in \S\ref{obsmodes}, for which the primary recipes are: calibration (\recipe{hifa_cal}), imaging (\recipe{hifa_image}), and calibration+imaging (\recipe{hifa_calimage}). 
Table~\ref{tab:calRecipe} contains the ordered list of pipeline tasks for the interferometric calibration recipe for the 2022 release (the corresponding recipe for imaging is presented in Table~\ref{tab:imgRecipe} of \S\ref{imgsection}).  
A set of analogous recipes are defined for full polarization MOUS and calibrator survey MOUS, as these datasets require additional calibration steps and/or control parameters.
The selected recipe forms the template for the pipeline processing request (PPR) XML file, which provides the primary mechanism by which CASA executes the pipeline \citep{Davis2012}.  
Alternatively, the pipeline can be executed via a python script containing an appropriate sequence of pipeline commands, known as the \file{casa_pipescript.py}.  Beginning in \S\ref{importdata}, we describe the various pipeline stages.

\begin{table}[ht]
\caption{The stage names and stage numbers of the standard calibration recipe of the 2022 release, which includes continuum imaging of the calibrators (cals) and \checksources\/ (chksrc).  Modifications of the recipe will result in different stage numbers, as typically happens with each new release. The corresponding standard imaging recipe is in Table~\ref{tab:imgRecipe}.\label{tab:calRecipe}}
\hspace*{-1cm}
\begin{tabular}{llll}
\textbf{\#} & \textbf{Task} & \textbf{\#} & \textbf{Task}\\
\hline
1 & \pcmd{hifa_importdata}  & 15 & \pcmd{hifa_gfluxscaleflag} \\
2 & \pcmd{hifa_flagdata}    & 16 & \pcmd{hifa_gfluxscale} \\
3 & \pcmd{hifa_fluxcalflag} & 17 &\pcmd{hifa_timegaincal} \\
4 & \pcmd{hif_rawflagchans} & 18 &\pcmd{hifa_targetflag} \\
5 & \pcmd{hif_refant}       & 19 & \pcmd{hif_applycal} \\
6 & \pcmd{h_tsyscal}        & 20 & \pcmd{hif_makeimlist} (cals) \\
7 & \pcmd{hifa_tsysflag}    & 21 & \pcmd{hif_makeimages} (cals) \\
8 & \pcmd{hifa_antpos}      & 22 & \pcmd{hif_makeimlist} (chksrc) \\
9 & \pcmd{hifa_wvrgcalflag} & 23 & \pcmd{hif_makeimages} (chksrc) \\
10 & \pcmd{hif_lowgainflag} & 24 & \pcmd{hifa_imageprecheck} \\
11 & \pcmd{hif_setmodels}   & 25 & \pcmd{hif_checkproductsize} \\
12 & \pcmd{hifa_bandpassflag}& 26 & \pcmd{hifa_renorm} \\
13 & \pcmd{hifa_bandpass}   & 27 & \pcmd{hifa_exportdata} \\
14 & \pcmd{hifa_spwphaseup} & & \\
\end{tabular}
\end{table}

\subsection{Input data}
\label{InputData}
\begin{figure*}   
\centering
\includegraphics[scale=0.77]{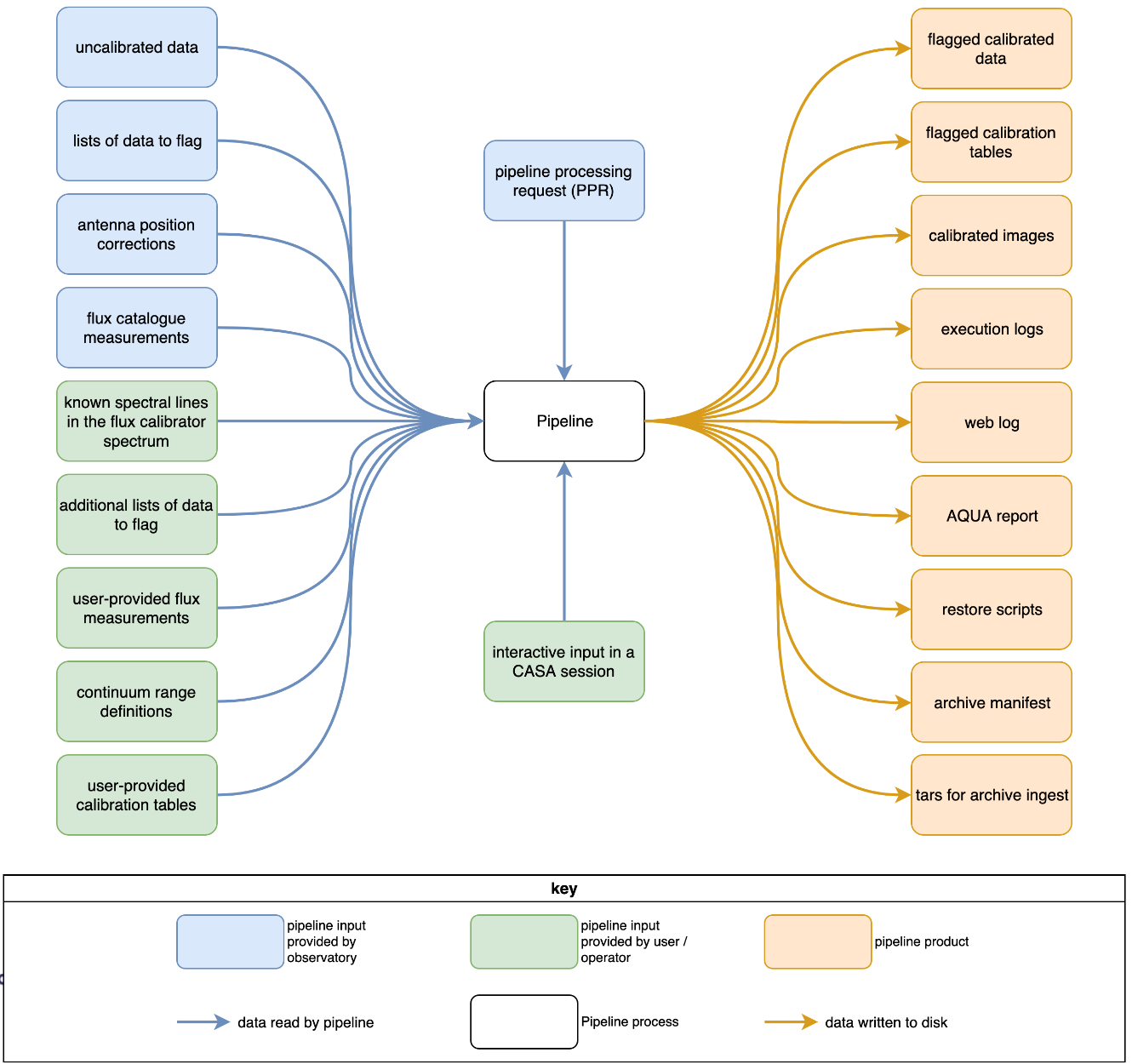}
\caption{Diagram showing the inputs and outputs of the pipeline. Green boxes indicate optional inputs that can be provided by the user of the pipeline in the rare case that it is necessary to override or augment the fully automated heuristics.  }
\label{fig:IO}
\end{figure*}

The ALMA data input to the pipeline consists of one or more successful, quality-assured observational executions of a single Scheduling Block (SB), which have passed initial quality assurance \citep[QA0, see chapter 11 of the ALMA Technical Handbook, updated\footnote{The most recent version is always available at \url{https://almascience.org/proposing/technical-handbook}} for each observing cycle,][]{Remijan2019}.
An individual SB execution, also referred to as an Execution Block (EB), results in an independent dataset referred to as an ALMA Science Data Model \citep[ASDM,][]{Viallefond2006}, which contains metadata and binary data. The collection of ASDMs (EBs) needed to reach the requested sensitivity level of the SB's science goal are held in a data structure called a Member Observing Unit Set (MOUS), which is the typical data unit on which the pipeline operates \citep{Davis09}, but the minimum data unit is a single EB.  As shown in Figure~\ref{fig:IO}, additional information is obtained from the ALMA observatory at the time of processing, via a web interface for updated calibrator flux density measurements and a revision-controlled text file containing the latest antenna position measurements.

\subsection{Products}
\label{Products}
As shown in Figure~\ref{fig:IO}, the pipeline produces the following four categories of products: (1) calibration products for each ASDM (including calibration and flagging tables), (2) imaging products (FITS images) made from the joint combination of all ASDMs (although not necessarily for all science targets, see \S\ref{mitigation}), (3) informative logs, reports, and reprocessing scripts, and (4) a set of nested JavaScript-enabled \citep{javascript} webpages containing diagnostic messages, tables, plots, images, and Quality Assessment (QA) scores, collectively known as the ``WebLog'' \citep{Masters2020}.
These products are reviewed as part
of the ALMA QA process \citep{Chavan16}, and, if satisfactory, are stored into the ALMA Science Archive \citep{Felix2014,Felix2012,Santander2012} and delivered to users. 

\subsection{Nomenclature and abbreviations}
\label{nomenclature}
In the rest of the paper, CASA task names, pipeline task names, and Python functions are written in bold monospace font (e.g., \ccmd{flagdata}) while task parameters are written in italics (e.g., \param{robust}) and string parameter values are in single quotes (e.g., \parvalue{auto}).   Pipeline recipes and task names begin with a prefix indicating its scope, either \ccmd{h_}, \ccmd{hif_}, or \ccmd{hifa_}, depending on whether the task is shared between the interferometric and single dish pipelines, is shared by only the ALMA interferometric pipeline and the Karl G. Jansky Very Large Array (VLA) pipeline \citep{Kent2020}, or is specific to the ALMA interferometric pipeline, respectively.  The rest of the name indicates the primary action of the task.  Names of text files that pass information to pipeline tasks are written in bold proportional space font (e.g., \file{antennapos.csv}). Scan intents and and data column names in the CASA MS are written in capital letters. 

The following non-acronym abbreviations are used for frequently-referenced terms:
spectral window (spw), reference antenna (refant), and calibration table (caltable).  The following expressions are defined
for frequently-referenced frequency quantities: the user-specified bandwidth for sensitivity (\bwForSensitivity), spw bandwidth (\spwbw), spw channel width (\spwchanwidth) in the topocentric frame, and cube channel width (\cubechanwidth) typically defined in the LSRK (kinematic Local Standard of Rest) frame \citep{Gordon76}.
Throughout this document, unless otherwise noted, the quantity termed MAD refers to the median absolute deviation from the median multiplied by 1/0.6745 to emulate the standard deviation of a noise signal with Gaussian statistics.  In equations, the MAD is written as \texttt{MAD} and the NumPy \texttt{median} function is abbreviated as \median\/ for brevity.

\section{Calibration Heuristics} 
\label{calsection}

One of the main products of the pipeline is calibrated visibilities, which have been generally accepted by the user community to be reliable and, as a result, form the input of several customized imaging pipelines developed for ALMA large programs \citep[e.g., PHANGS-ALMA and MAPS,][]{phangs2021,maps2021}.   This achievement demonstrates that the pipeline has established provenance in this area, which is essential to instill confidence in the scientific data products, ensure reproducibility, and spur further innovation \citep{Johnson2021}.
In this section, we describe the fundamental calibration strategy and heuristics.   Although one of the important final steps of the calibration pipeline is per-spw continuum imaging of the calibrators (and CHECK sources), discussion of this topic is reserved until \S\ref{calibratorImaging} (and \S\ref{checksource}), where it appears in the context of the imaging pipeline heuristics.

\subsection{Required scan intents}
\label{requiredintents}

In order to calibrate ALMA data via the canonical interferometric approach of instrumental gain measurement on point-like calibrators with subsequent transfer to targets of interest, the pipeline relies on the scan intent labels originating in the ALMA Observing Tool 
\citep[OT,][]{Liszt14,WilliamsBridger13} and attached to each scan 
during the execution of the SB by the DataCapturer 
module of the control system \citep{Lorente2012,Hafok2006}. A minimum set of scan intents (each denoted by a string in capital letters) is required for successful pipeline processing as indicated in Table~\ref{tab:intents}.  
For MOUS that contain science target spws with full-polarization correlation products, CALIBRATE\_POLARIZATION scans are required. An optional intent of OBSERVE\_CHECK\_SOURCE is included in some SBs, primarily those with angular resolution $<0.25''$, to provide a measure of the phase transfer quality and astrometric accuracy by observing an extra calibrator, termed a ``checksource'', to be calibrated as if it was another science target.  Standard SBs do not contain a delay calibration or focus calibration intent because instrumental delays and focus settings are verified prior to the execution of SBs and, if necessary, remeasured and updated in the telescope monitor and control database (TMCDB).
Hereafter, the intent names in Table~\ref{tab:intents} will be referred to by the shorthand text in capital letters that is used by the pipeline, which corresponds to the text following the first underscore but preceding the second underscore (if any), e.g., ATMOSPHERE, TARGET, CHECK, etc.

\begin{table*}
\caption{The observing intents in ALMA observations, their purpose, and their usage in the pipeline.\label{tab:intents}}
{\centering
\begin{tabular}{lll}
\textbf{Intent} & \textbf{Purpose} & \textbf{Pipeline usage} \\
\hline
CALIBRATE\_ATMOSPHERE & Measure \trx\/ and \tsys  & flagged and ignored\tablenotemark{a} \\
CALIBRATE\_BANDPASS & Measure bandpass shape & required\\
CALIBRATE\_DELAY\tablenotemark{b,c} & Measure delay corrections & 
OBSERVE\_CHECK\_SOURCE\\
CALIBRATE\_FLUX\tablenotemark{d} & Measure flux scale & required\\
CALIBRATE\_FOCUS\tablenotemark{b} & Measure local focus corrections & flagged and ignored\\
CALIBRATE\_POLARIZATION & Measure $D$ terms & required (in polarization recipe)\\
CALIBRATE\_POINTING & Measure local pointing corrections & flagged and ignored\\
CALIBRATE\_SIDEBAND\_RATIO\tablenotemark{b} & Measure sideband ratio & ignored\\
OBSERVE\_TARGET & Measure science target & required\\
OBSERVE\_CHECK\_SOURCE & Assess phase transfer quality & optional (see \S~\ref{requiredintents})\\
\hline
\end{tabular}
}
\tablenotetext{a}{The \tsys\/ values from the TelCal-generated SYSCAL caltable are transferred to \tsys\/ calibration tables.}
\tablenotetext{b}{Engineering intent that is not included in science SBs.}
\tablenotetext{c}{In early observing Cycles, the CALIBRATE\_DELAY intent label was temporarily used to indicate a scan on a ``checksource'' and is still interpreted as such in the pipeline.} 
\tablenotetext{d}{This intent was called CALIBRATE\_AMPLITUDE in early Cycles, so they are treated as synonyms.}
\end{table*}

\subsection{Observing modes supported}
\label{obsmodes}

\begin{figure*}  
\centering
\hspace*{-1mm}
\includegraphics[scale=0.32]{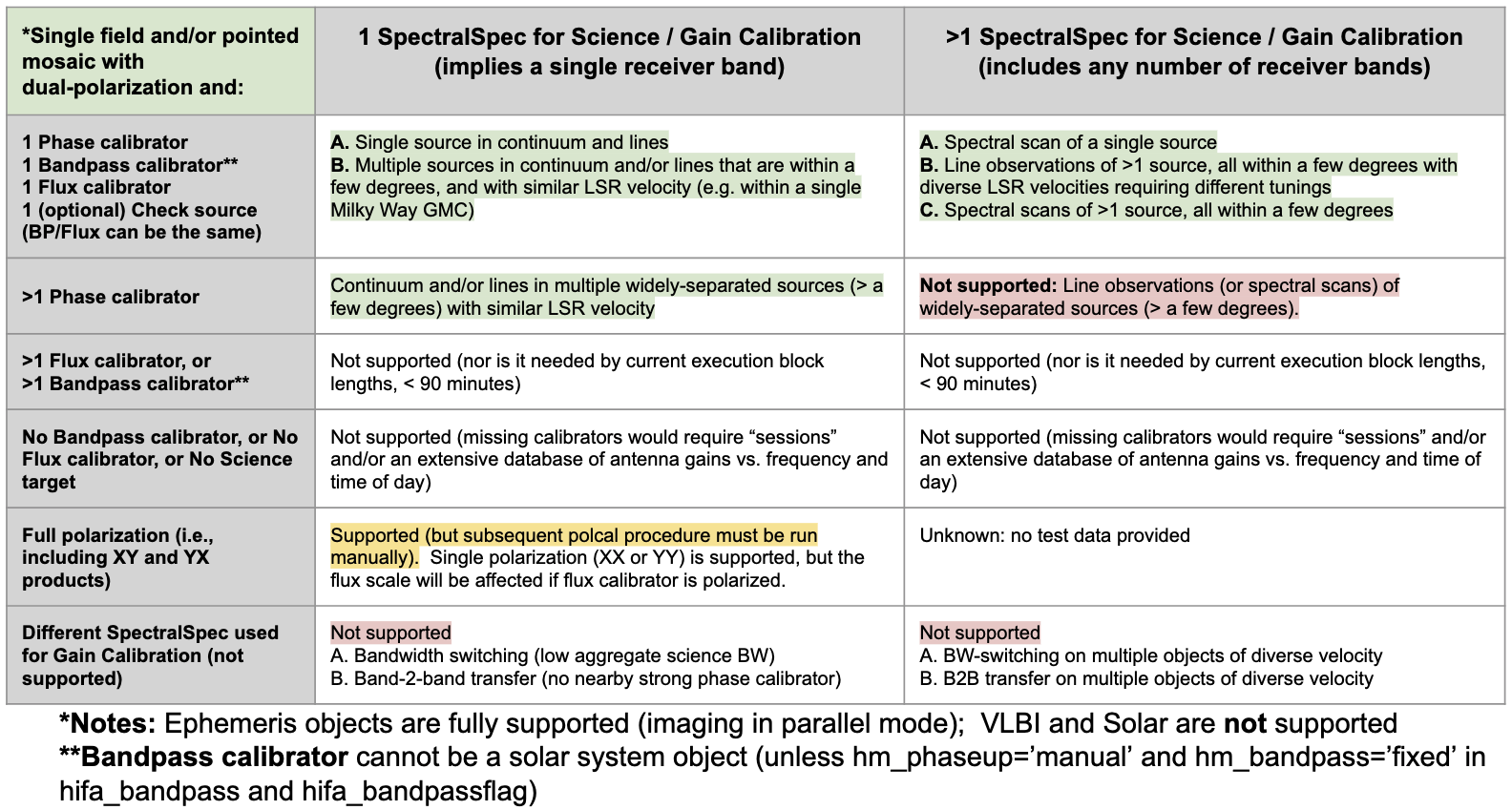}
\caption{Matrix summarizing which of ALMA's observing modes are currently supported by the ALMA pipeline releases for Cycle 8 (2021) and Cycle 9 (2022).
Modes with no text highlighted in green, yellow, or red are not expected to be attempted in normal science observations.
}
\label{fig:obsmode}
\end{figure*}

From the ALMA user’s perspective, an observing mode provides a unique science capability, requires an SB of a specific format, and produces a specific set of data products.  From the observatory's perspective, an observing 
mode comprises a set of rules for the cadence and duration of the observation (calibrators and science targets) and for the contents of the resulting ASDM that every subsystem must follow in order for the data product to be successful.
The observing modes supported (and some of those not supported) by the pipeline are summarized in the observing mode matrix in Figure~\ref{fig:obsmode}.  All science target spws must be observed toward appropriate calibrators with the aforementioned calibration intents, which means that the more complicated calibration methods of bandwidth switching and band-to-band (B2B) transfer under development \citep[e.g.,][]{Maud20} are not (yet) supported, as indicated in the final row of the matrix in Figure~\ref{fig:obsmode}.

\subsubsection{Types of spws supported}

The two types of science data spws supported by the pipeline correspond to the major operating modes of the BLC: time division multiplexing (TDM) spws and frequency division multiplexing (FDM) spws.   In TDM mode, the correlator functions with an XF architecture and the least significant bit of each 3-bit digitized intermediate frequency (IF) baseband signal is dropped, making the data fundamentally 2-bit in nature. 
Spectral windows have 128 channels per 2\,GHz baseband in dual-polarization mode. In FDM mode, the 3-bit digitized IF signals first enter the digital tunable filter banks (TFBs) for processing in (up to 32) 62.5\,MHz subbands per baseband, yielding an FXF architecture in which the output data are resampled to 2-bits just before correlation \citep{Baudry2012}.  
From the user's perspective, the primary difference between these modes is that FDM spws support higher frequency resolution and larger numbers of spectral channels (up to 3840).  Further details on the possible spectral setups can be found in 
\citet{Remijan2019}. The ACAC, having an FX architecture (with 4-bit correlation), has no such distinction in spw type, but for consistency in appearance in the WebLog, the pipeline labels ACAC spws with TDM-like characteristics as TDM.

\subsubsection{Observation content requirements}

In addition to the presence of all required scan intents (\S\ref{requiredintents}), the pipeline requires that there be a limit to the number of calibrator targets of a given
type, which is generally one per intent.  There can be multiple phase calibrators as long as each one is observed in all science spws and there is only one science spectral tuning specification (``SpectralSpec'').  This limitation precludes using a different phase calibrator for different groups of objects that have a substantially different systemic velocity relative to the narrowest \spwbw\/ as those require different ``SpectralSpecs''.  Instead, such observations must be split into separate MOUS.  

Another limitation is on which scan intents can be shared by an individual object and/or scan -- currently only BANDPASS and FLUX intents can be combined into the same scan.  Finally, although ALMA EBs currently observe science targets in only one receiver band, this decision is not fundamentally driven by the pipeline.  The pipeline will run successfully on multi-band data as long as the FLUX target observed is the same in all spws of all bands. However, to better support such multi-band data in the future, some minor improvements to the WebLog would be helpful.  

\subsubsection{ASDM content requirements}
There are two major requirements on ASDM content. First, approximate flux density values for all calibrators, except solar system objects (SSOs), must be written by the online control system into the ASDM Source.xml. These flux densities are used by several subsequent heuristics: to set the visibility model of the FLUX calibrator (\S\ref{setmodels}), to set the frequency interval of the bandpass solution on the BANDPASS calibrator (\S\ref{bandpass}), and to choose the spw mapping scheme for each PHASE calibrator and CHECK source (\S\ref{spwmapping}). While all of these values will be updated to the best available value at the time of pipeline execution (see \S\ref{importdata}), they must be initially populated in order to be updated.  Second, the names attached to specific spws must be consistent across different EBs of the same SB, even though the order in which their metadata is recorded in the SpectralWindow.xml file (which sets the spw number) is free to vary between EBs (\S\ref{importdata}).  

\subsubsection{Exclusion of specialized modes}

The pipeline does not process the less common interferometric observing modes, including solar observing and VLBI observing, for which separate analysis processes have been developed \citep[see, e.g.,][]{Bastian22,Goddi19,Blackburn19}.  Other calibration schemes still under development are also not supported by the pipeline, such as bandwidth switching and band-to-band phase transfer \citep{Maud20}, along with other observing modes that have not yet completed the observatory's commissioning process \citep[ObsMode,][]{Takahashi21}.  

\subsubsection{Partial support for full polarization data}

Full polarization datasets are processed through the calibration pipeline using a consistent reference antenna (\S~\ref{refantpolcal}) to calibrate the parallel-hand products within each multi-EB observing session, and all the normal Stokes\,I images are produced by the imaging pipeline. If a POLARIZATION scan intent is present in the data, then the total parallactic angle coverage of the polarization calibrator observed across all EBs of each polarization observing session is shown graphically in the WebLog, and reported quantitatively.  The minimum threshold is $60^\circ$, below which a warning is generated.  Also, special consideration is applied for edge channel flagging (\S~\ref{edgechannelflagging}).
However, additional manual steps are currently needed to run the CASA \ccmd{polfromgain} and \ccmd{polcal} tasks to calibrate the cross-hand products and subsequently produce full Stokes images, but future pipeline support is planned (\S\ref{futurePolcal}).

\subsection{Data import}
\label{importdata}
In the first stage of the pipeline, \pcmd{hifa_importdata}, each ASDM of the MOUS is imported into a CASA MS \citep{MS2015}.  All subsequent processing operates on these MS. By default, the import process fills the visibility data into the DATA column of the MS and the associated metadata in the form of subtables of the MS. It also populates the SIGMA column of the MS using the following relation based on the radiometer equation which describes how the uncertainty in the strength of a radio frequency (RF) signal declines as the number of independent samples increases \citep{Hunter2015}:
\begin{equation}
\label{sigmaColumnInitialized}
\sigma_{\rm s,v} = 1 / \sqrt{2\Delta\nu_{s,chan}\Delta{t_v}}
\end{equation}
where $\Delta\nu_{s,chan}$ is the channel width of spw $s$, $\Delta{t_v}$ is the integration time of visibility $v$, and the factor of 2 accounts for the cross correlation of two independent RF signals.  The associated WEIGHT column is then initialized to 1/$\sigma_{\rm s,v}^2$. Ultimately these weights are used in the imaging process, and to estimate the theoretical noise level in an image (\S\ref{sensitivity}).

In each MS, the spws are numbered by integer values beginning with zero, which correspond to the row numbers in the SPECTRAL\_WINDOW subtable of the MS.  Each spw also has a corresponding string name, which is gathered from the SB and written to the ASDM by the ALMA control system \citep{Marson16}.  As the numerical ordering of the spws are not guaranteed to be the same in successive EBs, the pipeline uses the spw names to match the same spw across EBs within an MOUS, then labels them with a common virtual spw number (equal to the number in the first imported EB). Another important subtable loaded into the MS is the SYSCAL subtable, which contains the system temperature (\tsys) measurements computed online by the TelCal software module from the autocorrelation data in the ATMOSPHERE scans and written to the ASDM.  The \file{SBSummary.xml} file of the ASDM
is also read into the ASDM\_SBSUMMARY table to record the SB name, the angular resolution requirement, and the representative target, frequency, and spw name.

After importing the data and metadata, for each non-SSO calibrator, if \param{dbservice}=True (the default value), then the pipeline queries the ALMA observatory's calibrator web service to retrieve the best estimate of the flux density and spectral index at the center frequency of each spw for the date of observation of each EB.  For FLUX calibrators, these values establish the overall fluxscale (\S\ref{setmodels}), while for PHASE and CHECK sources, the values are used to predict their SNR in each spw and choose their calibration strategy accordingly (\S\ref{lowsnrheuristics}). If the service is not available, and the backup service also does not respond, then the flux density values used in later pipeline stages are the older estimates written in the Source.xml file of the ASDM by the online system at the time of observation. In either case, the flux density and spectral index information is also recorded in a text file \file{flux.csv}, which can be manually adjusted if desired.

The \pcmd{hifa_importdata} task can alternatively load data directly from an MS, which allows users to define new recipes, such as imaging recipes that use data that may have been calibrated manually.  In this use case, the \param{datacolumns} parameter allows the user to specify what type of data (raw, calibrated, continuum-subtracted, etc.) is contained in the DATA and CORRECTED columns.

\subsection{Earth rotation parameter check}

The date of each EB is compared to the date range of the tables in the CASA data repository originating from the International Earth Rotation and Reference System Service \citep[IERS,][]{IERS2004}, which can be set up for periodic updates in CASA installations.  If the date of any EB is not fully covered by the IERSeop2000 table, then CASA will use the IERSpredict table instead, and a warning will be written under that EB on the main WebLog page.  If the date of an EB is beyond the final entry in the IERSpredict table, then the following warning message is generated:
``MS dates not fully covered by IERSpredict. Please update your data repository.''

Coordinated Universal Time (UTC) is adjusted from International Atomic Time (TAI) by the addition of occasional leap seconds to account for the difference between the definition of the second and the rotation of Earth \citep{McCarthy2008}. In the CASA data repository,  the TAI\_UTC table tracks the date of each leap second. Recently, these corrections have only been added every few years, and the next one might need to be negative, which has never happened. Regardless of whether a leap second was needed, CASA checks the VS\_DATE in this table and if it is too old relative to the standard dates when leap seconds are introduced (July 1 / Jan 1), then it becomes concerned that there might have been a recent leap second unaccounted for and it emits this message: ``Leap second table TAI\_UTC seems out-of-date.''  However, it is only a problem if there was a leap second. Since there have not been any lately, and the leap second is being phased out \citep{leapSecond2022}, the pipeline is not currently checking this condition or propagating any warning into the weblog (but the CASA message will still appear in the CASA log).

\subsection{Reference Antenna Selection}

The antenna-based calibration solves performed by the tasks \ccmd{gaincal} and \ccmd{bandpass} in subsequent stages are always computed with respect to a reference antenna, selected from a prioritized list generated automatically 
in the \pcmd{hif_refant} stage 
using the following heuristics.

\subsubsection{Antenna scoring}
\label{antennaScoring}
In \pcmd{hif_refant}, the prioritized list of preferred reference antennas (refants) is calculated on a per-EB basis, with preference given to antennas closest to the center of the array and those with a low flagging fraction. The following antenna-based score ($S_i$) is comprised of a distance term plus a flagging term:

\begin{equation}
\begin{aligned}
S_{\rm i} & = n_{\rm ant}(1-d_{\rm i}/\texttt{max}(d)) + n_{\rm ant}v_{\rm i}/\texttt{max}(v),
\end{aligned}
\end{equation}
where $n_{\rm ant}$ is the number of antennas, $d_{\rm i}$ is the distance from antenna $i$ to the center of the array defined by the median values of the list of antenna latitudes and longitudes, and $v_{\rm i}$ is the number of unflagged visibilities including antenna $i$.  The distance term will be zero for the most distant antenna and as high as $n_{\rm ant}$ if the nearest antenna coincides with the median location, while the flagging term will be zero if all data are flagged and $n_{\rm ant}$ if no data are flagged.  The two components can be individually disabled via the input parameters \param{geometry} and \param{flagging}.
The data that contribute to $v_{\rm i}$ are all scans with primary calibration intent (BANDPASS, FLUX, PHASE, POLARIZATION), but only from the science spws, which are defined as those spws that are observed with science target intent at some point in the MS. Note that $v_{\rm i}$ is based on channel count, so spws with more channels count more than others, which can be problematic when a single spw dominates the channel count and an spw with low channel count is flagged.   A future improvement will employ equal weighting per spw in the calculation of $v_{\rm i}$.   

At the point in the recipe where \pcmd{hif_refant} is called (see Table~\ref{tab:calRecipe}), $v_{\rm i}$ accounts for deterministic flags (\S\ref{deterministicFlags}), SSO line flags (\S\ref{ssoLineFlags}), and channel-based flags of raw data (\S\ref{rawchanflags}), if any. Subsequent flagging stages that analyze the calibrated data column (CORRECTED) with respect to the source model  (\S\ref{correctedampflagtext}) will demote antennas to the end of the list if they receive significant additional flagging.  Antennas are removed from the list if all spws are flagged.  

\subsubsection{Manual adjustment option}

To circumvent the automatic heuristics, a single reference
antenna can be defined in the PPR by setting the \param{hm_refant} parameter to `manual' and \param{refant} to the antenna name, but this will apply to all EBs.  A future improvement is to allow \param{refant} to be a list of comma-delimited strings so that an independent refant sequence can be fully specified for each individual EB.

\subsubsection{Reference antenna in polarization sessions}
\label{refantpolcal}
In the full-polarization calibration recipe (\recipe{hifa_polcal}), a subsequent stage \pcmd{hifa_session_refant} selects a single refant for each entire session (for ALMA full-polarization observations, a session is often composed of 2-3 EBs). Antennas are first ranked by the product of their per-EB ranking originating in \pcmd{hifa_refant} (\S\ref{antennaScoring}). The task next performs a phase-only \ccmd{gaincal} with solution interval equal to a single integration time (\param{gaintype}=`p', \param{solint}=`int') on all scans with PHASE intent for each EB starting with the highest ranking antenna as sole refant, and checks the resulting caltable to see if the refant ever changes, which can happen if the solution fails on an individual integration. It chooses the first antenna that does not result in any change of refant. If none of the top 3 antennas qualify (which should be rare), then the antenna with the most solutions as refant is chosen, and a message is displayed with the number of phase outliers, where the word ``outlier'' is wherever the refant phase was non-zero, which indicates that a different refant was chosen for some integrations. The total number of possible solutions, $N_{\rm sol}$ is:
\begin{equation}
N_{\rm sol} = N_{\rm EBs} \times N_{\rm spws} \times N_{\rm integrations} \times N_{\rm pol}.
\end{equation}

The subsequent stage \pcmd{hifa_lock_refant} sets the refant to the single specified antenna and sets \param{refantmode}=``fixed'' for all subsequent calibration tasks. The refant can be ``unlocked'' with the \pcmd{hifa_unlock_refant} stage, but that stage is not needed in the standard polcal and polcalimage recipes. Note that in the polcal and polcalimage recipes, \pcmd{hifa_bandpass} and \pcmd{hifa_spwphaseup} are called a second time following the locking of the reference antenna, to ensure that the bandpass and spw-offsets calibration tables are using that fixed reference antenna. 

While the current algorithm has proven successful, it is not infallible.  If an antenna is fully flagged on only one of the calibration intents, such as can happen with shadowing, the antenna location aspect of the score can still dominate, particularly on the ACA.  If it happens on a full polarization dataset,
this antenna might get selected as the (single) refant, and then a subsequent pipeline stage (\pcmd{hifa_gfluxscale}, \S\ref{fluxscale}) will fail due to no valid solutions on that intent.  A further improvement to avoid this situation
is being implemented.

\subsection{Scaling by system temperature (\texorpdfstring{\tsys\/}{})}
\label{tsyscal}

In order to convert the amplitude scale of ALMA raw data to antenna temperature units and to correct for atmospheric spectral features, the visibility data are scaled by the ALMA \tsys.
The \tsys\/ as a function of frequency is calculated for both individual polarization signals at the time of observation by the TelCal process \citep{Telcal2011} using the baseband autocorrelation data from the ATMOSPHERE scans. These spectra are imported to a subtable of the MS by the \pcmd{hifa_importdata} stage. In \pcmd{hifa_tsyscal}, these spectra
are copied into a CASA calibration table by the \ccmd{gencal} task, which also sets flags in the caltable for all channels with zero or negative \tsys. \pcmd{hifa_tsyscal} subsequently generates spectral plots of \tsys\/ vs. frequency. Because the BLC has no mechanism to compute and apply a 3-bit quantization correction, all ATMOSPHERE scans on the BLC are observed with TDM spws in order to record these data  with the proper 2-bit linearity correction. As a result, these scans are recorded in different spws from the science spw when the science spw is FDM.  In contrast, the ACA correlator added the capability of observing \tsys\/ in high spectral resolution mode beginning in Cycle~3, as the proper 3-bit quantization correction is applied in this FX correlator. For these reasons, the WebLog indicates the mapping of \tsys\/ spws to science spws, both in a table and by labeling, when relevant, the frequency ranges of FDM science spws in the individual TDM \tsys\/ spectral plots generated by \ccmd{plotbandpass} with \param{showfdm}=True.

\subsection{Flagging of \texorpdfstring{\tsys\/}{} caltables}

The subsequent \pcmd{hifa_tsysflag} stage searches for erroneous
\tsys\/ measurements of several different kinds and applies flags to the \tsys\/ caltable.  
There are six separate flagging metrics, where each metric creates its own flagging view and has its own corresponding flagging rule(s). A flagging view is a two dimensional matrix of a specific statistic, with typical dimensions of antenna or baseline vs. time or frequency. The metrics are evaluated in the order set by the parameter \param{metric_order} (default order: nmedian, derivative, fieldshape, edgechans,  birdies, toomany).  Each one is explained in depth in the following sub-subsections.

The first three metrics are based on ``time vs. antenna'' matrix views, where some value is calculated for each (antenna, timestamp) point. These views are typically per spw, and per polarization. Data are selected for a given scan intent of interest (ATMOSPHERE). Under the assumption that there will only ever be a single \tsys\/ spectrum for one field per timestamp, this means that creating data points per timestamp effectively means the same as creating separate data points per field number.  Flags generated from these views are of the entire spectrum of the solution, not individual channels.

The flagged \tsys\/ tables produced by these stages are pre-applied to the data in all subsequent calibration solves with linear interpolation in time and frequency, including interpolation of any flagged channels.  As with all amplitude solution tables, they are applied by \ccmd{applycal} with \param{calwt}=True.  This \param{calwt} setting triggers the visibility WEIGHT column to be recalculated each time from the SIGMA column (which remains fixed after import, see \S\ref{importdata}) combined with the \tsys\/ values. Although the \tsys\/ tables are spectral in nature, the weight value used in \ccmd{applycal} for an spw is based on the mean \tsys\/ value of that spw.  The use of channelized weighting is a future improvement (\S\ref{WSU}).

\subsubsection{\texorpdfstring{\tsys\/}{} median heuristic}

This \tsys\/ flagging heuristic is designed to eliminate antenna/spw combinations with grossly different \tsys\/ magnitude from the rest, on a per-spw basis.
A separate view is generated for each polarization and each spw with axes {\textquotedbl}time{\textquotedbl} vs. {\textquotedbl}antenna{\textquotedbl}. Each point in the matrix is the median value of the \tsys\/ spectrum for that antenna/time.
The views are evaluated against the {\textquotedbl}nmedian{\textquotedbl} matrix flagging rule, where data points are identified as outliers if their value is larger than a threshold factor (\param{fnm_limit}: default=2.0) times the median of all non-flagged data points.

With the default setting of \param{fnm_byfield}=True, individual sources are evaluated separately in order to prevent elevation differences between targets from causing unnecessary flags, which mostly affects high frequency data. Flagging commands for the \tsys\/ table(s) are generated for each of the identified outlier data points.

\subsubsection{\texorpdfstring{\tsys\/}{} derivative heuristic}

This \tsys\/ flagging heuristic is designed to eliminate antenna spws which show a ringing pattern across the bandpass.
A separate view is generated for each polarization and each spw, with axes of time vs. antenna, and value $v$ as follows:
\begin{equation}
v_{\rm time,ant} = 100\times \text{MAD($\Delta$\tsys)}, 
\end{equation}
where $\Delta$\tsys\/ is the channel-to-channel difference (\ccmd{numpy.diff}) in \tsys\/ for that antenna/timestamp
considering all unflagged channels in the solution. The views are evaluated against the ``max abs'' matrix flagging rule, where data points are identified as outliers if the $v_{\rm time,ant}$ exceeds the threshold (\param{fd_max_limit}: default=5\,K\,channel$^{-1}$). Flagging commands for the \tsys\/ table(s) are generated for each of the identified outlier data points, each of which correspond to flagging the solution for a scan, not individual channels.

\subsubsection{\texorpdfstring{\tsys\/}{} fieldshape heuristic}

In a normally functioning receiver/IF system, 
changes in airmass will raise or lower the overall sensitivity, but the spectral shape of \tsys\/ vs. frequency will remain similar over time as long as the local oscillator (LO) tuning and IF attenuator settings remain identical. In early Cycles, stochastic changes in attenuator settings due to automatic, unconstrained re-optimization between scans
led to significant bandpass shape changes, which adversely impacts the calibration.  As a result, the online system now (mid-Cycle 1 onward) holds attenuator settings fixed after initial optimization on the BANDPASS calibrator at the beginning of an EB.  A second set of attenuator settings is determined for ATMOSPHERE scans, and the system switches between them accordingly as the EB progresses.  In order to detect unexpected instrumental changes to the spectral shape of \tsys, the fieldshape heuristic searches for deviations, $v_{\rm fld}$, computed as the mean absolute deviation of the median-normalized \tsys\/ spectrum from the median-normalized reference \tsys\/ spectrum of each spw in percentage units:  
\begin{equation}
v_{\rm fld} = 100 \times \overline{|T_{\rm sys}/
\median(T_{\rm sys}) - T_{\rm sys,ref}/\median(T_{\rm sys,ref})|}.
\end{equation}
The reference spectrum is the median of all \tsys\/ spectra recorded on the target with intent specified by the {\it ff\_refintent} parameter (default=BANDPASS).
The threshold for flagging is a deviation $>13$\%, which was
increased from 5\% prior to Cycle 7 in order to avoid excess flagging due to large variation of atmospheric transmission across an spw.  An improved method that adjusts the threshold based on the variation in transmission would restore the original sensitivity of this heuristic.

\subsubsection{Edge channels heuristic}
\label{edgechannelflagging}
Due to the baseband anti-aliasing filter ahead of the digitizers, the edge channels of BLC TDM spws have reduced sensitivity and hence the measured \tsys\/ value typically rolls sharply upward.  The effect is exacerbated if the spw is placed at the edge of the receiver RF band. The edge channels heuristic is designed to flag the solution in the least sensitive channels.
Data are selected for the usual scan intent (ATMOSPHERE) but also for BANDPASS and FLUX. The metric for this heuristic is based on a ``median'' \tsys\/ spectrum view, calculated from spectra for all antennas, both polarizations, and all fields / timestamps. Flags are generated on a per-channel, per-spw basis, but across all timestamps (i.e., all fields).

\subsubsection{\texorpdfstring{\tsys\/}{} birdies heuristic}
Narrow signals generated by the instrument (historically called ``birdies''), such as LO leakage from the water vapor radiometer (WVR) on some antennas, can produce spectral features in the \tsys\/ spectra, as they are based on autocorrelation data.  While these features can sometimes calibrate out in later stages, this heuristic was designed to flag them. The metric for this heuristic is based on a view per spw and per antenna. It first calculates, for a given spw, a median \tsys\/ spectrum based on all antennas, both polarizations, and all fields / timestamps. It also calculates, for a given spw, the median \tsys\/ spectrum for each individual antenna, based on both polarizations and all fields / timestamps. Each view is then calculated as the difference between the ``per antenna median'' -- ``median for all antennas''.  Flags are generated on a per spw, per antenna, per channel basis, but across all timestamps (i.e., all fields).

\subsubsection{\texorpdfstring{\tsys\/}{} ``too many'' heuristic}
The philosophy of this heuristic is to prevent the downstream creation of image products (\S\ref{imgsection}) for heavily-flagged spws with few surviving antennas because they will have drastically different properties from the rest of the spws and thus be useless for science purposes.  The metric for this heuristic first evaluates for each timestamp if too many antennas are flagged (set by \param{tmf1_limit}) and if so it flags the \tsys\/ solutions for all antennas for that timestamp.  Subsequently, it considers data across all timestamps (i.e., multiple fields). Any spws for which a large fraction ($>2/3$) of the antennas have their \tsys\/ solutions flagged in all timestamps due to prior heuristics will get completely flagged here.  

\subsection{Antenna position corrections}
\label{antposcorr}

Sometimes the positions of one or more antennas included in a particular observation are refined after the science data were recorded. The \pcmd{hifa_antpos} task searches the pipeline working directory for a text file (\file{antennapos.csv}), which contains a list of position corrections in the geocentric frame for a subset of antennas in each EB. The corrections are listed in the WebLog, and are incorporated into a calibration table for each EB by the CASA \ccmd{gencal} task.  
These tables are pre-applied in all subsequent calibration solves.  Currently, the origin of the text file is the \ccmd{correctMyAntennaPositions} function of the analysisUtils Python package \citep{aU} developed and maintained by ALMA science staff.  The package contains text files recording the date and time of every antenna move, and the subsequent geocentric position solutions derived from so-called ``baseline tracks'' in which numerous quasars observed across the sky \citep{Hunter2016}.  For each EB, the contents of these files are inspected to find all position measurements of the antenna during the continuous period that it resided on the pad it occupied in that EB. These positions are compared to the value stored in the ASDM and if the magnitude of the vector difference is larger than 50\,$\mu$m and more than 5$\sigma$ compared to the measurement uncertainty, it is added to a list of potential corrections.  The corrections are sorted by date, and the one with the latest date is written to the position corrections file.   The corrections are shown in two tables in the WebLog, one sorted by antenna name and one sorted by the vector total correction.  Values larger than \param{threshold} (default=1.0\,mm) are highlighted in bold in these tables.
In the future, the observatory plans to provide an antenna position service to remove this dependence on analysisUtils. 

\subsection{WVR corrections}
\label{wvrcorr}
Fluctuations in the water vapor content of the atmosphere give rise to path length fluctuations in astronomical signals which degrade observations particularly at longer baselines and higher frequencies \citep{Nikolic2013}. 
Radiometers are specialized receivers that measures
relative changes in sky brightness temperature versus time, either in a broadband continuum channel or in several narrow channels within an atmospheric spectral line \citep{Brogan2018}.
The front end receiver assembly of each 12\,m antenna contains a WVR that measures the total power brightness temperature in four double sideband filters offset from the 183\,GHz water line \citep{WVR2010}.  These WVR measurements of the water vapor emission are converted by the CASA task \ccmd{wvrgcal} into path lengths and subsequently into a differential phase correction table for each spw that can be applied to the science data \citep{Nikolic2012}.  The atmosphere is assumed to be non-dispersive (\param{disperse}=False), which is a good approximation to the few percent level for most ALMA frequencies (those not near a strong atmospheric absorption line).

In the stage \pcmd{hifa_wvrgcalflag} stage, after the phase correction table has been created, gain solutions are determined on the BANDPASS and PHASE calibration scans, both with and without the WVR correction pre-applied. After applying these solutions, the resulting antenna-based phase root mean square (RMS) is measured for both cases, and the improvement ratio is used to determine if the WVR correction helps overall. Note that in cases where the phase solutions are noisy (phase RMS $>$90$^{\circ}$) due to a low signal-to-noise ratio (SNR) for the PHASE calibrator scans, the improvement assessment is made only using the BANDPASS source. A poor ratio on an individual antenna, rather than the majority of antennas, can signify an instrumental problem with the WVR units, in which case the solution is discarded and \ccmd{wvrgcal} forms an interpolated solution as long as there are at least \param{minnumant} (default=2) within \param{maxdistm} (default=500\,m) of the WVR-flagged antenna). The plots shown under the flagging metrics heading of the WebLog show the ratios of the phase RMS with WVR corrections applied / without WVR corrections applied per antenna in greyscale (values $<$1 indicate an improvement). If an antenna is flagged during this stage, then plots of before flagging and after flagging are created. Note that for datasets with heterogeneous antenna diameters, there must be at least \param{ants_with_wvr_nr_thresh} (default=3) antennas with WVRs (i.e., 12\,m antennas) {\bf and} at least \param{ants_with_wvr_thresh} (default=0.2) fraction of antennas with WVRs (e.g., 3 out of 15, such as 3 PM antennas included with 12 CM antennas to increase the SNR of the calibration), otherwise this stage will not attempt to determine any WVR solutions. 

\begin{figure}   
\centering
\gridline{\fig{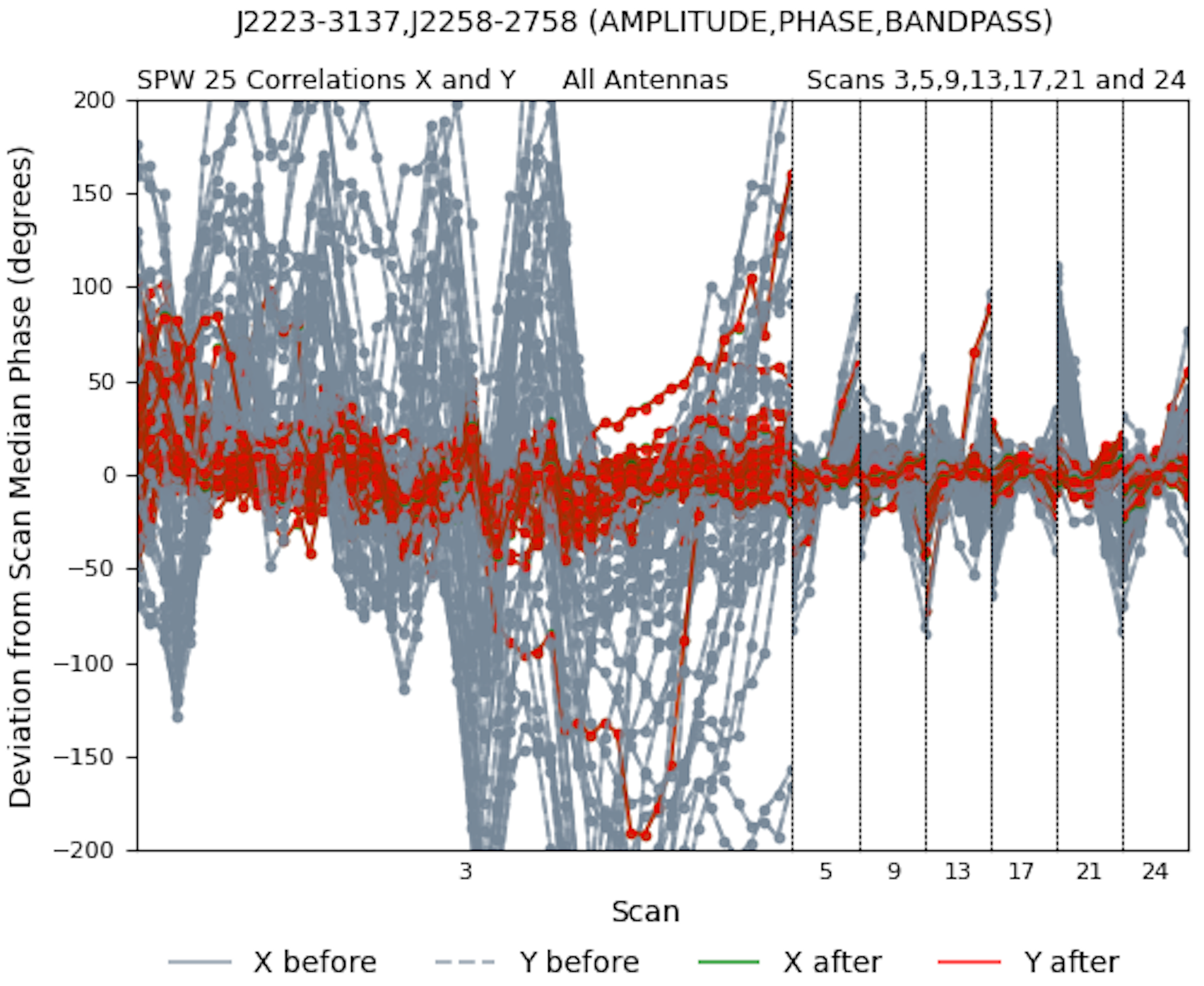}{0.47\textwidth}{(a)}}
\gridline{\fig{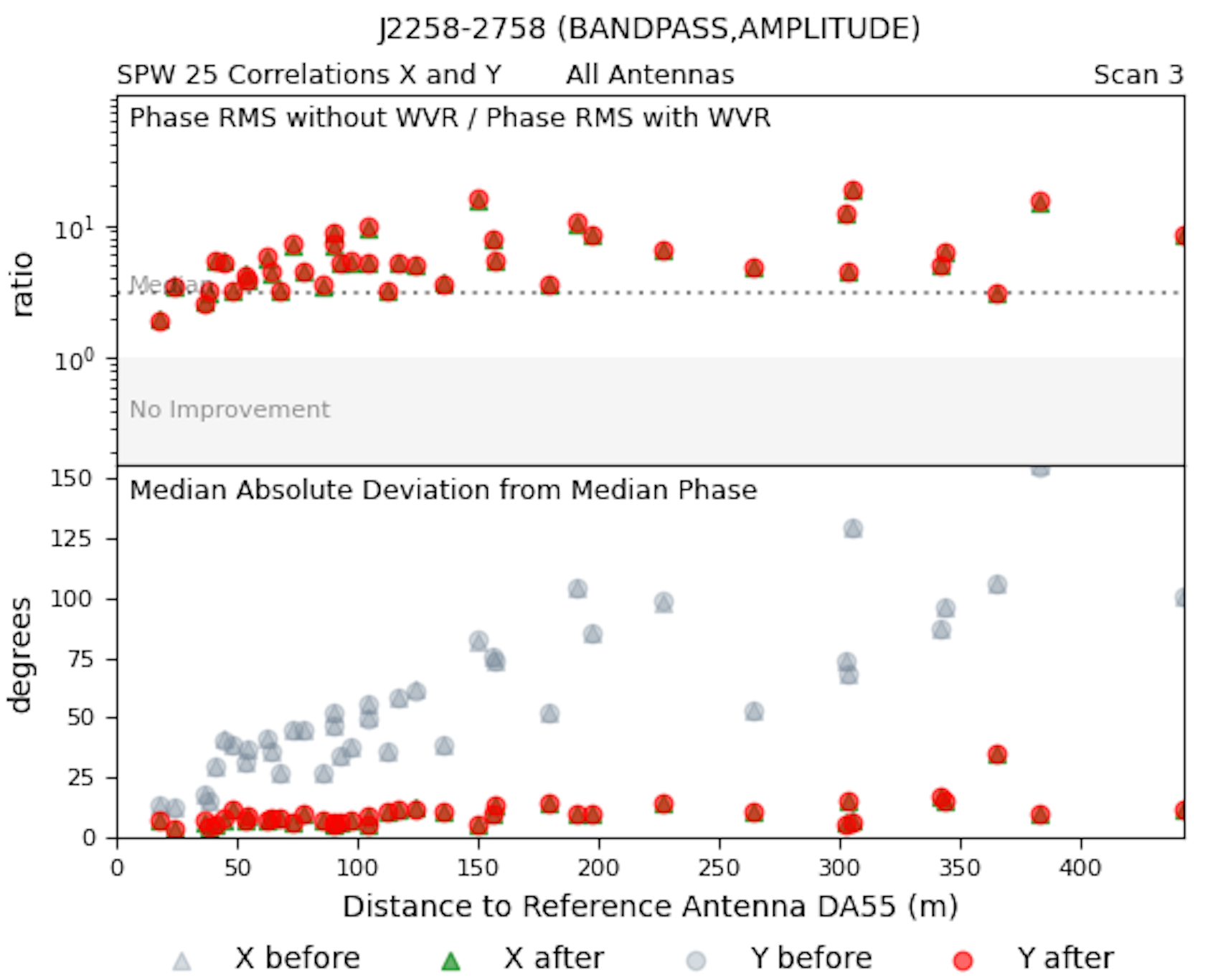}{0.47\textwidth}{(b)}}
\caption{Example of phase correction improvement factors established in the \pcmd{hifa_wvrgcalflag} stage. Panel (a) shows scan-based improvement factors per polarization (after correction in red and green, with before correction in gray), while the panel (b) indicates the improvement ratio and phase RMS in degrees as a function of antenna distance from the reference antenna for the BANDPASS scan. In both panels, the red points (Y polarization) overlay and typically obscure the green points (X polarization).}
\label{fig:hifa_wvrgcalflag_plots}
\end{figure}

There are two additional types of diagnostic plots to visualize the phase correction (Figure \ref{fig:hifa_wvrgcalflag_plots}): first, the deviation of the phase, per data integration, from the median phase is plotted for all calibrator scans as to highlight the reduction in phase spread; second, the ratio of improvement and the median deviation of the phase for each antenna as a function of distance from the reference antenna. Typically, improvement ratios $>$1 are expected and the median phase deviations should be visually reduced after WVR application. If the ratio is $<1$, then the WVR solution table is rejected and will not be added to the CalLibrary, so subsequent stages proceed without it. The per-antenna values of path length RMS (``RMS'') and channel-to-channel discrepancy (``Disc'') are also provided in the WebLog table.  Further information on the meaning of these columns can be found in \citep{Nikolic2012} and the CASA documentation. 

\begin{figure}  
\centering
\includegraphics[scale=0.5]{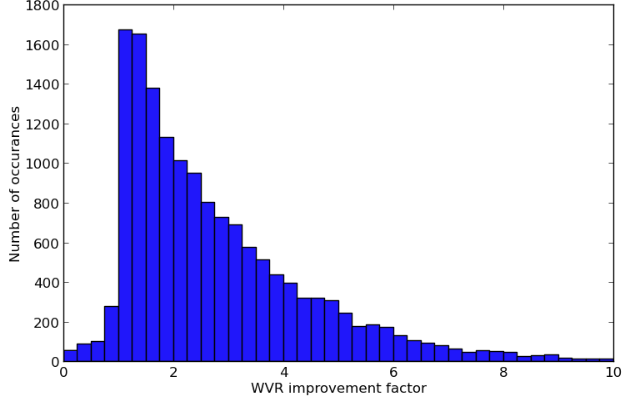}
\caption{Histogram of the WVR improvement factor for all science EBs obtained in Cycles 3 through 7 and passed QA requirements.}
\label{fig:wvrHistogram}
\end{figure}

While the WVR corrections do help reduce the phase RMS in the majority of cases, in some cases by a large factor (see the histogram in Figure~\ref{fig:wvrHistogram}, $>$90\,\% of data have an improvement $>$1, while $\sim$50\,\% have an improvement $>$2), there are cases where they do not, which typically happens in the presence of clouds where continuum emission from liquid water droplets (hydrosols) rather than water vapor dominates the WVR brightness temperature \citep{Maud23}.  In this case, the WVR data need to be fit first with a continuum component of emission, which can be done with an experimental program called \pcmd{remove_cloud} available in CASA by importing the \pcmd{casarecipes.remove_cloud} module\footnote{\url{https://help.almascience.org/kb/articles/what-is-remcloud-and-how-could-it-reduce-phase-rms}}.  The output of this program is a table of offsets which can be pre-applied to the WVR data in a pipeline run for one (or more) EBs via the \param{offsetstable} parameter of \pcmd{hifa_wvrgcalflag} either via PPR or \file{casa_pipescript.py}.

\subsection{Flux calibrator model}
\label{setmodels}

ALMA observations choose from two types of flux calibrators depending on the angular resolution of the data.  Low resolution projects typically observe one object, if available, from a list SSOs for which accurate models exist in CASA \citep{ButlerALMAMemo594}.  Otherwise, one of the quasars is observed from the list of 49 ``grid'' calibrators, which are bright sources in the ALMA Calibrator Source Catalog\footnote{\url{https://almascience.org/alma-data/calibrator-catalog}} \citep{Fomalont2014}, the majority of which can be classified as blazars, typically having flat spectra in the 1--5\,GHz range \citep{ALMACAL2018}.
In the case of grid calibrators, the \pcmd{hif_setmodels} stage will create a point source model of the specified Stokes\,I flux density for the observed visibilities using CASA task \ccmd{setjy} with \param{usescratch}=True, which will fill the MODEL column of the MS. The spectral index is also set if a value was obtained in \pcmd{hifa_importdata}. If the BANDPASS calibrator is an independent object, then its spectral index is also set, so that its slope is accounted for in the bandpass solution (\S~\ref{bandpass}), while its flux density is left at 1\,Jy until \pcmd{hifa_gfluxscale} (\S~\ref{fluxscale}). In the case of SSOs, it will fill the model column appropriately for each visibility using the source size and brightness temperature.  Tables of the frequency-averaged, per-spw model flux density are provided in the WebLog as are plots of these flux densities vs uv distance.  The plots are primarily useful for the SSO case.  This task assumes that calibrators observed with FLUX intent are observed in all science spws, as is the case with all ALMA observations that pass QA and are delivered to the user.  If a FLUX target is observed in only a subset of spws, then the generated \ccmd{setjy} command will crash.

\subsection{Flagging of antenna/spw combinations with outlier amplitude gain solutions}
\label{lowgainflag}
In the \pcmd{hif_lowgainflag} stage, antenna/spw combinations with persistently highly discrepant amplitude gains are detected and flagged.  Such data typically remain obvious outliers after calibration, so the best course is to flag it early so that it does not compromise the subsequent calibration solutions, nor reduce the sensitivity of the more sophisticated visibility flagging heuristics (\S\ref{correctedampflagtext}). The bandpass calibrator scan is used for the assessment as it typically provides the highest SNR solutions in the dataset. Prior to assessment, temporary solutions are obtained and carried in the local context for pre-application of the subsequent solve. 

First, \ccmd{gaincal} is used to phase-up the bandpass calibrator (see \S\ref{bandpass} for the definition of this concept) on integration-based timescale.  Second, \ccmd{bandpass} performs a solve using a fixed \param{solint} of `inf' (one solution per scan) for the time interval and 7.8125\,MHz for the frequency interval (corresponding to 1/8 of the 62.5\,MHz bandwidth of the individual TFBs of the BLC).  Third, \ccmd{gaincal} re-solves for integration-based phases using pre-application of the bandpass shape. Finally, \ccmd{gaincal} solves for scan-based amplitudes with pre-application of the bandpass shape and phases.  This final amplitude gain caltable is used to identify antennas with outlier gains, for each spw separately. A two dimensional matrix of this scan-based amplitude gain is formed with axes of scan number vs. antenna number calculated from the \ccmd{gaincal}.  The matrix is evaluated against the `nmedian' matrix flagging rule, where data points are identified as outliers if:
\begin{enumerate}
\item Their value is smaller than a threshold factor times the median of all non-flagged data points, where the threshold is \param{fnm_lo_limit} (default: 0.7), or
\item Their value is larger than a threshold factor times the median of all non-flagged data points, where the threshold is \param{fnm_hi_limit} (default: 1.3).
\end{enumerate}
Flagging commands are generated per-spw for each of the identified outlier data points, and are applied to the entire ms.  The WebLog contains per-spw greyscale images formed from the matrix both before flagging and after flagging, and are color-masked to distinguish between previously flagged data and newly flagged solutions.

\subsection{Bandpass}
\label{bandpass}

The ALMA correlators normalize the cross-correlation spectral visibilities by the geometric mean of the autocorrelation spectra of the two contributing antennas.  This step removes the bulk of the bandpass shape, particularly that due to the atmosphere.  The residual bandpass shape arises from the independent receivers and IF equipment on the individual antennas, and must be calibrated.
Observed once near the beginning of each EB, the bandpass calibrator is typically chosen dynamically by the calibrator query mechanism of the ALMA online scientific software (SSR) to be strong enough to measure the antenna-based complex bandpass of the interferometer in all spws in a reasonable amount of observing time.  Dynamic values are taken into account, including the number of antennas, the estimated system temperature, and any shadowing \citep{TechHandRemijan19}.
Because the largest variations in the bandpass shape are on scales $>$100\,MHz, the minimum flux density threshold is set to guarantee antenna-based solutions with SNR$>$50 in a single polarization using a bandwidth equal to the smaller of 128\,MHz or the narrowest \spwbw\/ in 15 minutes of integration.  In practice, such SNR can be achieved over much smaller bandwidth particularly in the lower frequency bands when the brightest calibrators are available to select.  

In the \pcmd{hifa_bandpass} stage, the flux density values retrieved from the calibrator service (see \S\ref{importdata}) are used to estimate the SNR in each spw at the science channel bandwidth.  The estimate accounts for the measured \tsys\/ values, integration time, and number of antennas flagged by less than 90\% on the BANDPASS scan.  More general, array-wide time-based flagging is not accounted for as it is not a common failure mode. If the estimated SNR is $>50$, then the \ccmd{bandpass} task will run at full spectral resolution, otherwise the frequency part of the \param{solint} parameter is set to a wider value that achieves the SNR goal and corresponds to an even number of channels.  However, the minimum number of solutions intervals across each spw is 8 in order to be able to measure (and thereby remove) a slope or coarse ripple.  A warning is generated if the SNR threshold cannot be reached at that bandwidth.  
The time part of the \param{solint} is set to `inf', \param{minsnr}=3, \param{minblperant}=4, and \param{combine} is set to `scan', even though ALMA bandpass observations are usually obtained in a single scan, i.e. the repeat interval set in the OT is longer than one EB. Note that \ccmd{bandpass} task has no \param{refantmode} parameter and is internally hard-coded to use \param{refantmode}=`flex', meaning that if the first antenna in the refant list drops out for an integration, the next antenna in the list is used.  To minimize residual atmospheric line features, a lower S/N threshold of 20 is used if the spw contains a strong atmospheric line, which is considered true if the normalized difference between the median $T_{sys}$ spectrum on the bandpass calibrator and a smoothed (by 1/16 of its bandwidth) version of itself is greater than 1.1.

Prior to computing the bandpass, the temporal variation of phase during the bandpass scan is measured using \ccmd{gaincal} to obtain a per-spw phase-up self-calibration solution assuming a point source model with \param{solint}=`int', \param{minsnr}=3, and \param{minblperant}=4. This table is pre-applied when solving for the bandpass. Both this phase-up table and the resulting bandpass table are loaded into the context for pre-application on all subsequent \ccmd{gaincal} solves.

If the BANDPASS calibrator scan contains line features, then the solution will contain these lines, adversely affecting the science data product.  Quasars with line emission are labeled in the ALMA calibrator catalog and avoided by the online query mechanism.  Nevertheless, the \param{fillgaps} parameter of \ccmd{bandpass} will be exposed in \pcmd{hifa_banpdass} in the next pipeline release (Cycle 10) in order to allow cases of unexpected line contamination to be manually flagged executed in the pipeline with a modified PPR.

\subsection{Spectral window phase offsets}

The ALMA analog IF downconversion system splits the broadband front-end signal into four independent 2\,GHz wide basebands \citep{Hills2008}.  After digitization and data transport, the four signals enter the correlator where one can define up to 4 spws for correlation within each baseband.  Perhaps the most common user setup is to define one spw per baseband, in which case each spw ultimately arises from a different analog signal. Therefore it is natural to expect different path lengths for each spw, and hence spw-to-spw phase offsets.  These offsets are measured in the \pcmd{hifa_spwphaseup} stage and the resulting caltable is loaded into the context for pre-application on all subsequent \ccmd{gaincal} solves.  The WebLog for this stage includes a table showing the mapping from each phase calibrator to its science targets and \checksources\/.

\subsection{Low SNR heuristics}
\label{lowsnrheuristics}
In addition to the channel pre-averaging heuristic in \pcmd{hifa_bandpass} (described in \S\ref{bandpass}), the pipeline implements heuristics to adjust the calibration strategy depending on the expected SNR of the gain solutions on each of the PHASE calibrator and CHECK source objects.  Both the spw mapping scheme and the solution time interval are chosen in \pcmd{hifa_spwphaseup}.  In contrast, the other calibrators (BANDPASS, FLUX, and POLARIZATION) always use integration-based, per-spw solutions as they are assumed to be sufficiently bright in all spws.  

\subsubsection{Spw mapping and combination}
\label{spwmapping}
\begin{figure*}
\centering
\includegraphics[scale=0.98]{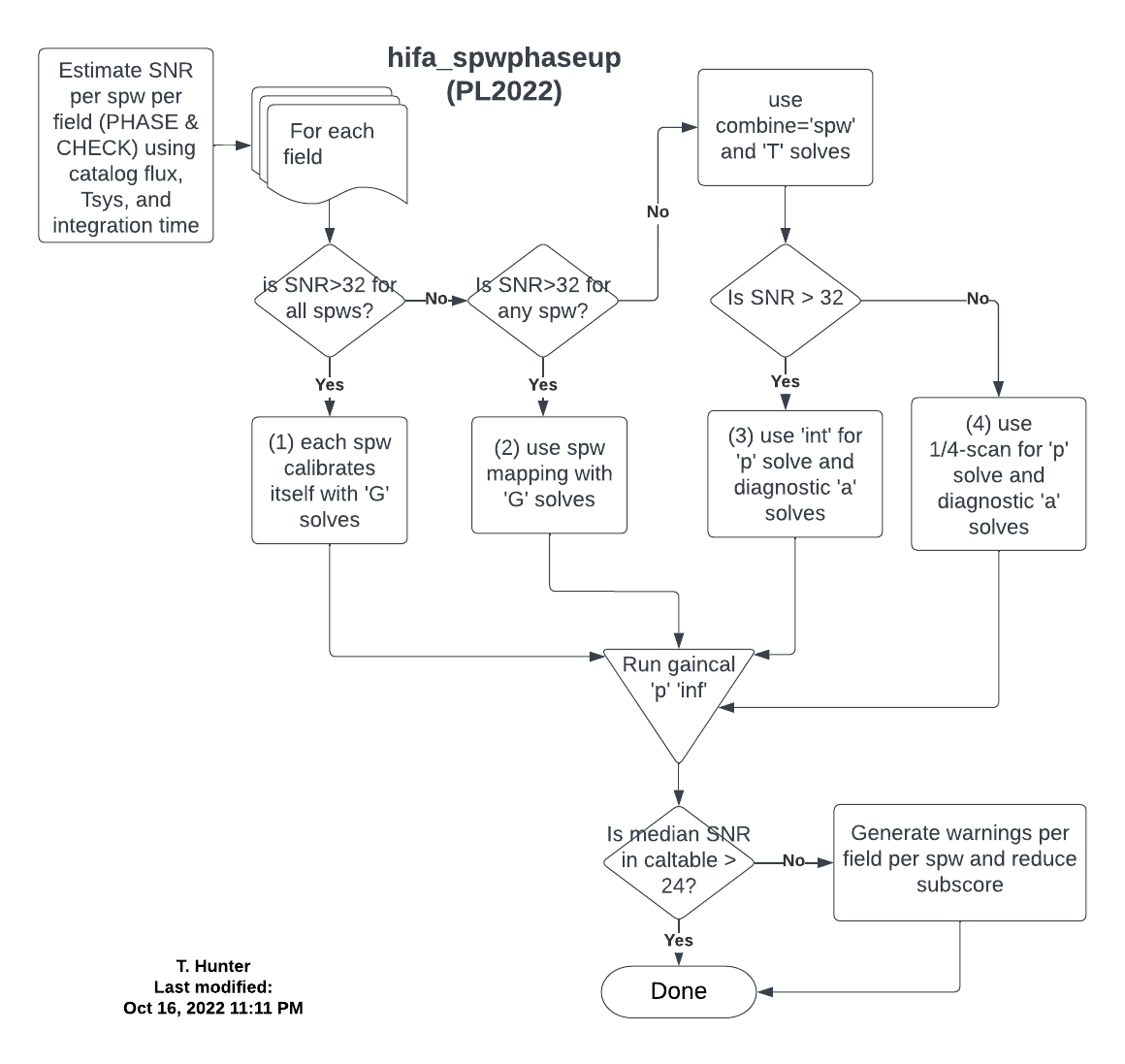}
\caption{The logical workflow of the \pcmd{hifa_spwphaseup} task that computes for each PHASE calibrator and CHECK source the \ccmd{gaincal} parameters (\param{spwmap}, \param{gaintype}, and \param{solint}) to be used in subsequent stages.}
\label{fig:hifa_spwphaseup}
\end{figure*}

The logical workflow of the low SNR assessment aspect of the \pcmd{hifa_spwphaseup} stage is shown in Figure~\ref{fig:hifa_spwphaseup}. Each PHASE calibrator and CHECK source is assessed independently to produce a pre-estimate of the expected SNR of the gain solution for each spw independently.  
For each field, the SNR is computed for a length of time matching the scan length of the scan
closest to the first \tsys\/ scan used for that field. For TDM spws, the bandwidth used is limited to the maximum effective bandwidth of a baseband  (1875\,MHz). A pre-computed list of per-Band sensitivities for a nominal \tsys\/
value with sixteen 12\,m antennas, 8\,GHz bandwidth, and 1\,minute integration with dual polarization is consulted
and scaled for the actual median \tsys, actual total collecting area of antennas, 
actual bandwidth, and a single polarization.  The collecting area factor ($f_{area}$) is defined as:
\begin{equation}
f_{area} = (n_{\rm 12m} + (12/7)^2n_{\rm 7m}) / (n_{\rm 12m} + n_{\rm 7m}),
\end{equation}
where $n_{\rm 2m}$ and $n_{\rm 7m}$ are the number of 12\,m and 7\,m antennas in the data, respectively.
If the SNR prediction does not meet the threshold of 32 for a given spw, then that spw is ``mapped'' to the spw with the highest SNR prediction, meaning that the subsequent phase solutions for the highest SNR spw will also be used to calibrate the less sensitive spw(s).  However, if no spw meets the SNR threshold, then the data from all spws are combined (vector averaged) in all subsequent phase solves.   In all cases, the
amplitude solves are still performed on the less sensitive spws with pre-application of the high-SNR phase solution, because this pre-application typically raises the SNR sufficiently above the \param{minsnr} value even for narrow spws.

Based on the decision in this stage, the workflow of the subsequent calibration stages: \pcmd{hifa_gfluxscale} and \pcmd{hifa_timegaincal} are altered with respect to the default high SNR case.

\subsubsection{Solution time intervals}

The standard solution time interval for phase solves on calibrators is one integration, which for ALMA ranges from 2--6 seconds \citep{Hunter2012}, and for amplitude solves is one scan length (ranging from $\sim$12--60\,seconds). However, if the combined solution using data from all spws is predicted to still not meet the SNR threshold, then the solution interval for phase solves is increased from one integration to one quarter of the scan length.

\subsubsection{Phase RMS  structure plots}
\label{ssf}
\begin{figure}[ht]
\centering
\includegraphics[scale=0.25]{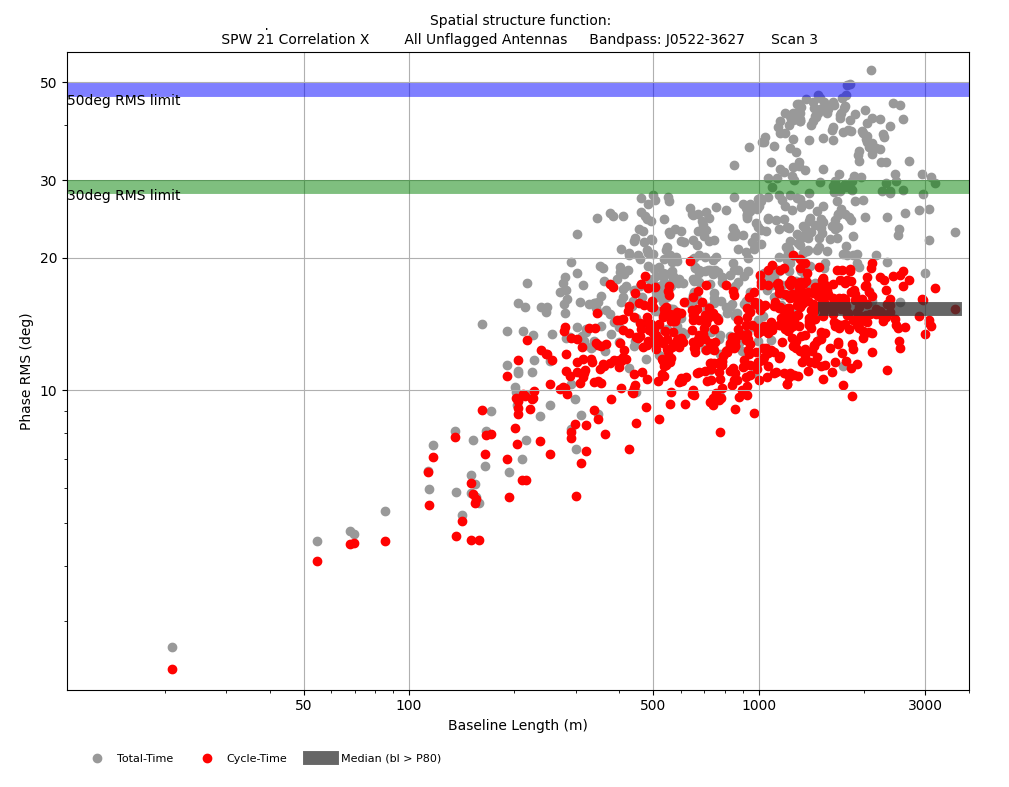}
\caption{The spatial structure function-like plots (\S\ref{ssf}) included on the \pcmd{hifa_spwphaseup} WebLog page to indicate the phase RMS  as a function of baseline length using the BANDPASS scan. Long timescale (BANDPASS scan duration) is shown in gray, while that calculated over the phase referencing cycle-time is shown in red. For longer baselines the phase RMS  usually flattens. The median phase RMS  from baselines longer than the 80$^{\rm{th}}$ percentile is shown as the black bar.}
\label{fig:hifa_spwphaseup_SSF}
\end{figure}

The diagnostic plots at the bottom of the WebLog \pcmd{hifa_spwphaseup} stage provide an insight into the observing conditions based on phase stability. Using the phase solutions of the BANDPASS scan from \pcmd{hifa_bandpassflag}, which  typically have high SNR, the baseline based phase-time streams are reconstructed and the phase RMS values are calculated and plotted as a function of baseline length (Figure \ref{fig:hifa_spwphaseup_SSF}), as so-called Spatial Structure Function plots \citep[see,][and references therein]{Matsushita17}. The phase RMS values are calculated over the total observing time of the BANDPASS scan, which are 5-30 minutes in length, to provide a longer time variability overview (gray symbols), and over a timescale equal to the phase referencing cycle-time (red symbols - the time taken to observe the phase calibrator, slew to and observe the science target, and slew back to the calibrator again to repeat the cycle), but only if the cycle time is shorter than the total duration of the BANDPASS scan. The cycle-time phase RMS acts as a proxy for the short-term ($<$cycle-time) phase noise that will remain in the science target after the phase referencing correction \citep[see,][]{Maud22}. Note that for longer baseline configurations the phase RMS typically flattens for the cycle-time estimate because the long-timescale large variations are well-corrected. 

Phase RMS  levels of $<$30$^{\circ}$, 30-50$^{\circ}$, 50-70$^{\circ}$, and $>$70$^{\circ}$ can be considered as corresponding to ideal, good/stable, poor/unstable, and bad conditions, respectively.  These labels also apply to the possible levels of decoherence in the target data. The median phase RMS  reported in the WebLog table considers baseline lengths greater than the 80$^{\rm{th}}$ percentile. Outlier antennas are also identified if they exceed 180$^{\circ}$ phase RMS, if they are more than double the median phase RMS, or if they exceed 6 or 4 times the median absolute deviation from the median phase RMS  for observations where the median phase RMS  is $<$50$^{\circ}$ or $>$50$^{\circ}$, respectively. In the latter case the median phase RMS  is recalculated excluding the outlier antennas. No flagging is made by this diagnostic. 

Note that there is no elevation-based scaling and that if the BANDPASS source is at a significantly higher (or lower) elevation than the science target, then the phase RMS in the science data could be underestimated (or overestimated). 

\subsection{Fluxscale}
\label{fluxscale}
\begin{figure*}
\centering
\includegraphics[scale=0.9]{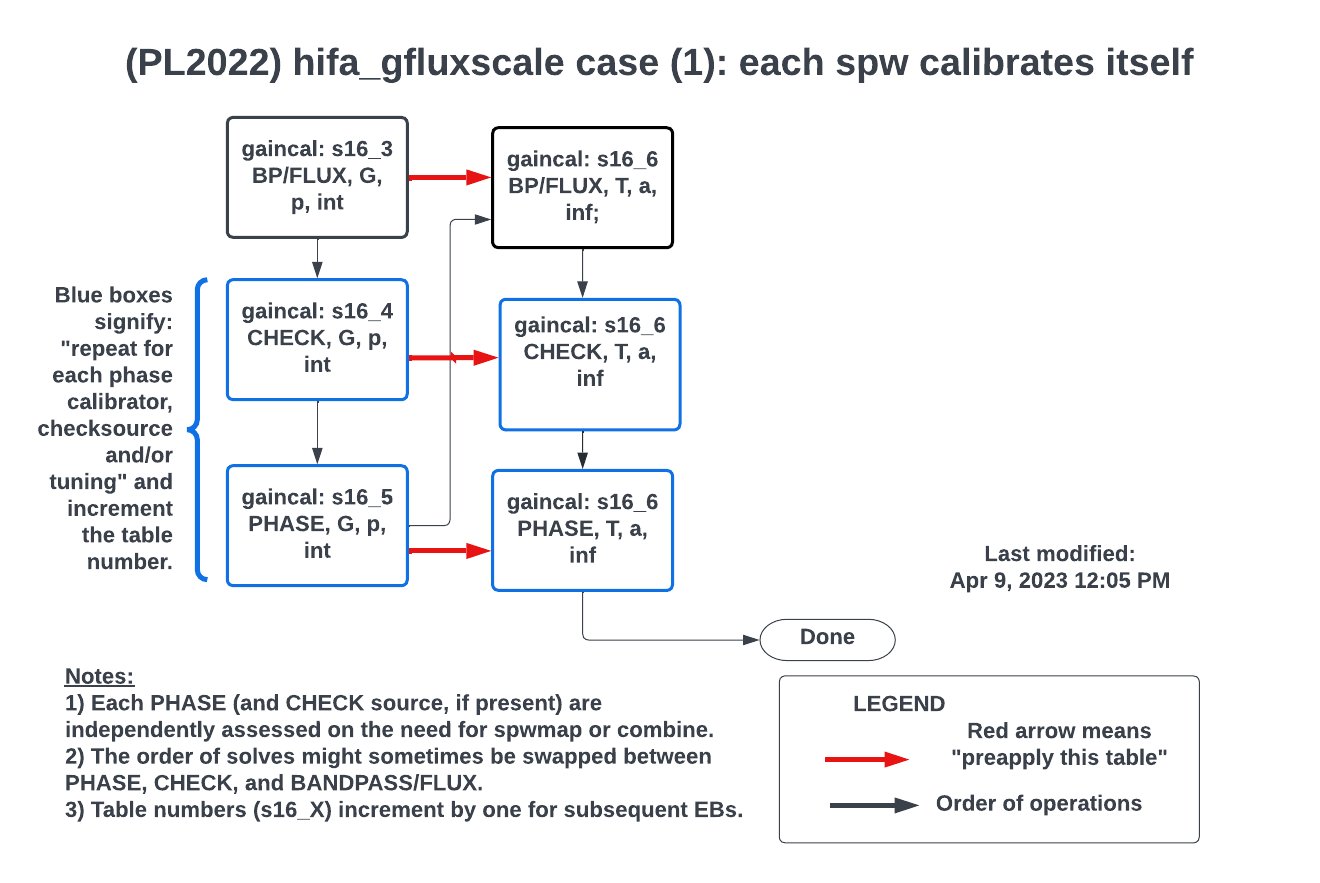}
\caption{Gain solution workflow of the \pcmd{hifa_gfluxscale} stage.  Each box represents one call to the \ccmd{gaincal} task for one EB, each of which produce a caltable. The content of the boxes indicate the key parameters of the task, including the table number (``s$<$stage\#$>$\_$<$table\#$>$'') embedded in the resulting \param{caltable} name, the calibrator \param{intent}, the \param{gaintype}, the \param{calmode}, the \param{solint}, and (where appropriate) the values of \param{spwmap} and \param{combine}.}
\label{fig:gfluxscale}
\end{figure*}

The fluxscale of the data is solved in the \pcmd{hifa_gfluxscale} stage, and the workflow of its \ccmd{gaincal} commands for the high SNR case is shown in Figure~\ref{fig:gfluxscale}.  First, the gains are solved on the FLUX calibrator scan (and the BANDPASS calibrator scan if it is independent) with \param{solint} set to the integration time for the phase solves and to the scan time for the amplitude solves. The amplitude gains provide the flux scaling factor.  The policy of using of `T' amplitude solves, which pre-average the XX and YY data, in all calibration stages effectively calibrates to the Stokes I flux densities measured by the ALMA calibrator survey program, and allows polarized calibrators to have differing flux densities in their calibrated XX and YY visibilities.  This technique thereby avoids introducing any false polarization into the science targets.  If interstellar absorption lines are present in the flux calibrator data, the fluxscale will be
affected accordingly.  There is no heuristic to identify and flag such lines, but ALMA labels objects with significant contamination in the calibrator catalog and SSR excludes them in queries for bandpass and flux calibrators.  The fluxscale is not affected by lines in the phase calibrator data.

For the case of an SSO as the FLUX calibrator, this stage first checks if it is partially resolved. The angular size of the object is compared to the available baseline lengths, and antennas whose distance from the refant is beyond a calculated threshold are excluded from the solution.  An alternative strategy that excludes data based solely on uvrange instead of antenna selection can sometimes lead to cases where there is very little data included for some antennas, which can bias the results for all antennas. For this reason, this alternative is not implemented. The exact steps followed are:
\begin{enumerate}
\item Retrieve the major axis of the angular size of the solar system object calibrator from the CASA data repository for the date of observation.
\item Determine longest observing wavelength in the science spws.
\item Determine the shortest unprojected baseline to the reference antenna.
\item Estimate the peak intensity of the visibility pattern at that baseline.
\item Find the baseline length where the Fourier transform of a uniform disk drops to 20\% of that peak.
\item Select antennas whose separation from the reference antenna are all within that limit. \label{step}
\item If the number of qualifying antennas $<$ 3, then default back to selecting all antennas.
\end{enumerate}

\begin{figure*}
\centering
\gridline{\fig{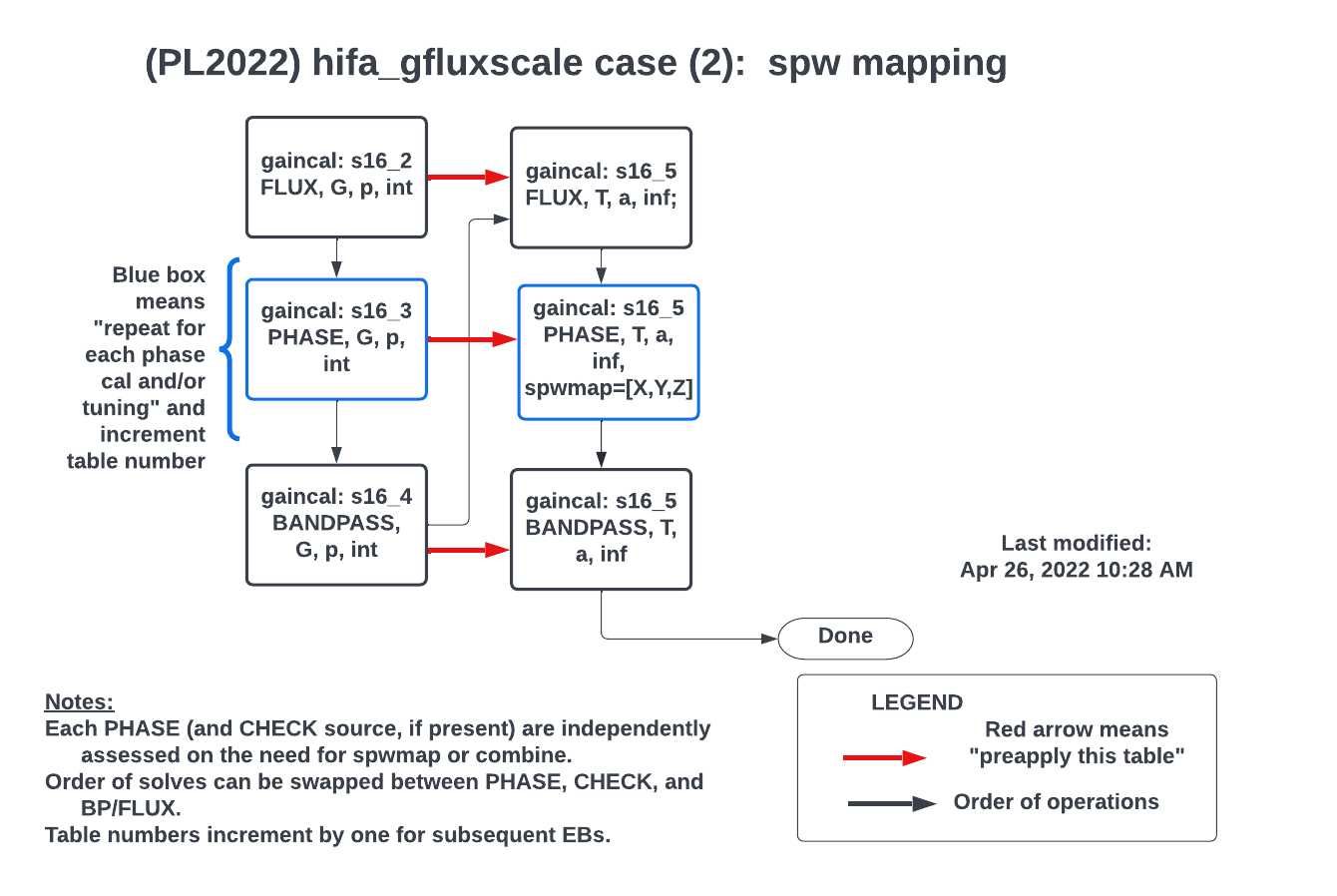}{0.84\textwidth}{(a)}}
\gridline{\fig{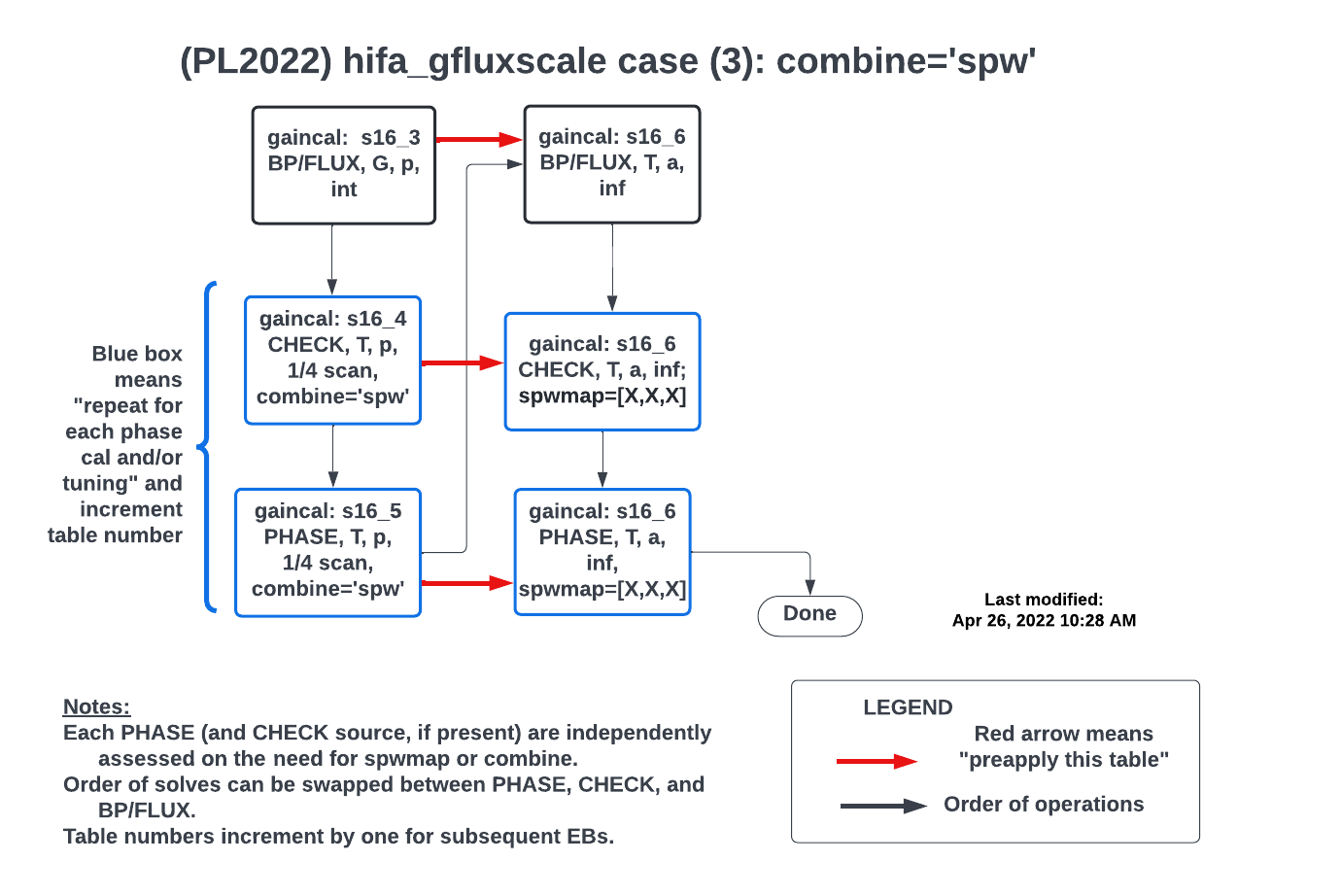}{0.84\textwidth}{(b)}}
\caption{Analogous to Figure~\ref{fig:gfluxscale}, these diagrams show the modified workflow
of the \pcmd{hifa_gfluxscale} stage for the separate cases of spw mapping (a) and spw combination (b).}
\label{fig:gfluxscaleLowSNR}
\end{figure*}

For all gain solves in this stage, \param{minsnr}=2.
Regardless of whether the FLUX calibrator is an SSO, \param{minblperant}=2 for all gain solves on the FLUX calibrator in case step \ref{step} drastically reduces the number of antennas.
Next, the gains are solved for the PHASE, CHECK, and POLARIZATION calibrator(s), if present.  These solves 
use \param{minblperant}=4 and 2 for the integration-based 'p' and scan-based 'a' solves, respectively, although here
there is no fundamental reason for using 2 instead of 4.
The separate workflows for the low SNR cases of spw mapping and spw combination are shown in Figure~\ref{fig:gfluxscaleLowSNR}.

Once the gain solutions are complete, the \ccmd{fluxscale} task is called to compute the flux densities of these calibrators on a per-spw basis using the flux scale factor that it derives from the FLUX calibrator.  Finally, the \ccmd{setjy} command is called to fill the model column for these calibrators with their corresponding flux scale factors without any reconciliation with a power law model. By not applying a power law fit, we ensure that spws with low SNR, which suffer noise bias, will yield images with a point source flux density comparable to the expected trend established by the solutions of the higher SNR spws. 
Note that even for bright calibrators with high SNR, the vector average calibrated visibility, which is also displayed in the WebLog table, is typically slightly lower than the value written to the model column due to some (unavoidable) level of residual phase calibration error.  Finally, if the bandpass calibrator is an independent source from the flux calibrator, the same catalog spectral index is once again set for each spw as in \pcmd{hif_setmodels} (\S~\ref{setmodels}), while the other calibrators are left at zero since they typically do not have catalog measurements taken in close proximity in multiple ALMA bands.

On rare occasions, \pcmd{gaincal} can produce amplitude solutions with abnormally large apparent SNR ($\sim$100 times typical values), as indicated in the SNR column of the caltable.  The corresponding gains are hugely aberrant, up to several thousand sigma above the median, and clearly in error.  Until this phenomenon is understood and fixed, the pipeline is susceptible to it.  A future improvement is to examine the caltable and flag these solutions.  In the cases encountered so far, flagging the gain outliers rather than SNR outliers would be a more distinguishing and hence safer technique.

\subsection{Time dependent gain}
\label{timegaincal}
\begin{figure*}
\centering
\includegraphics[scale=0.9]{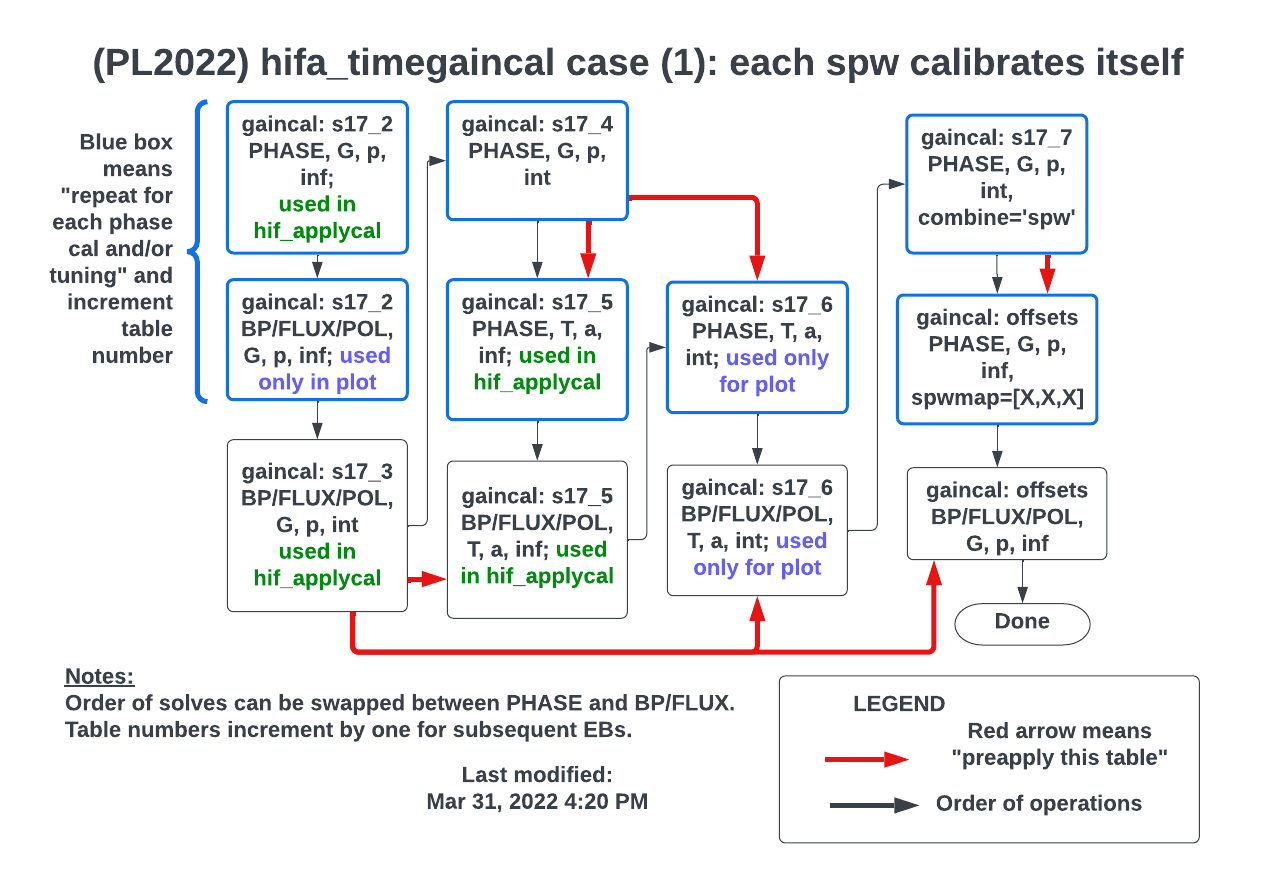}
\caption{Same as Figure~\ref{fig:gfluxscale} but for the \pcmd{hifa_timegaincal} stage.  An additional piece of information in each box is the subsequent usage of the resulting caltable, either for application to the data in \pcmd{hif_applycal} or merely for diagnostic plotting (\ccmd{plotms}).}
\label{fig:timegaincal}
\end{figure*}

The time variation of the instrumental plus atmospheric gain is solved for in the task \pcmd{hifa_timegaincal}.   In the case EBs with at least three 12\,m antennas, this gain is the post-WVR correction residual gain.
The workflow of the sequence of \ccmd{gaincal} calls is shown in Figure~\ref{fig:timegaincal} for the high SNR case, and in Figure~\ref{fig:gfluxscaleLowSNR} for the spw mapping and spw combination cases (\S~\ref{spwmapping})).  Phase and amplitude solutions are computed for each scan (\param{solint}=`inf', \param{minblperant}=4, \param{minsnr}=3) on each of the calibrators (BANDPASS, FLUX, POLARIZATION, PHASE) and plotted in the WebLog on a combined plot vs. time per spw. 
When registering the PHASE solutions, the \param{fldmap} parameter of the CalLibrary is populated correctly to prevent the \ccmd{applycal} commands in the later stage \pcmd{hif_applycal} from using solutions from the polarization calibrator when time-interpolating solutions to the TARGET and CHECK sources.  This detail is necessary because POLARIZATION scans can sometimes be observed in between scans of PHASE and these other intents.

For diagnostic purposes, integration-based solutions are also computed (with \param{minsnr}=2) and plotted separately. Finally, the PHASE calibrator data is processed again to search for temporal variation in the spw-to-spw phase offsets. The phase offsets are computed by pre-applying an spw-combined phase-only solution and computing a new phase solution for each spw. If the instrument is stable, the new phase solutions should scatter about zero with no drift.  One plot is shown for each spw, with phase offset plotted per antenna and polarization as a function of time.   If any antenna exhibits significant jumps away from zero, then a warning message and a poor QA score will be generated (\S\ref{timegaincalQA}).

\begin{figure*}
\centering
\gridline{\fig{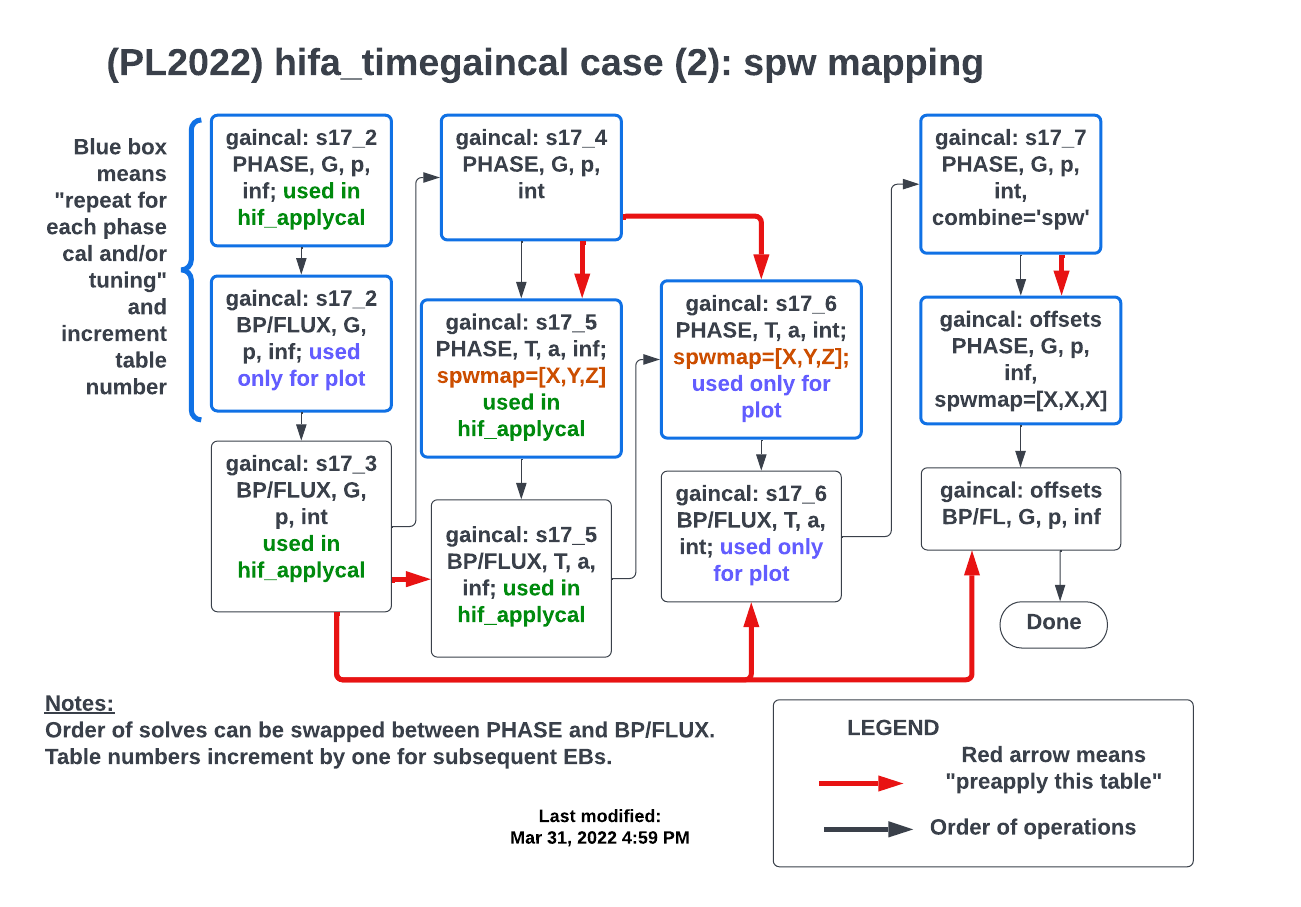}{0.82\textwidth}{(a)}}
\gridline{\fig{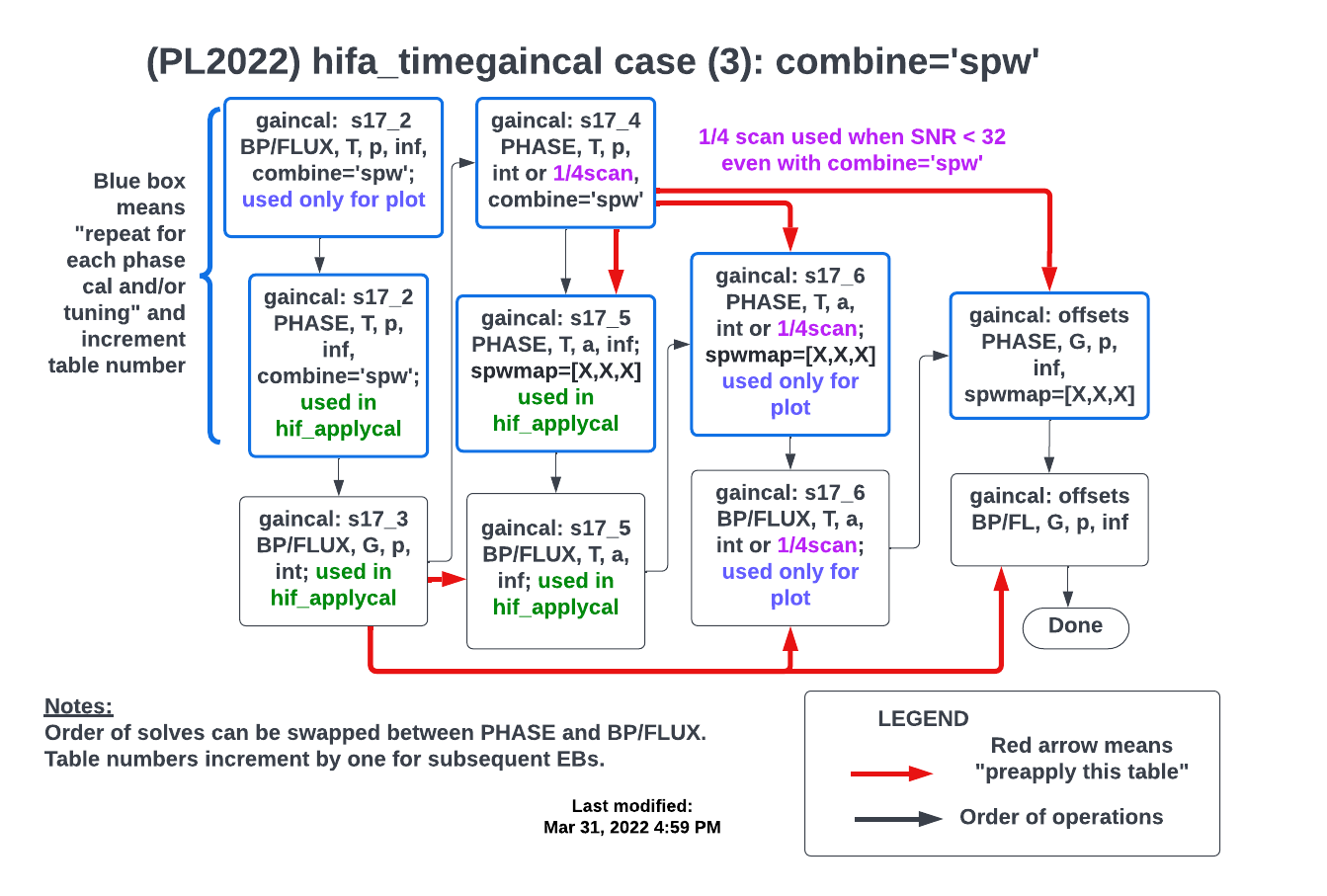}{0.82\textwidth}{(b)}}
\caption{Analogous to Figure~\ref{fig:timegaincal}, these diagrams show the modified workflow
of the \pcmd{hifa_timegaincal} stage for the separate cases of spw mapping (a) and spw combination (b).}
\label{fig:timegaincalLowSNR}
\end{figure*}

\subsection{Application of calibration tables}
\label{hif_applycal}
In the stage \pcmd{hif_applycal}, all of the calibration tables created and stored by prior stages into the pipeline CalLibrary within the pipeline context are finally applied using the calibration library (``callibrary'') mechanism of CASA.  
The callibrary is a means of specifying calibration instructions in a text file, rather than via the traditional \param{gaintable/gainfield/interp/spwmap/calwt} parameters that often become clumsy when many caltables are involved, and which have rather limited flexibility. 
The real power of the callibrary arises from the ability to specify calibration instructions for a caltable per MS selection.  This mechanism enables consolidating what would be multiple applycal executions using the traditional parameters into a single execution. The application of calibration solutions from the callibrary is invoked by setting \param{docallib}=True and passing the name of the library text file via the \param{callib} parameter in the \ccmd{applycal} task.   The callibrary mechanism is used throughout the pipeline stages that generate and apply temporary caltables (such as \pcmd{hifa_bandpassflag}), while any new permanent caltables created by a task are registered into the pipeline CalLibrary for each MS as each calibration task completes. 

On the WebLog page for \pcmd{hif_applycal}, a summary table contains a link to the text file version of the callibrary file for each MS, followed by a series of visibility plots of the calibrated data.  In addition to various plots of the calibrator data, the calibrated amplitude vs. frequency spectra of the representative source is shown for each MS. For mosaics, the representative field is identified as the field with the highest median channel-averaged amplitude, calculated over all science spectral windows.

\subsection{Renormalization}
\label{renormalization}

During the design phase of ALMA, it was decided that the cross-correlation products for each baseline and spw would be normalized by the autocorrelations of the corresponding antennas prior to storing the visibilities. This scaling minimizes the impact of many antenna-based issues on a given cross-correlation, and enables easy conversion of the normalized cross-correlations to antenna temperature units
via multiplication by the corresponding \tsys\/ spectrum (\S\ref{tsyscal}). It also results in flatter bandpasses to be solved online in TelCal and offline in \pcmd{hifa_bandpass}. However, for
this temperature calibration to work properly, it is necessary that the \tsys\/ spectrum accurately represents all of the
signal included in the autocorrelation power and ideally is measured with the same spectral setup (channel resolution). Because
\tsys\/ measurements as currently performed by ALMA only account for the autocorrelation power in a reference position located off source (typically offset by $2'$ in cross-elevation\footnote{The sky position of the \tsys\/ spectrum is indicated correctly in the MS POINTING table in the azimuth/elevation frame,  but a separate field direction is not generated in the MS as it is not fixed in equatorial coordinates, and thus it does not appear in \ccmd{listobs} output.}), the normalized cross-correlations will be improperly calibrated by an amount proportional to the ratio of the power from the target source vs. the reference position.
To address this discrepancy, the pipeline stage \pcmd{hifa_renorm} assesses, and optionally corrects, science target data suffering from incorrect amplitude normalization (\S\ref{bandpass}) caused by bright astronomical lines detected in the autocorrelations of some target sources.\footnote{A more complete description of ALMA's amplitude normalization and the effects of bright emission lines can be found here: \url{https://help.almascience.org/kb/articles/623}.} In brief, the effect occurs when there is sufficiently bright line emission to be strong in a single-dish spectrum of the source.

The main component of the code is to extract the autocorrelations that were already used for normalizing the cross-correlation visibilities in the correlator system at the time of observation and to create renormalization spectra (or scaling spectra). When applied, the scaling will compensate for the previously under-scaled amplitudes by rescaling the channels affected at a granularity of spw, scan, field, antenna, and polarization.

\begin{figure*}
    \centering
    \includegraphics[width=0.77\textwidth]{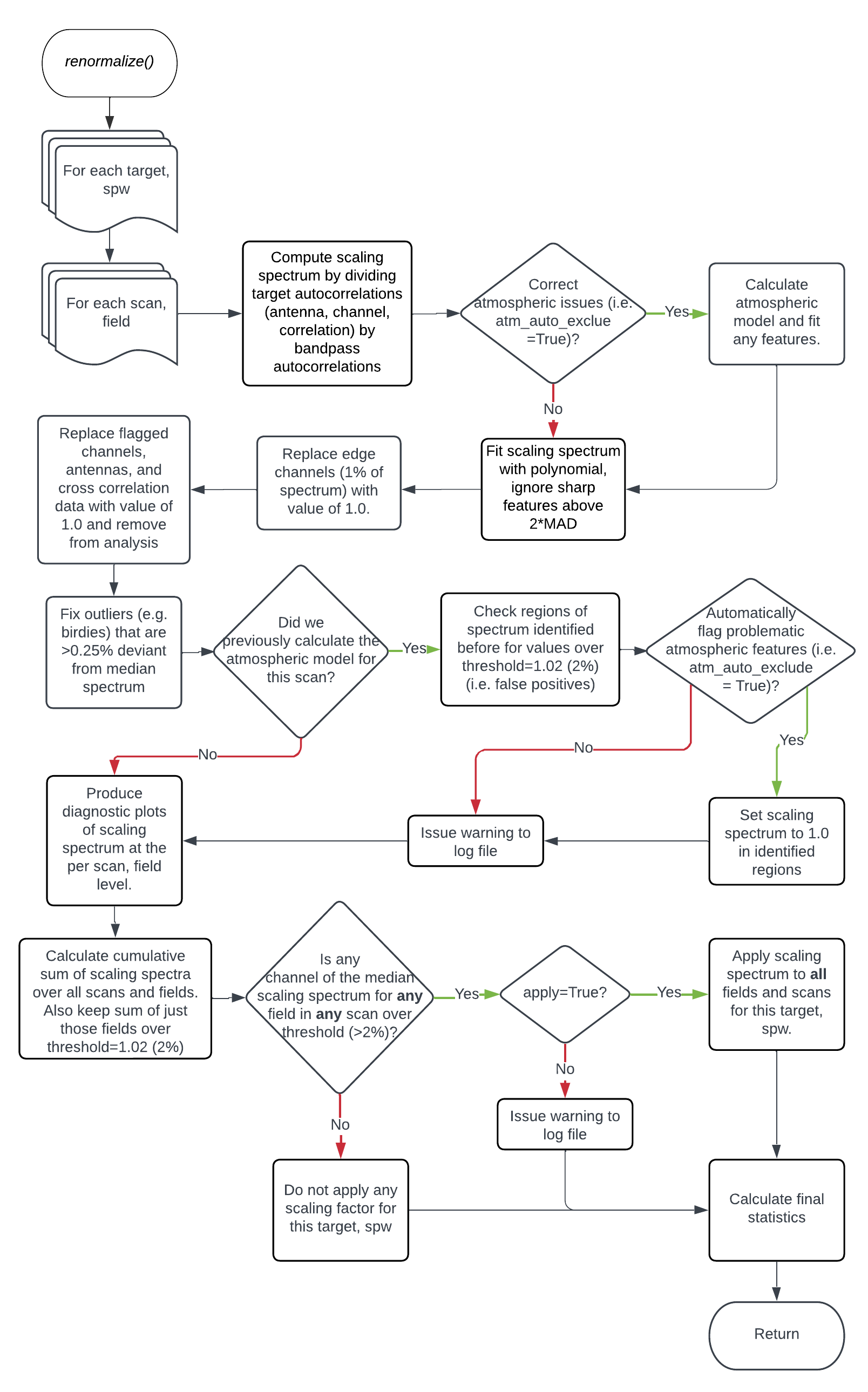}
    \caption{Logic flow for the \pcmd{renormalize()} method utilized within \pcmd{hifa_renorm} to calculate and apply an amplitude correcting scaling spectrum to the cross correlation data of the science targets (\S\ref{renormalization}).}
    \label{fig:renormalize}
\end{figure*}

In detail, the \pcmd{hifa_renorm} stage uses the \pcmd{almarenorm.py} code to evaluate each EB independently. For each EB, it creates an \pcmd{ACreNorm} object from the \pcmd{almarenorm.py} code and runs the \pcmd{renormalize} method on it. The full workflow of the \pcmd{renormalize} method is shown in Figure \ref{fig:renormalize} and works as described below. Once finished, it then uses the \pcmd{plotSpectra} method to create the summary spectra which shows the cumulative sum of the scaling spectra over all scans and fields for each antenna and correlation for each target and FDM spw\footnote{Because this effect is highly dependent only on strong line emission, we only analyze and fix FDM spws.}. It also packages all the diagnostic plots (displaying the scaling spectra at the per target, spw, scan, field level) into PDFs that are displayed in the WebLog. This relies on the \ccmd{poppler-utils}\footnote{See \url{https://pypi.org/project/poppler-utils/}} Python module to produce the PDFs.

The main premise of the renormalization correction implemented in the \pcmd{renormalize} method is to utilize an autocorrelation spectrum uncontaminated by celestial line features and compare this with a suspected contaminated autocorrelation spectrum from the science target. Simply, the bandpass calibrator's autocorrelations are assumed to be line-free and are divided from the science target autocorrelations leaving a unitless scaling per channel. If the bandpass calibrator and science target autocorrelations were identical, except for astronomical line detections on the target, then this division would already create the scaling spectrum. However, due to differences in observation time and airmass, the respective autocorrelations differ slightly and must be baseline fitted in order to establish whether there is any contaminant astronomical line and by how much each contaminated channel should be rescaled. This is done through iteratively fitting the baseline with increasing polynomial orders up to a fifth order polynomial. Due to the differences in airmass between the bandpass calibrator and the science target, the scaling spectra of some observations may exhibit residual scaling features related to an inability to properly fit residual atmospheric profiles. These features can be automatically removed from the analysis via the \pcmd{hifa_renorm} parameter \param{atm_auto_exclude}=True (the pipeline default value) and thereby eliminate false triggers.  Alternatively, channel ranges can be manually excluded using the older \param{excludechan} parameter.

Throughout the renormalization assessment, a number of heuristics are run and act to repair any incorrect values in the scaling spectra due to, e.g., birdies in the autocorrelation spectra, divergent baseline fitting near edges of segments, and poorly performing or flagged antennas. Deviations from the median spectrum of more than 0.25\% are identified and replaced by values from the median spectrum of the matching correlation. 

Ultimately, if \param{apply}=True (the default) and if any scaling spectrum is found to be above the \param{threshold}=1.02 level (the default), then the scaling spectrum is directly applied to the cross correlation data. For mosaic data, if any field is found to be above the threshold, then all fields have their associated scaling spectrum applied. 

\section{Visibility flagging heuristics}
\label{visibilityFlagging}
A variety of visibility flags are generated and applied at different places in the calibration pipeline workflow.  We describe them in the following subsections.

\subsection{Deterministic flags}
\label{deterministicFlags}
In the \pcmd{hifa_flagdata} stage, which runs immediately after \pcmd{hifa_importdata}, flags are applied to particular visibility data solely on the basis of their metadata, with no dependence on their value.  Each category described below has a corresponding boolean control parameter to enable it.  By default, all are set to True.

\subsubsection{Correlator binary data flags}

Both the ACA and BLC generate occasional flags due to a list of temporal hardware or software failure conditions. Some examples are the failure to complete correlator configuration requests in the allotted time \citep{Amestica20} or to compute and distribute the necessary calibration coefficients to align the signals output by the TFBs in the BLC. Flags for the affected spws and time intervals are recorded as binary data file (BDF) flags which are applied during the \pcmd{hifa_importdata} task (\S\ref{importdata}) when \param{bdfflags}=True (the default value).  The percentage of data flagged by BDF flags is visible (to the nearest 0.001\%) in the ``Before task'' column of the table in the WebLog page for this stage.

\subsubsection{Online flags}

The control system adds flags to the data to denote time periods where other hardware systems on a particular antenna or its particular receiver band are not in a ready state. Some common conditions are:
\begin{itemize}
\item ``Mount is off source''
\item ``Calibration device (ACD) is not in the correct position''
\item ``FrontEnd Local Oscillator Amps not optimized''
\item ``The Warm Cartridge Assembly is not locked''
\item ``Power levels are being optimized.''
\end{itemize}
A color-coded chart showing these condition names vs. time is included in the \pcmd{hifa_flagdata} WebLog page. Neither the BDF flags nor the online flags produce any channelized flags.

\subsubsection{Unwanted scan intents and correlation products}

Because the pipeline does not use the visibility data from any of the following CALIBRATE\_* scan intents, this heuristic flags all scans with these intents in order to prevent inadvertent usage of the data: ATMOSPHERE, FOCUS, and POINTING (see Table~\ref{tab:intents}).  Although the pipeline requires the ATMOSPHERE intent to be present, the visibilities from those scans are not used because the \tsys\/ values computed online from these data by TelCal are written to the SYSCAL table of the MS.  In addition, this heuristic flags the channel-averaged spw that is paired with each full resolution spw, and it flags the square law detector (SQLD) spw that is associated with each baseband. The data from these spws is used primarily for engineering purposes (the SQLD data is used for the fast-scanning mode of solar continuum observing, which is not supported by the pipeline.)

\subsubsection{QA0 flags}

The ALMA QA0 process \citep{Chavan16} attempts to analyze the data from each EB soon after it is completed in order to find any obvious problems in the system that require immediate action by the telescope operator to either fix or perhaps to remove an antenna from operation before the next EB. It also produces ``suggested'' flags, many of which are disabled and do not reach the ASDM as they are still experimental.  The only two flavors of QA0 flags that are applied by the pipeline {\it and} have any effect on the processing are for poor performance in the online reference pointing solutions and for large discrepancies of the receiver-related quantities (\trx\/ or \tsys) between the two polarizations of a given antenna/spw combination.  

The pointing-related QA0 flags include flags for large local pointing corrections ($\theta_{\rm lpc}$) and large local pointing correction uncertainties ($\delta\theta_{\rm lpc}$), either of which will flag all the data from an individual antenna.  For EBs with a single pointing scan, the respective thresholds are half the primary beamsize ($\theta_{\rm PB}$) and a quarter of $\theta_{\rm PB}$, where $\theta_{\rm PB}$  is computed for the representative frequency of the SB, which is set by the user in the OT.  For EBs with multiple pointing scans, a fairly common case, only the results from the second (and any following) scans are assessed but using the stricter thresholds of:
\begin{equation}
\begin{aligned}
|\theta_{\rm lpc}| >& \texttt{max}([4'', \theta_{\rm PB}/5]), {\rm or }\\
\delta\theta_{\rm lpc} >& \texttt{max}([1'',\theta_{\rm PB}/20]).
\end{aligned}
\end{equation}
This strategy of not assessing the ``first of N'' pointing scans will avoid flagging a well-performing antenna merely because it needed a large correction due to its thermal conditions having changed greatly since the previous pointing measurement during some earlier SB execution. 

The receiver-related QA0 flag is triggered if the absolute difference of the median values for polarization X and Y is more than half their mean value.  It is currently applied to all scans of the dataset on a per-antenna, per-spw basis.  Because the performance of most receivers is well-matched between X and Y, this flag appears fairly rarely and is likely to be deprecated from pipeline usage in a future observing Cycle, because the normal calibration process will properly handle the difference if the flag is not applied.

\subsubsection{QA2 flags}

There is a mechanism for adding manual flags to be processed by the calibration pipeline. It consists of a text file per EB (\file{<MS_name>.flagtemplate.txt}) containing flagging agent commands which will be applied by \pcmd{hifa_flagdata}.
This method is used in ALMA operations for the rare cases where the automated visibility flagging stages (\S\ref{correctedampflagtext} and \ref{correctedampflagScience}) fail to identify and flag all the bad data.  There is an analogous mechanism in the imaging pipeline to add flags to the science targets prior to imaging in the task \pcmd{hifa_flagtargets}.

\subsubsection{Autocorrelations}

All autocorrelation data are flagged in all spws that contain it, as these data are not used in any pipeline stage, except for \pcmd{hifa_renorm} (\S\ref{renormalization}).  This flagging will prevent these data from inadvertently appearing in any \ccmd{plotms} plots without needing to explicitly
exclude them via the antenna selection parameter.

\subsubsection{Partial polarizations}
\label{partialpol}
Occasionally, the BDF flags will flag only a subset of the polarization products on a particular integration on a particular antenna.  In many cases, the surviving polarization product(s) is also affected by the same real hardware problem but to a lesser extent. Also, the CASA imaging task (\ccmd{tclean}) will not use an integration at all if any of the contributing products (meaning XX and YY for Stokes\,I imaging) have been flagged.  For these reasons, this heuristic identifies integrations with partially flagged polarization products and flags the surviving products.  This issue was more prevalent in earlier observing Cycles (prior to Cycle 6) and is rarely observed now.

\subsubsection{Shadowed antennas}

In this heuristic, the \param{mode} = `shadow' option of \ccmd{flagdata} is used to flag all baselines to an antenna that suffers any shadowing from another antenna in the active array.  That mode has a sub-parameter \param{tolerance} with units of meters and a default value of zero. A positive number will allow antennas to overlap by that amount in projection.  A modest positive value may be set in a future release to recover some short spacings, particularly in 7\,m array data, but even the large configurations of the 12\,m array have some 15\,m baselines which can suffer unnecessary shadow flagging due to a very small amount of overlap.  The flags generated in the CASA 6.4 version of \ccmd{flagdata} could sometimes differ by a fraction of a meter from the requested \param{tolerance}.   This logic has since been fixed in CASA 6.5 to produce the exact result expected from the vector magnitude of the UVW column of the MS, and this version will be used in the 2023 release of the pipeline (Cycle 10).

\subsubsection{Edge channels}
\label{edges}
The edge channels of TDM spws (each 2 GHz wide) are flagged to avoid the roll-off in sensitivity caused by the baseband anti-alias filters, which becomes significant beyond a passband of width $\sim$1.875\,GHz. In order to flag channels outside this range, the number of channels flagged on each edge is controlled by the parameter \param{fracspw}, which has a default of 0.03125.  In dual-polarization mode, TDM spws have 128 channels, thus 128*0.03125 = 4 channels are flagged on each edge.  In full-polarization, there are only 64 channels, so 2 channels are flagged on each edge.  In single-polarization mode, there are 256 channels, so 8 channels are flagged on each edge.  In early observing Cycles, the ACA correlator sometimes produced 2\,GHz spws with 124 channels (dual polarization) or 248 channels (single polarization).  The additional parameter \param{fracspwfps}, with a default value of 0.048387, defines the number of edge channels to flag in these cases, e.g. 124*0.048387 = 6.

When FDM spws are placed up against the edge of the baseband anti-aliasing filter, the noise in the edge channels is very high, causing noisy channels in the cube which lead to deviant beamsizes and foil the common beam mechanism (\S\ref{tcleanCubes}).  A frequent case occurs on the ACA correlator when ``FDM'' windows of 1000\,MHz bandwidth are defined, which are significantly wider than their 937.5\,MHz counterparts on the 12\,m array, and have a greater chance of having channels too close to the edge of the baseband. The `edge' heuristic flags the problematic edge of these spws, which are selected by their bandwidth without consideration of the correlator type, because the BLC does not produce spws with these bandwidths.  The amount of bandwidth flagged is given in Table~\ref{tab:acaEdgeFlag}. As of Cycle 7, the OT does warn the user when an spw approaches within 30\,MHz of a baseband edge.

\begin{table}
\caption{Amount of bandwidth trimmed on ACA correlator spws that reach (or approach) one (or both) baseband edges (\S\ref{edges}).\label{tab:acaEdgeFlag}}
\begin{tabular}{cc}
\textbf{\spwbw} & \textbf{trimmed bandwidth}\\
\textbf{(MHz)} & \textbf{(MHz)} \\
\hline
2000 & 62.5\tablenotemark{a}  \\
1000 & 62.5 \\
500  &	40\\
250  &	20\\
125  &	10\\
62.5 &	5 \\
\end{tabular}
\tablenotetext{a}{Amount trimmed from both baseband edges}
\end{table}

\subsubsection{Low transmission}
\label{lowTransmission}

The sensitivity to celestial signals at ground-based submillimeter observatories decreases precipitously as the atmospheric transmission decreases \citep{Kawamura2002}, and some of the ALMA receiver bands include narrow frequency ranges with extremely low transmission on Chajnantor \citep{Paine2000,Pardo2022}. In order to avoid situations where the user has placed an spw on an unusable portion of the spectrum, this heuristic identifies spws where more than 60\% of the channels (considering all integrations of all scans on the science targets) are predicted to have transmission less than \param{mintransnonrepspws}=10\% and completely flags those spws on all scans of the MS. The transmission is computed using the atmospheric model tool in CASA, which is based on the package Atmospheric Transmission at Microwaves \citep[ATM,][]{Pardo2001} with periodic updates applied. The following initialization values are used: pressure=563\,mb, altitude=5059\,m, temperature=273\,K, relative humidity=20\%, scale height of water vapor=1.0 km, highest altitude layer=48\,km, initial pressure step between layers=5.0\,mb, multiplicative pressure step=1.1, and atmType=``mid-latitude winter''.

In order to allow intentional searches for the celestial lines at challenging frequencies, a lower threshold of \param{mintransrepspw}=5\% is used for the representative spw (repSpw), which is the spw that contains the representative frequency specified by the user in the OT.    This heuristic uses the Astropy package \citep{astropy2022} to compute the elevation of targets based on celestial coordinates and time. Because Astropy no longer recognizes the J2000 coordinate frame (in favor of ICRS), ALMA data older than Cycle 3 currently requires this heuristic to be disabled by setting the Boolean value \param{lowtrans} to False in the PPR.  This constraint will be removed for the Cycle 10 release by converting all J2000 labels to ICRS in \pcmd{hifa_importdata}

\subsection{Flagging of known spectral features (in SSOs used as flux calibrators)}
\label{ssoLineFlags}
If the flux calibrator is a SSO, known strong lines in the object (e.g., CO in Titan's atmosphere) are flagged by the stage \pcmd{hifa_fluxcalflag}.   
If a large fraction (more than \param{threshold}=0.75) of a given spw is flagged for this reason, then a reference spw map is calculated and stored in the pipeline context. Later in \pcmd{hifa_gfluxscale}, the \param{refspwmap} parameter of the \ccmd{fluxscale} task will be set to this value in order to transfer the flux scale from the nearest other spw. The WebLog shows if any flagging or spwmap was required. In Mars, Venus, Titan, and Neptune, $^{12}$CO is flagged in all ALMA bands. In Mars, Venus, and Titan, $^{13}$CO is also flagged. In Titan, HCN, H$^{13}$CN, and HC$^{15}$N are also flagged, as is HCN v$_2$=1. Finally, in Titan, CH$_3$CN is flagged up through Band 8.   The latter two species are not present in the CASA model (see \citet{ButlerALMAMemo594} for details). 

The general policy of flagging all strong lines in SSO objects is a holdover from earlier years when the SSO models in CASA and the mechanism for computing their model visibilities was less well developed and tested.  Future pipeline development may revisit this flagging policy, at least for the combinations of species and objects that are accurately modeled, in particular those that are less sensitive to planetary seasonal variations.

The choice of bandwidth to be flagged around each line center was based on either published spectra or ALMA commissioning data of 1 or 2 transitions of each species. For other transitions, these frequency widths are scaled to maintain the same velocity widths at other line frequencies. A detailed list of the topocentric frequency ranges flagged is given in Appendix~\ref{appendixB}. Note that the flags must be applied in the native (topocentric) frequency frame of ALMA observations, and the central frequency of the line will vary throughout the year and the day.  Rather than computing the exact shift for each EB, for simplicity the frequency width to be flagged is simply broadened by the relative velocity extrema between the object and the geocenter computed once over a decade (2013-2023).

Because not all SSOs with atmospheres have been surveyed for spectral lines in all ALMA bands, additional spectral ranges to be flagged can be manually specified to augment the built-in list by specifying the name of a text file in the \param{linesfile} parameter.  The simple format for each line of this file is four values (space-delimited): field name, line name (user-defined), start frequency (GHz), and stop frequency (GHz). Finally, the built-in list of lines can be ignored in favor of the user-specified ranges by setting \param{appendlines}=False.

\subsection{Channel-based flags}
\label{rawchanflags}
In the stage \pcmd{hif_rawflagchans},
the raw visibilities are inspected for severe outliers in amplitude in specific channels or ranges of channels that correspond to a quarter of the \spwbw. Typically these outliers only occur due to hardware problems that persist throughout the EB, thus only the bandpass calibrator scan is evaluated.  As the ALMA observatory has matured, this heuristic is not frequently triggered, and was more relevant during early Cycles of observations when the online correlator calibration and flagging mechanism was less well developed. Currently, the stage sometimes triggers on strong atmospheric ozone lines, and a mechanism to avoid placing flags in these known ranges is planned for a future release.  If RFI becomes a problem in future years, this stage may once again become more relevant.  

\subsection{Flagging of visibilities that are outliers in calibrated amplitude}
\label{correctedampflagtext}
Substantial effort was applied in the early observing Cycles of ALMA to develop and refine strategies to identify and flag outlier visibilities.
While flagging on the basis of outliers in gain {\it solutions} is helpful to remove obviously bad data, as is implemented in \pcmd{hif_lowgainflag} (\S\ref{lowgainflag}), we found that this method will ultimately generate unnecessary flags as the threshold is lowered. Instead, the philosophy we adopted is that only outlier data points that have remained outliers {\it after} calibration should be flagged. This philosophy requires solving and applying temporary calibration tables for the purpose of identifying and flagging bad data, prior to performing the final calibration stages.

\subsubsection{Calibrators}
\label{correctedampflagcal}
The \pcmd{hifa_bandpassflag} stage performs a preliminary phased-up bandpass solution and an `int'-based gain solution (see \S~\ref{refantpolcal}), both using \param{minblperant}=4 and \param{minsnr}=3. It temporarily applies these solutions, then computes the flagging heuristics by calling \pcmd{hif_correctedampflag} which looks for outlier visibility points by statistically examining the scalar difference of the corrected amplitude minus model amplitude, and then flags those outliers. This method replicates a human examining a plot of amplitude vs. either time or {\it uv}-distance and looking for deviations from the expected pattern of a point source calibrator, i.e. equal flux on all baselines and all times.  For data with heterogeneous antenna diameters, the groups of baselines are assessed independently (12\,m-12\,m, 7\,m-7\,m and 12\,m-7\,m) due to their differing noise spread. Note that the phase of the data is not assessed, as it is fairly rare for an outlier in calibrated phase to not also impact the calibrated amplitude.  For multi-scan calibrators (phase and polarization), only the Stokes I product (XX+YY) is assessed rather than individual polarization products, as the parallactic angle rotation between scans can be significant.

\begin{figure*}
\centering
\hspace*{-5mm}\includegraphics[scale=0.666]{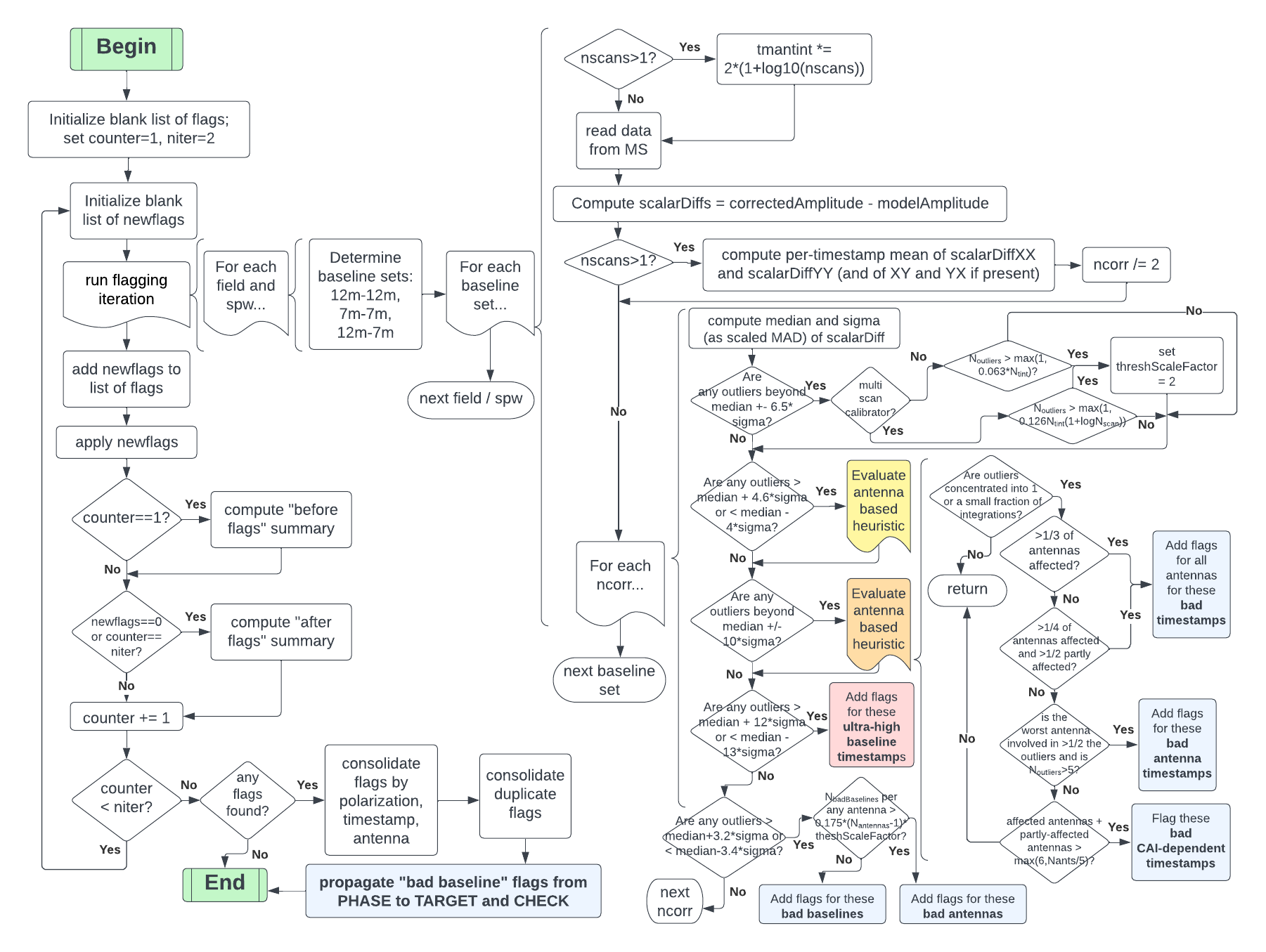}
\caption{Logic flow in \pcmd{hif_correctedampflag} for generating flags on calibrators.  The blue and red-colored boxes denote locations where flagging commands are generated.  Only the first of the qualifying antenna-based heuristics (yellow or orange boxes) are executed, via the sub-sequence of commands to their right. The logic for target sources flags only ultra high points (red box) computed within overlapping uv range bins, as described in \S\ref{correctedampflagScience}. }
\label{fig:correctedampflag}
\end{figure*}

There are several tiers of flagging: regular, high, very high, and ultra high. 
First, for each spw and polarization product to be assessed, the median and MAD ($\sigma$) are computed after vector averaging each visibility spectrum across unflagged channels.  Regular outliers are points that are $>+3.2\sigma$ or $<-3.4\sigma$ relative to the median.  If there are any points $<-6.5\sigma$, then the presence of decorrelation (which lowers the amplitudes) is considered likely and the lower threshold for regular outliers is doubled to $<-6.8\sigma$.  This decorrelation heuristic could be improved in the future by requiring outliers to be spread over a significant number of antennas rather than only one (or a few), as this relaxation can lead to failure to flag bad data when all the outliers are due to a single antenna. High outliers are points at $>4.6\sigma$ or $<-4\sigma$ and are used for identifying antenna-based flags. Very high outliers are points $>+10\sigma$ or $<-10\sigma$ and are used for identifying antenna-based flags if no high outliers were present.  Finally, ultra high outliers are $>+12\sigma$ or $<-13\sigma$ and they are flagged as individual baseline timestamps with no attempt to group them by time or antenna.  The logic flow is shown in Figure~\ref{fig:correctedampflag}.

Note that ultra high outliers are simply flagged and no assessment is made of antennas or timestamps involved.  In all cases, the polarization product is not included in the generated flag commands, so that no partial polarization flags are introduced, since these were all flagged in the earlier \pcmd{hif_flagdata} stage (see \S\ref{partialpol}).
Any antenna-based flags generated on the phase calibrator will naturally propagate to the science target and \checksource\/ via the application of the subsequently produced caltable solutions. However, baseline flags will not naturally propagate, so baseline flags on the phase calibrator that are present for all timestamps are explicitly propagated via flag commands to the science target and \checksource\/ data.

After the first iteration, flags are generated and applied, and another iteration is performed, which is sometimes necessary to identify a set of lower level outliers.   Plots are generated at two points in the workflow of this stage: after calibration but before flagging heuristics are run, and after all flagging heuristics have been run and applied. If no points were flagged, then the ``after'' plots are neither generated nor displayed in order to avoid unnecessary clutter of the WebLog page. The flagging statistics is also displayed. The ``before'' statistics are computed before applying the newly-generated flags, but after the temporary \ccmd{applycal} calls, while the ``after'' statistics are computed after applying the newly-generated flags.

The later stage \pcmd{hifa_gfluxscaleflag} operates in a similar manner on the flux calibrator (if it is different from the bandpass calibrator), the phase calibrator(s), and the checksource(s), but using  \param{minsnr}=2.  Because the final bandpass solution is already available to pre-apply, no temporary bandpass solution is needed. Finally, in polarization recipes, the stage \pcmd{hifa_polcalflag} operates analogously on the polarization calibrator data.

\subsubsection{Science targets}
\label{correctedampflagScience}
Because science targets are generally not point sources, the visibility flagging algorithm in \pcmd{hifa_targetflag} needs to be more clever than \pcmd{hifa_bandpassflag}, \pcmd{hifa_gfluxscaleflag}, and \pcmd{hifa_polcalflag}. First of all, rather than solving for gain solutions, it applies the interpolated solutions on the phase calibrator as registered into the CalLibrary in the prior stage \pcmd{hif_timegaincal}.  Because it lacks an {\it a priori} source model, it still simply assumes a point source model and computes the same difference in corrected minus model amplitude as for calibrators (\S\ref{correctedampflagcal}).  Then, for each science target, the algorithm identifies only the ultra high outliers, and it does this by examining statistics within successive overlapping radial uv bins, allowing it to adapt to arbitrary uv structure.  Outliers must appear to be a potential outlier in two adjacent bins in order to be declared an outlier.  This strategy reduces the chance that accurate high amplitude data from a rogue short baseline will be flagged unnecessarily.  The bins are defined by computing the overall minimum and maximum of the uv distance values that are unflagged, then build bins from the minimum ($r_{\rm inner}$) upward by a geometric factor ($f_{bin}$) until exceeding the maximum. Empirical tests show that the bins need to be closer together at short uv distances and can become wider at longer distances. We define the factor with 3 tiers: 
\begin{equation}
\begin{aligned}
f_{bin} &= 1.1, (r_{\rm inner} < 40\,\text{m})\\
    &= 1.2, (40\text{m} < r_{\rm inner} < 90\,\text{m})\\
    &= 1.26, (r_{\rm inner} \geq 90\,\text{m})
\end{aligned}
\end{equation}
Note that 1.26 is approximately the cube root of 2, which gives 3 bins per each doubling of scale beyond 90\,m.  For datasets with small numbers of uv points per field, which includes 7\,m mosaic snapshots, such fine binning results in too few points per bin for statistical flagging to work correctly (obvious outliers are missed). So if the number of uv points per field $<$ 1000, a fixed increment of $\sqrt{2}$ is used, leading to only 5 bins across the 7\,m array uv range. The main pitfall to avoid, which can happen for any dataset, is empty bins or bins with small numbers of points. Empty bins are simply skipped, while bins with small numbers of points ($<6$) are merged with the next larger bin, except for the final bin, which is merged with the prior bin and then reassessed.

Due to the typically larger variation in intensity vs. uv distance of science targets, particularly at short baselines, this stage uses the midhinge (the mean of the first and third quartiles) as the midpoint of the visibilities within each uv bin rather than the median.  Similarly, it uses the larger of two quantities that measure the spread of these data: the interquartile range (IQR) and a scaled version of the interdecile range (IDR). These two quantities ($s_1$ and $s_2$) are computed from the differences of the first and third quartiles (Q1 and Q3) and the first and ninth deciles (D1 and D9) as follows:
\begin{equation}
\begin{aligned}
s_1 &= 0.5*\text{IQR} = 0.5*(\text{Q3}-\text{Q1})\\
s_2 &= 0.5*\text{IDR}/1.9004 = 0.9502*(\text{D9}-\text{D1}).
\end{aligned}
\end{equation}
To guard against the case where a large fraction of outliers exist with very large amplitudes, which can skew the IDR to unreasonably high values, we limit how high $s_2$ can become to 2*IQR. 
As a guard against small number statistics, for example from mosaic snapshots, if there are fewer than 1000 points across all uv bins for a given field, then the standard median and MAD shall be used as the midpoint and spread.

Because this heuristic must operate on per-field basis (due to the possible large change in amplitude vs. science target or between the fields of a mosaic), this stage can add significant processing time, particularly in making the plots, even though each visibility is only read from the MS once. So to save time, the plots of amplitude vs. time are made only if flags are generated for a given target field, and the plots of amplitude vs. uv distance are made for only those spws that generated flags for that field. Note that in mosaic observations, these plots vs. uv distance only include data from fields in the mosaic with new flags, and these field numbers are indicated in the captions.

\section{Imaging Heuristics}
\label{imgsection}

The primary purpose of the imaging pipeline is to confirm that the user-requested angular resolution and sensitivity level can be achieved by the calibrated data. For many projects, however, the images produced can be used directly for science analysis and publication. The imaging strategy for the pipeline takes into account the properties of each dataset when determining the characteristics of the science target images to produce.  Table~\ref{tab:imgRecipe} contains the ordered list of pipeline tasks for the standard imaging recipe. 

\subsection{Data preparation}

The \pcmd{hif_mstransform} stage produces for each EB a new MS that contains data only for the science target(s), with the suffix \file{``_targets.ms''}.  Specifically, the \ccmd{mstransform} task reads the calibrated data from the CORRECTED column of the parent MS and writes them to the DATA column of the new MS.

\subsection{Image weighting selection}
\label{weights}

The \pcmd{hifa_imageprecheck} stage uses the CASA toolkit to estimate the synthesized beam and sensitivities for the representative science target and frequency for four possible values of the \pcmd{tclean} \param{robust} weighting parameter \citep{Briggs}: 0.0, +0.5 (default), +1.0, and +2.0.  The beam calculations are intrinsically multi-frequency synthesis continuum calculations, using the \bwForSensitivity\/ specified by the user in the OT, rounded to the nearest integer number of channels.
If \bwForSensitivity\/ $>$ \repSpwBW, then the beam is predicted using all spws, otherwise the beam is predicted for the repSpw alone. 
The \param{robust} value chosen is the one closest to the default (+0.5) that predicts a beam area (defined as simply major axis $\times$ minor axis) that is in the range of the beam areas corresponding to the user selection in the OT, known as the "Min / Max Acceptable Resolution"\footnote{This value is present in the ASDM metadata for data from Cycle 5 onward.} If none of these \param{robust} values predict a beam area that is in range, then +2.0 is chosen if the predicted beam area is too small, and 0.0 is chosen if the predicted beam area is too large.   The chosen value is used for all subsequent science target imaging.

In addition to a beam estimate for the repBW, an estimate for the aggregate continuum bandwidth (aggBW) is also computed assuming no line contamination but accounting for spw frequency overlap.   All of the estimates in this stage account for \tsys, the observed uv-coverage, and prior flagging, but they do not account for (1)  manually-requested flagging in the imaging pipeline (in \pcmd{hifa_flagtargets}); (2) loss of continuum bandwidth due to the avoidance of channels containing spectral lines (as identified in \pcmd{hif_findcont}) when producing the continuum images; (3) Issues that affect the image quality like (a) poor match of uv-coverage to image complexity; (b) dynamic range effects; (c) calibration deficiencies (poor phase transfer, residual baseline based effects, residual antenna position errors, etc.).  
Because the beam calculations are multi-frequency synthesis (mfs) continuum calculations (see \S~\ref{contmfs}), the synthesized beam for a single channel in a cube can be larger depending on the details of uv-coverage and \cubechanwidth, particularly for snapshot mosaics. The adoption of \param{perchanweightdensity}=True and \param{weighting}=`briggsbwtaper' with CASA 6.2.1.7 for Cycle 8 has partially mitigated this effect. 

In the WebLog, a table containing all the estimated values is presented and the row corresponding to the chosen \param{robust} value is highlighted in green. A message appears on the ``By Task'' view if a non-default value of \param{robust} (i.e., not +0.5) is chosen. Additionally, if the predicted beam is not within the PI requested range using one of the four values, then warning messages appear on this page.

\subsection{Product size mitigation}
\label{mitigation}

While the long-standing ALMA goal is to image every channel of every spw for every science target for all projects, with the current state of computing and CASA tools there are practical limits to the file size of individual cubes that can be cleaned \citep[see][]{memo623}.  The total data volume of all cubes can also be challenging to download by users when there is a large number of spatial pixels and channels per spw coupled with a large number of spws and/or targets.  Although parallel processing on multiple cores of a single cluster node using the Message Passing Interface (MPI) mechanism of CASA \citep{Castro17} was added in the Cycle 6 pipeline release, the processing time of large cubes remains prohibitive using the default imaging parameters.  For these reasons, the pipeline performs a product mitigation step in the \pcmd{hifa_checkproductsize} stage, which will modify the characteristics of the imaging products in order to decrease their size.

Datasets that have been mitigated will have imaging products with different characteristics than those that have not been mitigated. Full imaging products can be recreated by users, by modifying the \ccmd{tclean} commands that are in the \file{casa_commands.log} file, or by calling the appropriate \pcmd{hif_makeimlist}/\pcmd{hif_makeimages} pair of stages with all default values, which will attempt to make full imaging products without mitigations. The mitigations are done in a priority order, with the mitigation halted once the predicted sizes fall below the thresholds.   In ALMA Cycle 7, approximately 20\% of MOUS were mitigated, and 75\% of those reduced the number of products \citep{memo623}. Nearly all cases are from antenna configurations C43-4 and larger.

The parameters that control the mitigation and their default values are:
\begin{itemize}
\item \param{maxcubesize} (40 GB): cube size at which to start triggering mitigation
\item \param{maxcubelimit} (60 GB): maximum cube size that will be produced. Also controls the number of large cubes.
\item \param{maxproductsize} (500 GB): maximum size of products that will be produced
\end{itemize}
The pipeline recipes explicitly encode these values so they can be easily changed universally for all pipeline runs. The \file{casa_pipescript.py} also encodes these values explicitly, so they can be easily changed on a per-MOUS basis, and the pipeline re-run using the modified file. The size calculations (in gigabytes) for a dataset are based on the following definitions:
\begin{equation}
\begin{split}
mfssize & = 4n_{\rm x}n_{\rm y} / 1\text{e}9\\
cubesize & = 4n_{\rm x}n_{\rm y}n_{\rm chan} / nbin_{\rm mitigation} / 1\text{e}9\\
productsize & = 2\sum_{products} (mfssize + cubesize),\\
\end{split}
\end{equation}
where 4 is the number of bytes per pixel, $n_{\rm x}$ and $n_{\rm y}$ are the number of pixels in the right ascension and declination axes, $n_{\rm chan}$ is the number channels in the spw, $nbin_{\rm mitigation}$ is the factor of channel averaging to be applied during imaging (either 1 or 2), and the factor of 2.0 accounts for the size of the corresponding primary beam images/cubes which are delivered to the user along with the intensity images/cubes.

\begin{figure*}
\centering
\hspace*{-4.5mm}\includegraphics[scale=0.68]{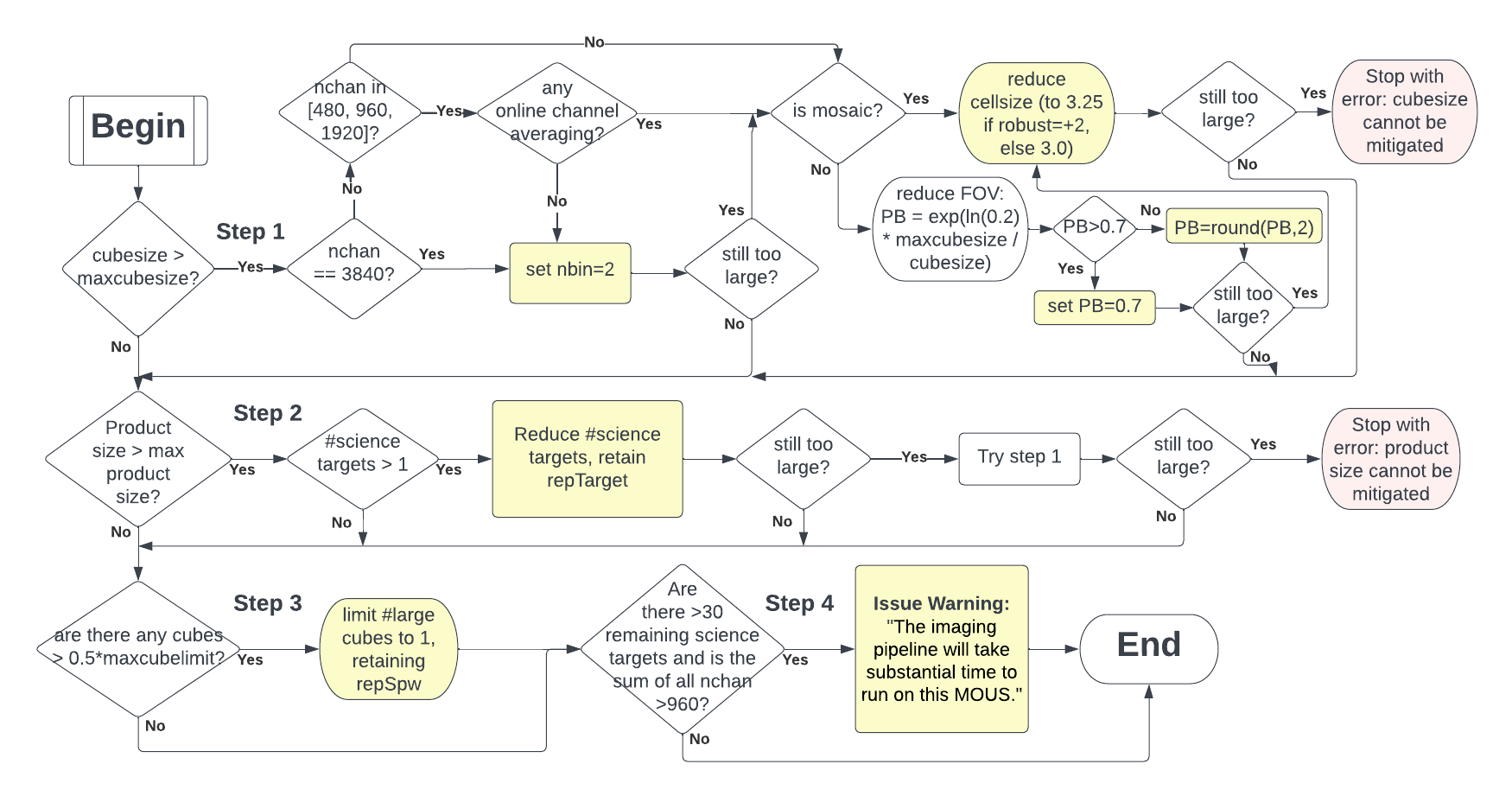}
\caption{The logical workflow of the cube size and product size mitigation procedure in \pcmd{hifa_checkproductsize} (\S\ref{mitigation}).}
\label{fig:mitigation}
\end{figure*}

The mitigation cascade, shown in Figure~\ref{fig:mitigation}, proceeds through the following four steps.  
\begin{itemize}
\item Step 1: If any of the cubes have a $cubesize$ larger than $maxcubesize$, then for each such cube,
\begin{enumerate}[(a)]
\item Check to see if online channel averaging was performed. If not, and the spw has 480, 960, 1920 or 3840 channels, then apply a factor of 2 averaging by setting $nbin_{\rm mitigation}$=2.  
\item If the new $cubesize$ is still too large, and the cube is not a mosaic, then reduce the field of view by calculating the Gaussian primary beam response level ($PB'$) at which the largest $cubesize$ of all targets is equal to the $maxcubesize$ using these equations:
\begin{equation}
\begin{split}
PB_0 & = \texttt{exp}(\texttt{ln}(0.2) * maxcubesize / cubesize),\\
PB' & = \texttt{round}(\texttt{min}(1.02PB_0,0.7), 2),
\end{split}
\end{equation}
where the latter equation ensures that the field of view encompasses (at least) the 0.7 response level.
\item If the $cubesize$ is still too large, then change the pixels per beam (\param{cell}) from 5 to 3.25 if \param{robust}=+2, or to 3.0 otherwise.
\item If the cube is still too large, then stop with an error: ``the largest size cube(s) cannot be mitigated.''
\end{enumerate}

\item Step 2: If $productsize > maxproductsize$, then
\begin{enumerate}[(a)]
\item If the number of science targets (single fields or mosaics) is greater than 1, then reduce the number of targets
to be imaged until $productsize < maxproductsize$. The representative target is always retained. Targets that have been mitigated are not imaged. They  are also not assessed in \pcmd{hif_findcont}, thus they are not continuum subtracted.
\item If $productsize$ is still too large, then repeat steps 1a, 1b, and 1c, recalculating $productsize$ each time.
\item If $productsize$ is still too large, then stop with the error: ``$productsize$ cannot be mitigated.''
\end{enumerate}

\item Step 3: For projects with large cubes that can be mitigated, restrict the number of large cubes that will be
cleaned, as they may each require long cleaning times:
\begin{enumerate}[(a)]
\item If there are cubes with sizes greater than $0.5 * maxcubelimit$, then limit the number of large cubes to be cleaned
to 1. The spw encompassing the representative frequency shall always be among the cubes that are cleaned. In any case, all spws are processed in \pcmd{hif_findcont}, \pcmd{hif_uvcontfit}, and \pcmd{hif_uvcontsub}.
\end{enumerate}

\item Step 4: For projects that have many science targets to image (i.e., still $>$30 after Step 3) and a large number of channels integrated over all spws ($>$960), then a warning is printed to the WebLog that ``the pipeline will take substantial time to run on this MOUS.''

\end{itemize}
When the cube size or product size cannot be mitigated, the following warning will appear at the top of the \pcmd{hif_checkproductsize} stage: ``QA Maximum cube size cannot be mitigated''
and then the pipeline will stop in the first \pcmd{hif_makeimlist} 
that creates cubes with the message: ``Error! Size
mitigation failed. Will not create any clean targets.''

\begin{table}[ht]
\caption{The stage names and stage numbers of the standard imaging pipeline recipe in the 2022 release. Modifications of the recipe will result in different stage numbers, as typically happens with each annual release.\label{tab:imgRecipe}}
\begin{tabular}{lll}
\textbf{\#\tablenotemark{a}} & \textbf{\#\tablenotemark{b}} & \textbf{Task name} \\
\hline
1 & ... & \pcmd{hifa_restoredata}                    \\
2 & 28 & \pcmd{hif_mstransform}                    \\
3 & 29 & \pcmd{hifa_flagtargets}                    \\
4 & ... & \pcmd{hifa_imageprecheck} \\
5 & ... & \pcmd{hifa_checkproductsize} \\
6 & 30 & \pcmd{hif_makeimlist} (per spw continuum) \\
7 & 31 & \pcmd{hif_findcont}                       \\
8 & 32 & \pcmd{hif_uvcontfit}                       \\
9 & 33 & \pcmd{hif_uvcontsub}                       \\
10 & 34 & \pcmd{hif_makeimages} (per spw continuum) \\
11 & 35 & \pcmd{hif_makeimlist} (aggregate continuum)\\
12 & 36 & \pcmd{hif_makeimages} (aggregate continuum) \\
13 & 37 & \pcmd{hif_makeimlist} (cubes)\\
14 & 38 & \pcmd{hif_makeimages} (cubes) \\
15 & 39 & \pcmd{hif_makeimlist} (representative cube)\\
16 & 40 & \pcmd{hif_makeimages} (representative cube)\\
17 & 41 & \pcmd{hifa_exportdata}\\
\end{tabular}
\tablenotetext{a}{Stage number in the imaging-only recipe}
\tablenotetext{b}{Stage number in the calibration+imaging pipeline recipe (as a continuation of the numbering in Table~\ref{tab:calRecipe}).}
\end{table}

\subsection{General imaging heuristics}
\label{makeimages}
Imaging is performed using the \pcmd{tclean} task within the pipeline task \pcmd{hif_makeimages}, which is used for all of the different types of imaging stages listed in Table~\ref{tab:imgRecipe}. The heuristics and details common to all imaging stages are described in this subsection. Details pertaining to specific imaging stages are then described in subsequent subsections.  For all image products, the full \pcmd{tclean} command that
produced it can be accessed easily from a link associated with each thumbnail image shown on the WebLog page of each \pcmd{hif_makeimages} stage.

\subsubsection{Theoretical sensitivity}
\label{sensitivity}
An estimate of the theoretical sensitivity of an image ($\sigma_{\rm theory}$) is essential for various heuristics in the imaging pipeline stages.
The theoretical noise RMS ($\sigma_{\rm i,j,s,v}$) for an interferometric visibility $v$ for a channel of spw $s$ 
with noise $\sigma_{\rm s,v}$ (described by Equation~\ref{sigmaColumnInitialized}) between two antennas $i$ and $j$ in the weak signal limit is given by:
\begin{equation}
\sigma_{\rm i,j,s,v} (Jy) = \frac{2k\sigma_{\rm s,v}} {\eta A_{\rm eff}}  \sqrt{T_{\rm sys,i} T_{\rm sys,j}} \times 10^{26} {\rm Jy},
\end{equation}
where $k$ is the Boltzmann constant, $\eta$ encompasses a product of efficiency factors due to quantization and other analog and digital losses, and $A_{\rm eff}$ is the effective collecting area of the antenna accounting for illumination, spillover, pointing, focus, polarization, and surface efficiencies \citep[see Equations 7-33 of][]{Crane89}.  

As described in \S\ref{importdata}, the SIGMA column of each MS is initialized upon data import only with the $\sigma_{\rm s,v}$ term.
The visibility weights are computed from SIGMA column and scaled by the product of the \tsys\/ values and the product square of the antenna amplitude gains ($g_i$ and $g_j$) in the final \pcmd{applycal} calls (\S\ref{hif_applycal}) because the \param{calwt} parameter is set to True by the pipeline whenever amplitude tables are applied.  The derived gain values will naturally reflect the relative sensitivity of antennas with different antenna diameters, and the flux calibration establishes the overall scale.  Therefore, in practice, the values of $k$, $\eta$, and $A_{\rm eff}$ never need to be set.  

To estimate $\sigma_{\rm theory}$, the \pcmd{hifa_imageprecheck} stage uses the \ccmd{apparentsens} function of the CASA \ccmd{image} toolkit, which accounts for both image weights and visibility weights.  However, this function does not properly compute the effective noise bandwidth of a collection of contiguous channels, it merely sums them.  The correction applied by the pipeline for this property is described in \S\ref{cleanThreshold}.

\subsubsection{Cleaning strategy}
\label{cleanStrategy}
As many imaging heuristics require knowledge about properties of the resultant image (e.g. dynamic range), the imaging process is broken down into two sequential \pcmd{tclean} calls, denoted `iter0' and `iter1'. The `iter0' call creates a dirty image, upon which statistics can be calculated, and then the `iter1' call utilizes this information and cleans the image. The `iter0' files are copied and used as the start point for the `iter1' call.  

On rare occasions when the emission is strong but very compact and the pruning criteria of the automask algorithm (\S\ref{automask}) removes the entire mask, another iteration is invoked (`iter2') which uses a clean mask set by the primary beam response level ($PB>0.3$).  This circumstance can also exist on occasion for spectral lines, especially masers, but due to the time expense of recreating the cube, `iter2' is not enabled for cubes.

\subsubsection{Cleaning threshold}
\label{cleanThreshold}
The cleaning \param{threshold} intensity level in the `iter1' call ($I_{\rm threshold}$) is nominally set to \param{tlimit}=2 times the expected image sensitivity ($\sigma_{\rm theory}$, see \S\ref{sensitivity}), but properly accounting for the effective noise bandwidth when a range of contiguous channels selected.  Since the ALMA correlators apply the Hann window \citep{Harris78} to the data (BLC: in the time domain; ACAC: via frequency profile synthesis to match the BLC data product), the effective noise bandwidth of a single isolated channel ($\Delta\nu_{\rm eff}$) is 8/3 (or 2.667) times the physical bandwidth of the channel ($\Delta\nu_{\rm chan}$)  when no online channel averaging is performed prior to data storage, i.e., $N_{avg}=1$.  Table~\ref{chanavgfactor} shows the noise bandwidth correction factor ($f_{\Delta\nu}$) for the other possible values of $N_{avg}$ selected by the user. Since Cycle 3, these correction factors are automatically applied to spw metadata in the ASDM by the control system and propagate to the EFFECTIVE\_BW column of the MS SPECTRAL\_WINDOW table. Thus, \texttt{apparentsens} will compute the correct single channel value of $\sigma_{\rm theory}$ for the case of cube imaging (\S\ref{tcleanCubes}). The value of $N_{avg}$ for each spw is included as a column (named "Online Spec. Avg.") in the table on the Spectral Setup Details page of the WebLog.  
\begin{table}
\caption{The effective noise bandwidth correction factor ($f_{\Delta\nu}$) due to the Hann window function for each possible level of user-selected online channel averaging ($N_{avg}$). \label{chanavgfactor}}
\begin{tabular}{cc}
$N_{avg}$ & $f_{\Delta\nu}$\\
\hline
1 & 2.667 \\
2 & 1.600 \\
4 & 1.231 \\
8 & 1.104 \\
16 & 1.049 \\
\end{tabular}
\end{table}

\begin{figure}
\centering
\includegraphics[scale=0.48]{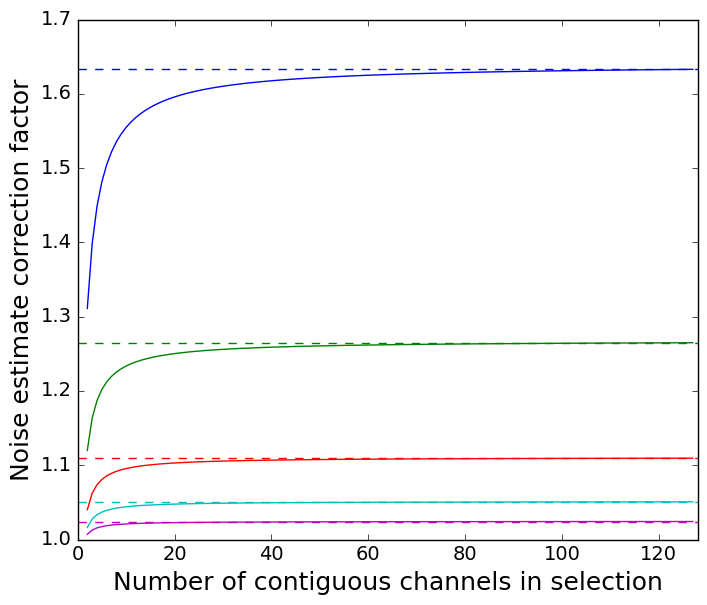}
\caption{The noise correction factor ($f_\sigma$) described by Equation~\ref{noiseEquation} for the case of $n_{\rm chan}=128$, for each value of $N_{avg}=1, 2, 4, 8, and 16$, from top to bottom (colored blue, green, red, cyan, magenta).  The dashed lines show the desired asymptotic value for large $w_{\rm chan}$, which correspond to $\sqrt{f_{\Delta\nu}}$ from Table~\ref{chanavgfactor}. }
\label{fig:noiseCorrection}
\end{figure}

In contrast to cubes, continuum images are built from collections of contiguous channel ranges, each of which has a different net effective noise bandwidth.  Thus, each range of channels found by \pcmd{hif_findcont} (\S\ref{findcont}) is assessed for its effective noise bandwidth using the following empirical formula in order to generate the noise correction factor ($f_{\sigma}$) to apply to $\sigma_{\rm theory}$ to raise its value above the optimistic one computed by \texttt{apparentsens}:
\begin{equation}
\label{noiseEquation}
\begin{aligned}
\Delta\nu_{optimistic} & = w_{\rm chan} \Delta\nu_{\rm eff}\\
\Delta\nu_{eff,approx} & = \biggl(w_{\rm chan} + 1.12 \frac{(n_{\rm chan} - w_{\rm chan})}{n_{\rm chan}N_{avg}}\biggr) \Delta\nu_{\rm chan}\\
f_{\sigma} &= (\Delta\nu_{optimistic} / \Delta\nu_{eff,approx})^{0.5},
\end{aligned}
\end{equation}
where $w_{\rm chan}$ is the width of the contiguous block of channels being considered. This function can be applied to arbitrary integer values of $w_{\rm chan}$ and is plotted in Figure~\ref{fig:noiseCorrection} for the five possible values of $N_{avg}$.

When the image dynamic range is significant, the predicted \param{threshold} is raised by a multiplicative factor, the dynamic range (DR) correction factor ($f_{DRC}$), to account for the fact that any residual phase error will scatter the science target flux around the image and raise the apparent RMS noise.  This factor is determined via the ``dirty dynamic range'' (dirty DR) of the dirty image (defined as the ratio of the image peak to theoretical noise). The $f_{DRC}$ values for calibrators are shown in Table~\ref{calibratorDR} and the factors for science targets are shown in Table~\ref{scienceDR}.  Once self-calibration has been implemented in the future (\S\ref{selfcal}), the $f_{DRC}$ value will effectively be reduced by those heuristics.

\begin{table}
\begin{tabular}{cc} 
\textbf{Calibrator} & \textbf{12m array dynamic range}\\
\textbf{Dynamic Range} & \textbf{correction (DRC) factor} \\
\hline
${\leq}$ 1000 & 1\\
1000 -- 3000 & DR/1000\\
${\geq}$ 3000 & DR/3000\\
\end{tabular}

\begin{tabular}{cc}
\hline
\textbf{Calibrator} 
& \textbf{7m array dynamic range}\\
\textbf{Dynamic Range} 
&
\textbf{correction factor -- 1EB}\\
\hline
${\leq}$ 200 & 1 \\ 
${\geq}$ 200 & DR/200\\ 
\end{tabular}
\caption{Dynamic range correction factors ($f_{\rm DR}$) to use when computing the \ccmd{tclean} noise \param{threshold} for calibrator imaging (\S\ref{cleanThreshold}).\label{calibratorDR}}
\end{table}

\begin{table}
\begin{tabular}{cc}
\textbf{Dynamic Range (DR)} &
\textbf{12m DRC factor}\\
\hline
${\leq}$ 20 & 1 \\
20 -- 50 & {1.5}\\
{50 -- 100} & {2}\\
{100 -- 150} & {2.5}\\
{${\geq}$ 150} & \texttt{max}(2.5, DR/150)\\
\hline  
\end{tabular}

\hspace{-1cm}  
\begin{tabular}{ccc}
{\bf Dynamic} & \multicolumn{2}{c}{\bf Dynamic range correction factor } \\
\textbf{range (DR)} &
\textbf{7m -- 1EB} &
\textbf{7m -- $>$1 EB} \\ \hline
${\leq}$ 4 & 1 & 1 \\ 
4 -- 10 & 1.5 & 1.5 \\ 
10 -- 20 & 2 & 2 \\ 
20 -- 30 & 2.5 & 2.5 \\ 
30 -- 55 & \texttt{max}(2.5, DR/30) & 2.5 \\ 
{55 -- 75} & \texttt{max}(2.5, DR/30) & 3.0 \\ 
${\geq}$ 75 & \texttt{max}(2.5, DR/30) & \texttt{max}(3.5, DR/55) \\
\end{tabular}
\caption{Dynamic range correction factors ($f_{\rm DR}$) used to compute the \ccmd{tclean} noise \param{threshold} for science target and CHECK source imaging.\label{scienceDR}}
\end{table}

\subsubsection{Cleaning iteration limit}
\label{cleanNiter}
The limit on number of cleaning iterations ($n_{\rm iter}$, specified in \ccmd{tclean} by \param{niter}) is computed based on the default loop gain ($g_{\rm loop}$=0.1), the ratio of the peak intensity in the `iter0' image ($I_0$) to the requested \param{threshold} level, and a value proportional to the number of synthesized beam areas enclosed by the mask ($r_{\rm mask}^2$), which is approximated by simply assuming a circular mask with a radius equal to 45\% of the width of the square circumscribing the image and that there are 5 pixels across a circular synthesized beam with radius equal to the major axis of the actual beam. Specifically:

\begin{equation}
\begin{aligned}
\label{fiter}
r_{\rm mask}^2 &= (0.45 \times \texttt{max}(n_{\rm x},n_{\rm y}) / 5)^2\\
f_{\rm iter} &= \texttt{int}\biggl(\frac{\kappa r_{\rm mask}^2}{g_{\rm loop}} \times \frac{I_0}{I_{\rm threshold}} \biggr),
\end{aligned}
\end{equation}
where $\kappa$=5 is an empirically-derived factor, which is conceptually related to the typical number of major cycles required to clean
ALMA images.  For a single-field image, this assumption translates to a mask filling $\approx25$\% of the typical area imaged (i.e., the region above PB=0.2 for science images not mitigated for size). This floating point value is then rounded to a convenient integer via:
\begin{equation}
\begin{aligned}
p &= -\texttt{int}(\texttt{log10}(\texttt{round}(f_{\rm iter}))),\\
n_{\rm iter} &= \texttt{round}(10^{p} f_{\rm iter}) / 10^{p},
\end{aligned}
\end{equation}
with a maximum limit of $2^{31}-1$.  When cleaning cubes, \ccmd{tclean} compares the stopping value (\param{niter}) against the running total summed across all channels after each major cycle.  In most cubes, only a minority of channels yield large area masks, so the pipeline does not scale the \param{niter} request by the number of channels. In practice, cleaning typically stops on \param{threshold} rather than \param{niter}.  But for other cases, further research into more optimal values of \param{niter} may be worthwhile.

\subsubsection{Automasking for cleaning}
\label{automask}
The continuum mfs and cube images are cleaned using \ccmd{tclean} with \param{cyclefactor}=1 and the \param{auto-multithresh} mode \citep{Kepley2020}. It takes a top-down approach to producing cleaning masks: it uses the residual images to identify significant peaks and then expands the mask to include emission associated with these peaks down to lower signal-to-noise noise. The control parameters are set by the pipeline on the basis of the representative baseline length, the diameter of the smallest antenna, the type of source (science target vs. quasar calibrator or checksource), and whether it is cube vs. mfs imaging.  
Details can be found in \citet{Kepley2020}; the only changes since that publication are that the \param{sidelobethreshold} was decreased. It was changed from 3.0 to 2.5 for  12\,m array configurations  with 75\% baseline lengths greater than $400$\,m 
and from 3.0 to 2.0 for 12\,m array configurations with 75\% baseline lengths between 300\,m and 400\,m. Cleaning continues until reaching the \param{threshold} noise level (see \S\ref{cleanThreshold}).

\subsection{MFS continuum imaging}
\label{contmfs}
All continuum imaging uses the multi-frequency synthesis (MFS) algorithm in \pcmd{tclean} \citep{Rau2011} with \param{deconvolver}=`hogbom' \citep{hogbom1974}. In cases where the fractional bandwidth of the total frequency range spanned by the image exceeds 0.1, higher order Taylor terms are needed, and the multi-term multi-frequency synthesis algorithm (\param{mode}=`mtmfs' with \param{nterms}=2) is utilized. With the current ALMA IF bandwidth, this criterion can be met for the aggregate continuum image in Bands 5 and below.  The selection of non-line contaminated channels for continuum imaging is detailed in \S\ref{findcont}.

\subsubsection{Calibrator imaging}
\label{calibratorImaging}
As part of the calibration pipeline (\S\ref{calsection}), per-spw mfs continuum images are made of the calibrators and CHECK sources, all of which are generally point-like quasars. For the case of SSOs used as FLUX calibrators, even if they are not point-like they will have small angular diameter disks compared to the primary beam.   Therefore, the field of view of all calibrator images is limited to the central \param{calmaxpix}=300 pixels (automasking is still invoked within that region).  By default, one image is made per calibrator for all of the data in the MOUS, including baselines with heterogeneous antenna diameters, if present, in contrast to science target imaging which uses only baselines involving the majority antenna diameter.  All channels are combined.  Setting the parameter \param{per_eb}=True will generate an image of each calibrator for each EB.  
In most cases, the \param{phasecenter} is set to the field coordinates, but for the case of an SSO used as a calibrator\footnote{The only calibration intent currently allowed for an SSO by ALMA is FLUX.}, it is set to `TRACKFIELD' to produce an image in the rest frame of the object. 

Because the number of iterations required to clean the image of a single point source is relatively small, the \param{niter} limit is intentionally set to 3000 in order to prevent \pcmd{tclean} divergence in the event that some very bad data escapes the flagging heuristics (\S\ref{correctedampflagcal}). This empirically-derived value is roughly consistent with Equation~\ref{fiter} for the special case of point-like calibrators ($r_{\rm mask}^2$=1) requiring only one major cycle ($\kappa$=1), as the equation then reduces to $SNR/(g_{\rm loop} * tlimit * DRC)$, which for $SNR$=1000 yields $f_{\rm iter}$=1000 for the 7\,m array and 5000 for the 12\,m array (see Table~\ref{calibratorDR} for DRC values).

\subsubsection{Image analysis}
\label{noiseAnnulus}
The noise level in all pipeline images is computed on a product that is not corrected for primary beam response using a standard annulus defined as
the area between the PB=0.2 and 0.3 response levels, but excluding pixels within the \pcmd{tclean} mask.  If the field size has been reduced due to data size mitigation (\S\ref{mitigation}), then the outer radius of the annulus is raised to be tangent to the edge of the image and the inner radius is set to the PB response level corresponding to a radius that is 15\% less than the outer radius, as this relation similarly holds for the 0.2 and 0.3 response levels of a Gaussian beam.

Additional quantities reported in the WebLog for each image 
are: the center frequency, fractional bandwidth, aggregate bandwidth, the synthesized beam parameters (FWHM major axis, FWHM minor axis, and position angle), theoretical sensitivity, cleaning threshold, the dirty DR and the corresponding DR correction factor (\S\ref{cleanThreshold}), the maximum and minimum of the primary beam corrected image, and the image QA score, designed to indicate how close the measured noise is to the theoretical noise, considering also the DR correction factor.

\subsubsection{Check source analysis}
\label{checksource}
In this imaging stage, per-spw, per-EB images of the CHECK source(s) are created using the dynamic range correction
as for science targets (Table~\ref{scienceDR}). Afterward, each image is fit for a point source using CASA \ccmd{imfit} task, as these objects are quasars chosen from the ALMA calibrator catalog.
To assess the astrometric quality of the phase transfer calibration, the angular offset of the fitted position from the
observed position (i.e., from ALMA calibrator catalog) is shown in the WebLog table in units of mas and synthesized beams.  The nominal positions returned by the catalog query are typically measurements from prior ALMA observations.
Warnings are generated for offset values $>0.30$ synthesized beams.
To assess the level of decorrelation and/or presence of resolved emission, the ratio of the fitted peak intensity
to the fitted flux density is computed and shown in the table.  Warnings are generated for values $<0.8$.
Warnings are also generated if the SNR of the fitted or \pcmd{hifa_gfluxscale}-derived flux densities are low ($<$20).  See \S\ref{checksourceQA} for further details.

\subsection{Continuum subtraction and science imaging}

The stage \pcmd{hif_findcont} creates a dirty cube for each spw using \param{robust}=1, regardless of the 
value chosen in \pcmd{hifa_imageprecheck}, in order to guarantee good sensitivity to faint line emission while avoiding the high sidelobes of natural weighting.  Next, it examines the cube to find the most appropriate channels for continuum.
The details of this process are described in \S\ref{findcont}.
The selected channels are converted to LSRK frequency ranges and written to a text file called \file{cont.dat}. If this file already exists before \pcmd{hif_findcont} is executed, then it will first examine the contents. For any spw that already has frequency ranges defined in this file, it will not perform the analysis described above (and not generate the dirty cubes) in favor of the {\it a priori} ranges. For spws not listed in a pre-existing file, it will analyze them as normal and update the file. In either case, the file \file{cont.dat} is used by the subsequent \pcmd{hif_uvcontfit} stage to define the channels to use to fit the continuum level, and by the \pcmd{hif_makeimages} (continuum) stages to select the channels to combine for imaging.

If no continuum ranges are found, then the corresponding continuum-subtracted cube will not be subsequently cleaned in the \pcmd{hif_makeimages} (cube) stage (\S\ref{tcleanCubes}).  Similarly, if only a single range is found that occupies $>=$ 92.5\% of the channels ($>91\%$ if the spw has fewer than 75 channels, i.e. 64-channel full-polarization TDM spws), then the spw is declared as ``AllContinuum'' and no cleaning will be attempted in the subsequent cube stage, in order to save on unproductive processing time.

For the \pcmd{hif_makeimages} (aggregate continuum) stage, if at least one spw does find a continuum range, but some others do not, then it will ignore the spws that found no continuum range, and will combine only the selected channel ranges from the successful spws.

Mitigated spectral windows (\S\ref{mitigation}) will still have continuum subtraction done for them, but mitigated sources will not have continuum subtraction done.

\label{uvcontsub}

\subsection{Cube imaging}
\label{tcleanCubes}

\subsubsection{\pcmd{tclean} parameters}

The \param{restoringbeam} is set to `common', which produces a constant beam size across all channels of the cube. The \param{gridder} is `standard' for single field images and `mosaic' for mosaics and any single fields that image data with heterogeneous antenna diameters, which currently occurs only in the calibrator imaging stage (\S\ref{calibratorImaging}).  The frequency frame used for cubes is LSRK, with the exception of SSO targets, which use the object's rest frame (REST).
The \cubechanwidth\/ for an spw is set by converting the frequency width of one topocentric channel into the frequency frame of the cube using the \ccmd{advisechansel} function of the \ccmd{synthesisutils} toolkit with \param{getfreqrange}=True.  This conversion is performed for each MS and the widest value is used. 

The number of channels is set by searching for the first non-flagged channel at either end of the spw per MS.  The weighted intersection of the ranges is then computed to avoid imaging edge channels that have drifted too much between MS.  The next step trims the cube frequency range not on the basis of flagged channels but on the physical boundaries of the spw in each MS in order to avoid failures with rare tunings located near the edges of LO ranges where successive EBs can be shifted from one another by tens of MHz due to tunability limitations of the LO hardware. 

Finally, one additional channel on each edge is dropped. For the REST frame case, \param{freqframe} is set to `SOURCE' and \param{phasecenter} is set to `TRACKFIELD'.
For spws that are found by \pcmd{hif_findcont} to contain no line emission (\S\ref{findcont}), the cube is not cleaned in order to 
save processing time.

\subsubsection{Representative bandwidth cubes}
\label{repBWcube}
If the \bwForSensitivity\/ requested by the user in the OT is larger than 4\cubechanwidth, then another cube of
the repSpw is produced using a coarser channel width in the final pair of \pcmd{hif_makeimlist} and \pcmd{hif_makeimages} stages.   The channel binning 
factor is rounded to the nearest integer number of channels.  This cube is useful for QA
purposes as it provides a sensitivity level at a frequency resolution close to the user-requested value.

\subsubsection{Additional QA images in WebLog}

For each \pcmd{hif_makeimages} stage, thumbnail images are created of the dirty image, the cleaned image, the PB-corrected cleaned image, the clean residual image, the clean mask image, the clean model image, and the PSF image.  If the extent of the image is larger than 123 pixels in either dimension, the PSF thumbnail image includes an inset in the lower right corner showing the central 41 pixels, with a dotted contour indicated the half power level.  For cubes, the dirty image and clean residual image are shown both as a moment 0 and moment 8, and five additional image products are created for the WebLog: the moment 0 and moment 8 images of the line-free channels, a spectrum of flux density within the flattened clean mask (union over all channel masks), a spectrum of the rms noise ($\sigma_\nu$) computed by \ccmd{imstat} using the standard noise annulus (\S\ref{noiseAnnulus}), and the beam size vs. frequency.  In order to enable quick visual evaluation of the whether line emission contaminates the continuum ranges, the intensity scale of moment 8 image ranges from a minimum of \texttt{Mdn}$-$\texttt{unscaledMAD} to a maximum of $10\times\texttt{Mdn}(\sigma_\nu)$.

\section{Continuum channel finding}
\label{findcont}

Due to the abundance of strong of thermal lines in the (sub)millimeter wavelength regime \citep{Coude2016,Groesbeck1995}, 
an essential component of the imaging pipeline is the determination of the channels in each spw of a calibrated uv dataset to consider as ``continuum only'' when performing continuum subtraction prior to producing image cubes. This step occurs in the \pcmd{hif_findcont} stage which calls the Python function \pcmd{findContinuum} contained in the $\sim$10,000-line code file \pcmd{findContinuum.py}.  

\subsection{Advantages of the image plane technique}
\begin{figure*}
\centering
\includegraphics[scale=0.38]{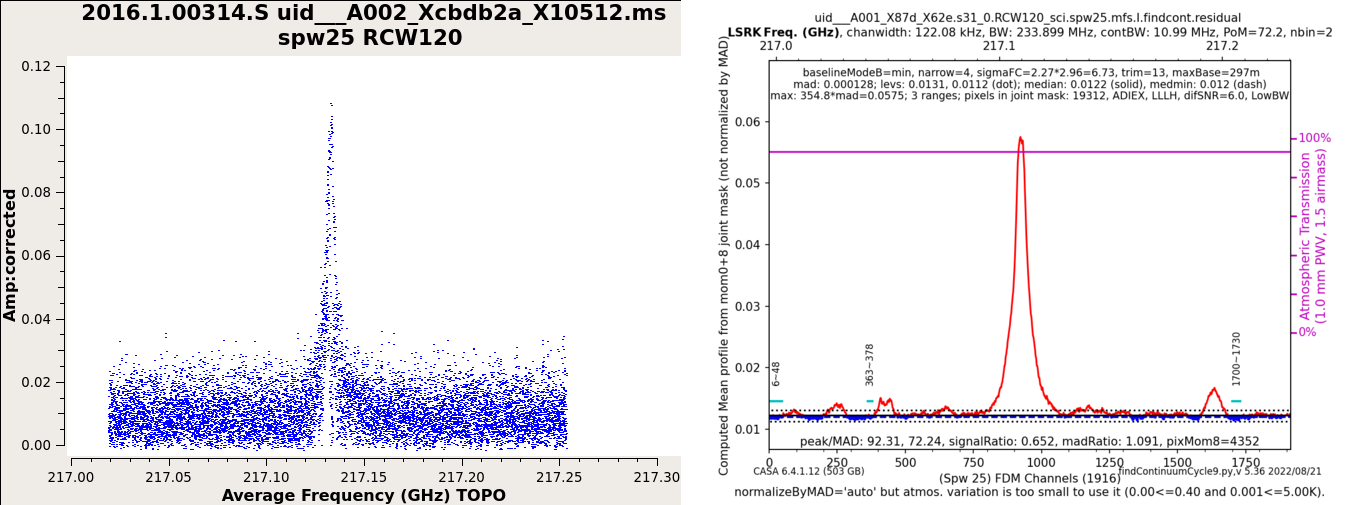}
\caption{Left panel) The time-averaged, baseline-averaged calibrated uv visibility spectrum of the single MS containing a 3-point mosaic of the massive star-forming region RCW120 centered on the bright SiO $J$=5-4 line. Both XX and YY polarization products are plotted.  Right panel) The mean spectrum (red trace) of the Stokes\,I dirty cube produced by the \pcmd{hif_findcont} stage of the pipeline, showing the three selected continuum channel ranges (as horizontal cyan line segments). In this case, the processing time required to generate and analyze the dirty cube is only about 5 times longer than to merely average and plot the spectrum of the calibrated visibilities.}
\label{fig:uvspecFindcont}
\end{figure*}

While the presence of line emission can be surmised from either the uv plane or the image plane, there are several complications and limitations of the uv plane approach.  First, in the uv plane, line emission that originates offset from the phase center will have a complicated phase structure vs. baseline length and orientation, which reduces the effectiveness of simply averaging the uv spectra.  In other words, because line emission can have vastly different strength on different baselines, a uv analysis would need to examine multiple uv ranges independently and aggregate them to produce a sensitive result. Second, in the case of multiple EBs, the visibility data would need to be assessed across multiple MS, which is not possible in \ccmd{plotms} without an expensive \ccmd{concat} operation.  Third, in the case of mosaic observations, visibilities from multiple fields would need to be assessed independently since simply averaging across fields will dilute the line signals that appear in only a subset of fields.   In light of these considerations, we developed a method to derive continuum channels by examining Stokes\,I image cubes, prior to cleaning, which have dimensions of longitude, latitude, and frequency/channel.  This approach naturally solves the latter two issues and converts the first issue into a matter of determining the appropriate spatial mask from which to construct a spectrum.  To illustrate the power of this technique, Figure~\ref{fig:uvspecFindcont} compares the visibility spectrum of a 3-point mosaic of a massive star-forming region \citep[RCW120,][]{Figueira2018} with the mean spectrum generated by \pcmd{findContinuum} in the pipeline.  While the SiO $J$=5-4 line is strongly detected by both techniques, the line wings and the fainter lines from hot core species (complex organic molecules) are buried in the noise of the uv spectrum, but are visible in the image plane spectrum.  

\subsection{Overview of algorithm}
\label{testset}
Development of the \pcmd{findContinuum} algorithm began in 2015 May, before the subsequent development of the auto-masking feature of \ccmd{tclean}.  Furthermore, because it is called prior to the image cleaning tasks, it must analyze dirty cubes without the benefit of an {\it a priori} emission mask.\footnote{For the manual use case, \pcmd{findContinuum} will accept an {\it a priori} mask via the \param{mask} parameter, which can lead to better performance, if constructed based on prior knowledge of the field or from an initial attempt at cleaning the cube.}  The algorithm first generates a representative spectrum designed to highlight all of the line emission in the cube and then identifies a list of channels from that spectrum.    While there are numerous control parameters and logic threshold values available for manual experimentation, all of them
default to an automatic heuristic designed and tuned to try to produce sensible results on all cubes. The default values were established based on experience with running the code on an extensive test set of $\sim
$200 ALMA image cubes, which was assembled from typical ALMA data as well as difficult cases encountered during ALMA operations.   Following the
development of the pipeline benchmark datasets (\S\ref{benchmark}), further development and regression testing of the code was expanded to include the cubes of those projects as well.

\begin{figure*}
\centering
\hspace*{-1.1cm}\includegraphics[scale=0.64]{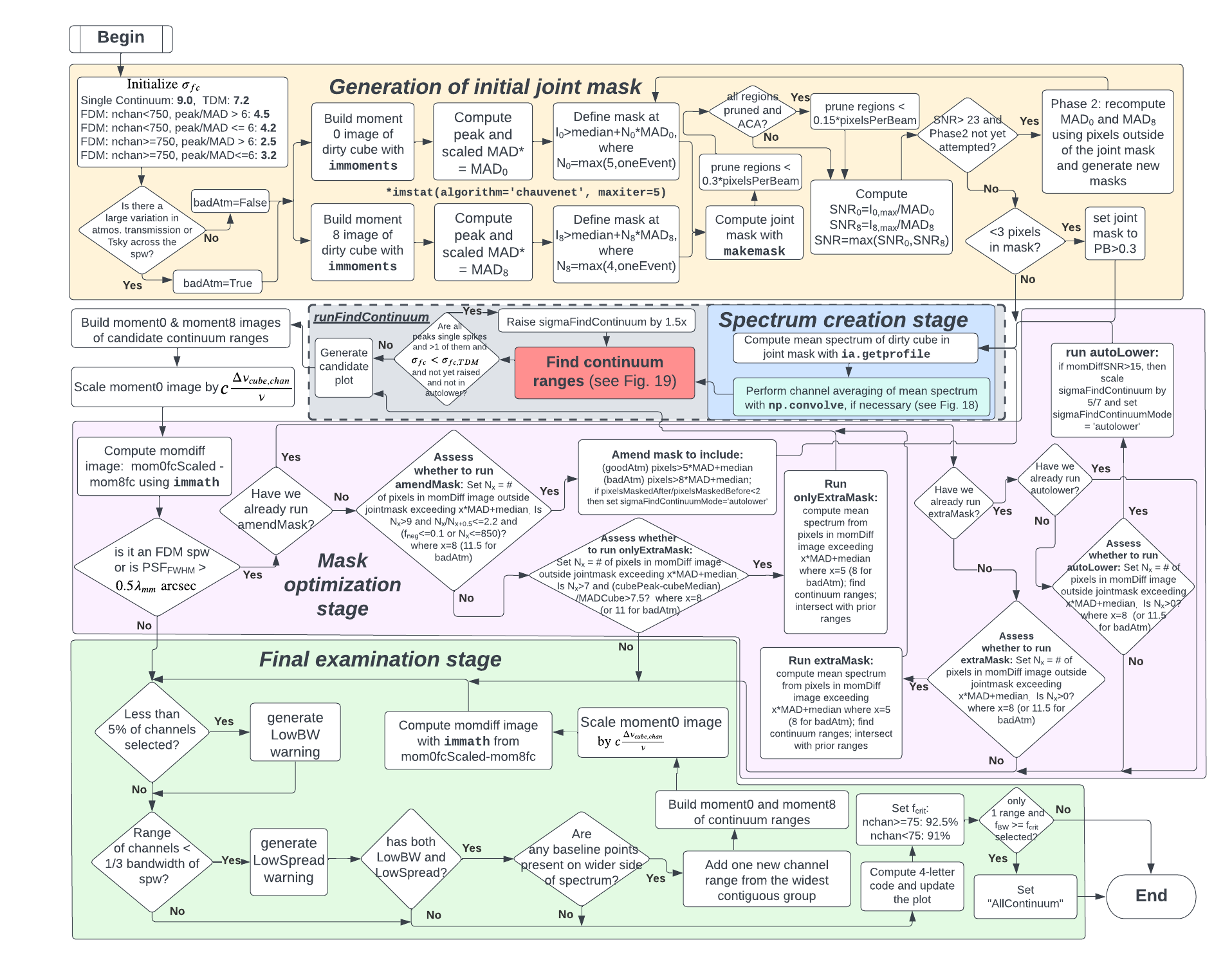}
\caption{Logic flowchart of the \pcmd{findContinuum.py} algorithm (\S\ref{findcont}). The grey box with a dashed border near the center denotes the core function \pcmd{runFindContinuum} which can be called multiple times as the mask optimization stage (pink box) proceeds. The channel averaging decision (cyan box) is further detailed in Figure~\ref{fig:findcont_nbin} and the continuum channel range selection (red box) is further detailed in Figure~\ref{fig:findcontFindChannels}.}
\label{fig:findcontOverview}
\end{figure*}

The structure of the \pcmd{findContinuum} program is provided in the flow chart in
Figure~\ref{fig:findcontOverview}.
The program is composed of seven major steps: (1) forming the initial joint mask, (2) forming the representative spectrum (\S\ref{repSpectrum}), (3) applying channel averaging when appropriate (\S\ref{averaging}), (4) selecting candidate continuum channel ranges from this spectrum (\S\ref{selection}), (5) plotting the results, (6) potentially optimizing the mask (\S\ref{momdiff}), and (7) generating warnings (\S\ref{warningRanges}).  Items (2) through (5) are contained with the function \pcmd{runFindContinuum}, which can be run multiple times if the mask optimization stage requires it.
The mask optimization stage, which runs for all FDM and some TDM spectra, makes an assessment of whether the candidate continuum channels contain line emission. It used the so-called ``moment difference analysis'' method, which analyzes the difference image computed from the peak intensity image (so called ``moment 8'' due to the definition of the \param{moments} parameter of the CASA task \ccmd{immoments}) and the moment 0 image scaled to the same bandwidth (\S\ref{momdiff}). This analysis can be expensive in processing time because of the need to generate moment images, but it is a powerful method of avoiding line contamination.

\subsection{Formation of the representative spectrum}
\label{repSpectrum}

We define the ideal ``representative'' spectrum of the dirty cube as one that provides a high signal to noise measurement of all the spectral lines present, even those that are located far away from the phase center.  There are various ways that one can try to compute a representative spectrum of a cube.  Three broad categories of approaches have been implemented in \pcmd{findContinuum} over the years: 
\begin{itemize}
\item \parvalue{meanAboveThreshold}: taking the mean of all spatial pixels above a threshold value;
\item \parvalue{peakOverMad}: dividing the peak value by MAD of all spatial pixels in each channel;
\item \parvalue{jointMask}:  generating a mask from the moment 0 and moment 8 images formed from all pixels with intensity above the median value by appropriate thresholds ($\sigma$=5 for moment 0 and $\sigma$=4 for moment 8) and then computing the mean spectrum within this mask. 
\end{itemize}
The \parvalue{meanAboveThreshold} technique works well if the noise characteristics of the data are uniform vs. frequency, while \parvalue{peakOverMad} works better if there is a variation in the noise, such as the presence of atmospheric lines, and is the method used by the ALMA Data Mining Toolkit \citep[ADMIT,][]{Teuben15}.  In ALMA Cycle~4 and 5, the default setting of the parameter \param{meanSpectrumMethod} was `auto', which was designed to automatically choose between the first two algorithms depending on the properties of the data. The current default option \parvalue{jointMask}  (added in the Cycle~6 release) combines the advantages of both approaches and uses CASA tasks and toolkit calls (\ccmd{imstat}, \ccmd{immoments}, and \ccmd{image.getprofile}) for the CPU-intensive calculations rather than Python.  

As shown in the yellow-highlighted box of Figure~\ref{fig:findcontOverview}, the jointMask option generates a joint mask by first using the \ccmd{immoments} task to build the ``moment0'' image (integrated emission) and the ``moment8'' image (peak).
It then computes the peak and MAD of these images using the \param{chauvenet} option in \ccmd{imstat}, which assumes that the underlying distribution is Gaussian and iteratively removes data that are inconsistent with that distribution.  The pixels of moment image $i$ with intensity higher than $N_i \times MAD_i$ above $Median_i$ forms its mask, where $N_0$=5 and $N_8$=4.  
The value of $N_i$ is adjusted upward if the total number of pixels in the cube is large enough that the expected number of random outliers beyond $N_i\sigma$ is greater than 1. In this case, $N_i$ is set to the $\sigma$ level for which one event is expected.
Next, the union of these two pixel masks forms the joint mask, which is then pruned of small regions that contain fewer than 0.3*pixelsPerBeam pixels, where pixelsPerBeam is defined as the ratio of the solid angle within the FWHM of the synthesized beam (obtained from the .psf image) to the solid angle of one pixel. If all regions are pruned and it is ACA 7\,m data, then the pruning threshold is reduced to 0.15*pixelsPerBeam.  Finally, if there are fewer than 3 pixels in the final mask, then the mask reverts to all pixels above the 0.3 level of the primary beam.

\subsection{Channel averaging of the mean spectrum}
\label{averaging}
\begin{figure}
\centering
\hspace*{-7mm}\includegraphics[scale=0.68]{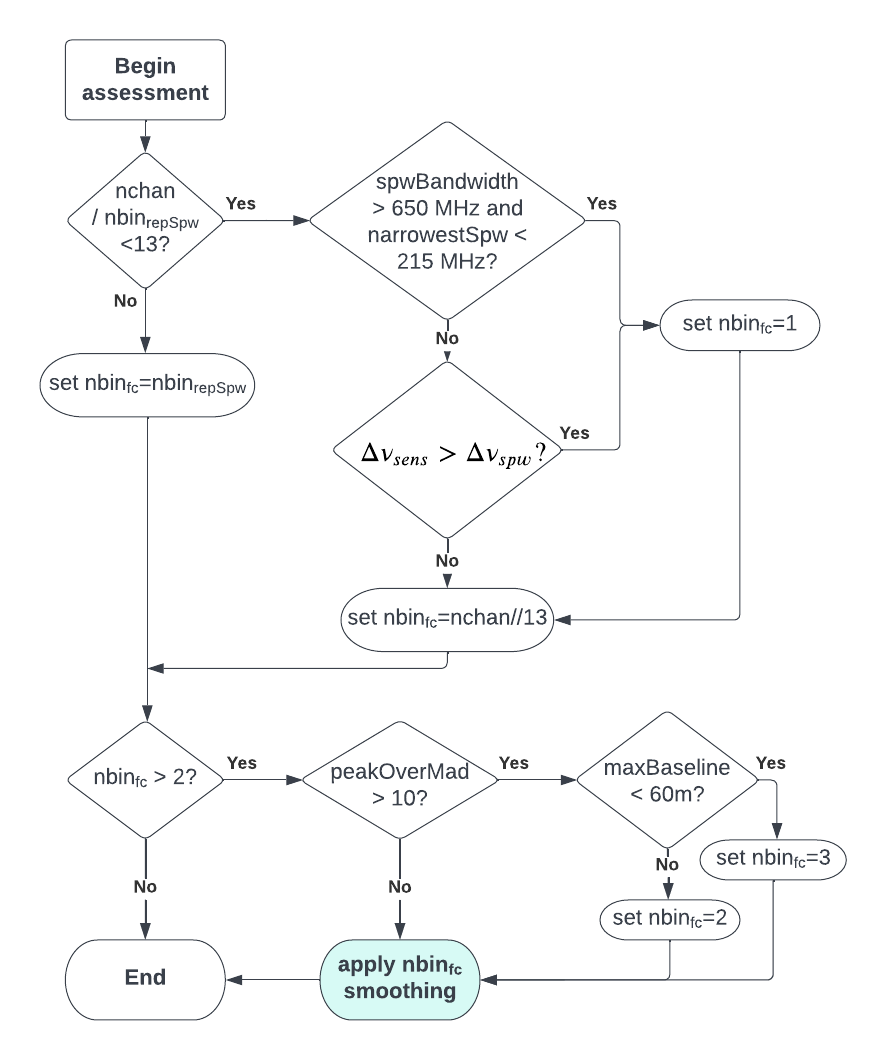}
\caption{Logic flow in \pcmd{findContinuum.py} for translating the channel binning factor (\nbinRepSpw) for the representative bandwidth cube (set by the user-requested bandwidth for sensitivity, see \S\ref{repBWcube}) into a channel averaging factor (\nbinFC) and applying it to the mean spectrum. }
\label{fig:findcont_nbin}
\end{figure}

In order to promote the detection of faint lines that are significantly broader than one channel, prior to
analysis of the mean spectrum of the joint mask
\pcmd{findContinuum.py} will apply a boxcar smoothing of this spectrum if user-specified bandwidth for sensitivity (\bwForSensitivity) is sufficiently wider than the \cubechanwidth\/ of the repSpw as to invoke
the creation of a repBW cube. The channel width of the spectrum is preserved in order to avoid the complication
of converting subsequently derived channel ranges back to the original channel width.
The width of the smoothing kernel (\nbinFC) is initially set to the \bwForSensitivity\/ rounded
to the nearest integer number of channels, which will correspond to the nbin value shown
in the "List of clean targets" table on the \pcmd{hif_makeimlist} (representative cube) stage (\nbinRepSpw).
Subsequently, as shown in Figure~\ref{fig:findcont_nbin}, it is limited to be $<$ \param{nchan}/13 in order to preserve
some large-scale structure in the spectrum if the user has set \bwForSensitivity\/ to be a large fraction of the \spwbw.

Unfortunately, the expected linewidth for the target as set by the user in the OT does not yet propagate to the ASDM, therefore additional
heuristics are needed to determine if any of the non-representative spws should not be similarly smoothed, as
the user may have set \bwForSensitivity\/ based on continuum RMS rather than line RMS. First, if strong line emission (peak
SNR $>$ 10) is present in the raw mean spectrum, then \param{nbin} will be further limited to 2 channels on the 12\,m
array and 3 channels on the 7\,m array. Second, if the \spwbw\/ $<$ \bwForSensitivity, then smoothing
is disabled on that spw. Third, because smoothing broad windows that
lack line emission can lead to reduced continuum bandwidth selections (due to ripple structure in the spectral
baseline), smoothing is disabled if the spw being processed is wide ($>$650\,MHz, i.e., 937.5 or 1875\,MHz) but only if
narrow spws ($<$215\,MHz, i.e., 58 or 117\,MHz) are also present in the MOUS. The latter combination also strongly
suggests that the user is expecting to detect narrow lines as well as continuum, so invoking any smoothing in the
wide windows could completely dilute any potential line emission in those windows, including serendipitous detections. This logic will be improved
in future Cycles once the linewidth specified by the user is propagated to the \file{SBSummary.xml} file of the ASDM via the \param{spectralDynamicRangeBandwidth} parameter, which is automatically set in the OT to 1/3 of the linewidth. Until then, there will be some cases where line-free spws that trigger smoothing will achieve less than optimal continuum bandwidth.

\subsection{Selection of continuum channels}
\label{selection}
\begin{figure*}
\centering
\hspace*{-4mm}\includegraphics[scale=0.73]{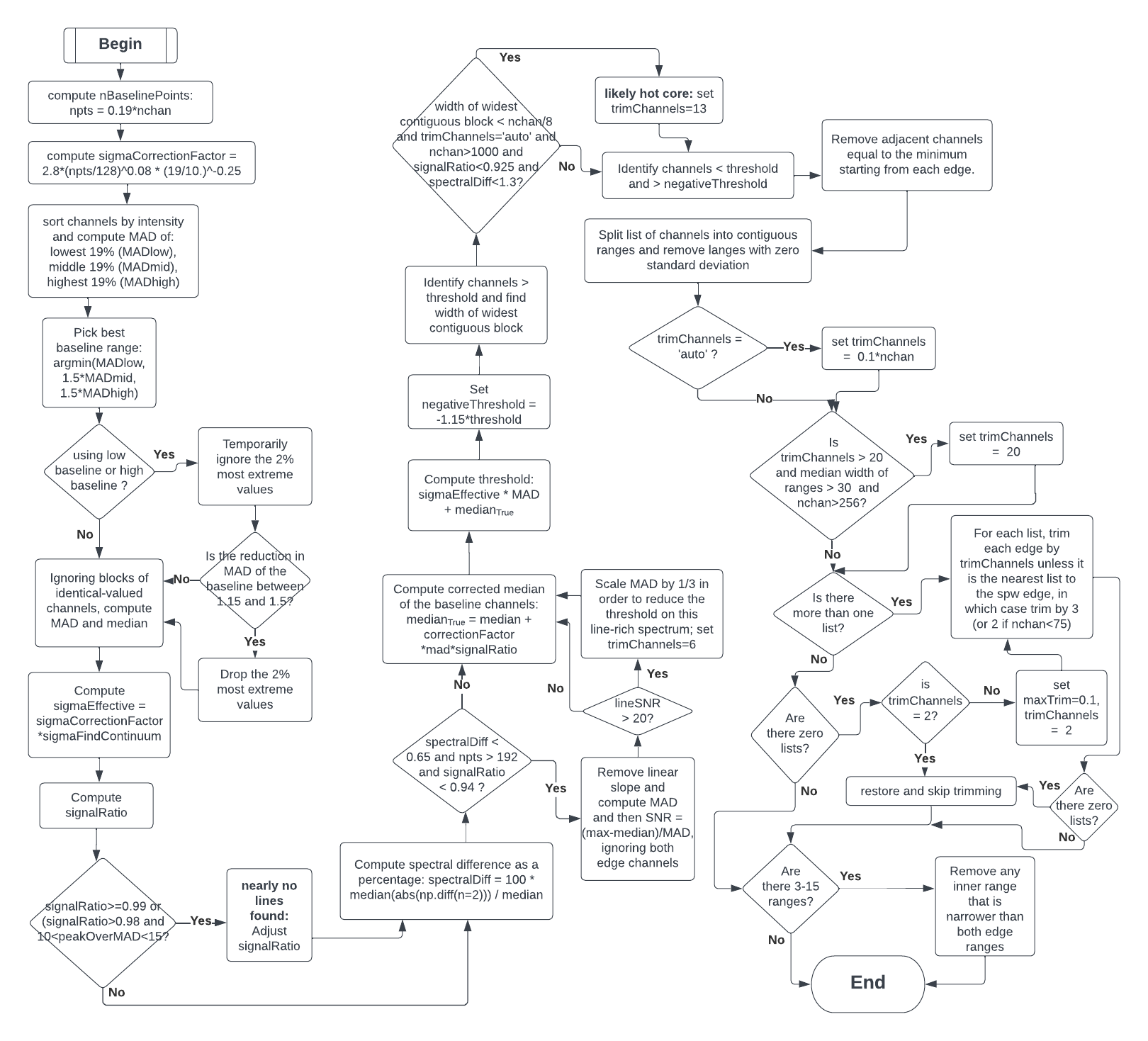}
\caption{Logic flow in the \pcmd{findContinuumChannels} function of \pcmd{findContinuum.py} (see \S\ref{selection}).  The number of channels in the spw ($n_{\rm chan}$) is denoted by ``nchan''. }
\label{fig:findcontFindChannels}
\end{figure*}

Once a representative spectrum ($S_{\rm rep}$) has been formed,
the functional algorithm to identify likely continuum channel ranges is run: \pcmd{findContinuumChannels}.  An overview of the bulk of the logic is shown in Figure~\ref{fig:findcontFindChannels}.  The basic idea is to: (1) identify an appropriate set of channels to use
as a baseline level, (2) set a positive and negative threshold level relative to that baseline, (3) identify channels with intensity
beyond those threshold, and (4) trim away narrow groups of channels, based on the number of groups and the width of the widest group, and their location relative to the edge groups.  

\subsubsection{Spectral baseline selection}

In order to define the spectral baseline -- the intensity level in a spectrum that corresponds to the continuum level -- 
a common practice  for well-understood spectra with spectral lines limited to the central
portion of the spw is simply to use a set of channels from both edges of the spw.  
However, there are various ALMA science use
cases where line emission can appear at numerous places across the spw, so relying on line-free edges is not a safe assumption.
Instead, we assume that there is some minority of channels that are mostly line-free (19\%), but that they might be scattered
across the spw. We proceed to sort the channels by intensity and compute the MADs of the lowest $n_{\rm pts}$ channels, 
middle $n_{\rm pts}$ channels, and highest $n_{\rm pts}$ channels, where $n_{\rm pts} = \text{int}(0.19*n_{\rm chan})$.  The median 
value of the group with the lowest MAD, where the median values from the middle and highest groups are first
multiplied by 1.5, is used as the baseline level.   The MAD of the selected group is taken to be the MAD of the baseline.
This technique allows the choice of the baseline level to adapt to spectra 
dominated by emission lines, absorption lines, both kinds of lines, or no lines.

The most common ALMA use case is emission lines and thus the group of channels with the lowest intensity is usually selected.
If the lowest group or highest group is selected as the set of the baseline channels $(B_{\rm rep}$), then we next filter out channels 
within that group that have large deviations from the rest to avoid biasing the median and MAD.
To implement this filter, we compute how much the MAD decreases by ignoring the $n_{ignore}$ most extreme-valued channels, where $n_{ignore}=\texttt{int}(0.02n_{\rm chan}$), and if this decrease factor is between 1.15 and 1.5, then those channels are dropped from the baseline.  The reason for employing an upper limit (1.5) is to avoid removing channels when 
the selected set of channels has unexpected statistical behavior, perhaps due to image artifacts.  Finally, if there are blocks of
identical-valued channels (such as exactly 0.0, as can happen in cube channels devoid of visibility data due to manual flagging), 
then we remove them from $B_{\rm rep}$ and recompute the MAD and median.

\subsubsection{Threshold setting}
\label{fcThresholds}
To set the threshold levels relative to the baseline median, we must correct for the fact that the
median computed from a sample drawn from an extreme low (or high) percentile group will under (or over) estimate the true median
of the population, and in both cases the MAD of the sample will underestimate the true MAD.  
For Gaussian noise, simple numerical simulations show that 
using the lowest $P$=19th percentile extreme will underestimate the 
the MAD by a factor of 2-3. Based on empirical experience with the extensive test set of 
\pcmd{findContinuum} cubes (\S\ref{testset}), we employ the following equation for the ``sigma correction factor'' ($k_{\rm corr}$),
\begin{equation}
\begin{aligned}
k_{\rm corr} &= 2.8(n_{\rm chan}/128)^{0.08} (10n_{\rm pts}/n_{\rm chan})^{-0.25}\\
              &= 1.07 n_{\rm chan}^{0.33} (p\times n_{\rm pts})^{-0.25},
\end{aligned}
\end{equation}
where $p$ is the percentile value as a fraction ($p=P/100$, e.g. 0.19), which yields a smoothly varying 
curve of values between 2.2-3.3 for the range of channel numbers (64..4096) inherent 
to typical ALMA spws (those recorded with 2 or 4 polarization products).  This equation also produces reasonable results for alternative
$P$ values in the range of 5-25, which can be employed in manual processing via the \param{nBaselineChannels} parameter.
Likewise, simulations show that the median will be underestimated by $\sim 3 \times$ the (uncorrected) MAD of the baseline channels (\texttt{MAD}($B_{\rm rep}$)).
To correct the median, we use the following equation for the correction ($\Delta M$) to add to the median, generalized to alternative values of $P$:
\begin{equation}
\Delta M = 6.3 (5.0/P)^{0.5} \texttt{MAD}(B_{\rm rep}) 
\end{equation}
For cases of line-rich spectra, the channels that ultimately get selected for continuum channels are likely
to be influenced by weak line emission, thereby raising the apparent variance of those channels, so we
temper the magnitude of the $\Delta{M}$ correction by a quantity termed the signal ratio ($r_{\rm signal}$), computed by:
\begin{equation}
r_{\rm signal} = (1 - (n'_{\rm chan}/n_{\rm chan}))^2
\end{equation}
where $n'_{\rm chan}$ is the number of channels with intensity above the median by $>f_{wl}k_{\rm corr}\sigma_{\rm fc}$ \texttt{MAD}($B_{\rm rep}$).
The weak line factor ($f_{wl}$) is initially set to 2.  The constant $\sigma_{\rm fc}$ is initialized to a value between 2.5 and 9 based on the 
properties of the spw and, for FDM spws, the strength of the peak spectral feature (see the ``Initialize $\sigma_{\rm fc}$ box in the  
upper left corner of the flow chart in Figure~\ref{fig:findcontOverview}).
The value of $r_{\rm signal}$ will be 1 if there
are no significant spectral features present, and 0.25 if half the channels have features, etc. When few line features are
found ($r_{\rm signal} \geq 0.99$) and the strongest is of only moderate strength (10 $<$ \texttt{max}($S_{\rm rep}$)/\texttt{MAD}($S_{\rm rep}$) $<15$), then $f_{wl}$ is reduced to 1.35 and $r_{\rm signal}$ is recomputed to look for additional weak lines.

With the four coefficients defined above, the thresholds ($T_+$ and $T_-$) for line channel detection can now be set:
\begin{equation}
\label{medianTrue}
\begin{aligned}
\text{Median}_{\rm true} &= \median(B_{\rm rep}) +  r_{\rm signal} \Delta{M}\\
T_+ &= k_{\rm corr} \sigma_{\rm fc}  \texttt{MAD}(B_{\rm rep}) + \text{Median}_{\rm true}\\
T_- &= -1.15 T_+.
\end{aligned}
\end{equation}
In the summary plot shown in the WebLog (\S\ref{findcontSummary}), an example of which is included in Figure~\ref{fig:uvspecFindcont}, the values of $T_+$ and $T_-$ are shown as black dotted horizontal lines.  The chosen baseline points ($B_{\rm rep}$) are marked by blue points, their mean is the black dashed horizontal line, and the corrected median is the black solid horizontal line.

\subsubsection{Adjustment of \texttt{MAD}}

The calculation of the thresholds is modified from the description in \S\ref{fcThresholds} in two specific cases.  
First, a quantity termed the spectral difference ($d_{spec}$) is 
computed over the $S_{\rm rep}$ using the NumPy \pcmd{diff} function to produce the second discrete difference in units of percentage ($d_{spec}$):
\begin{equation}
d_{spec} = 100\,\median(|\texttt{diff}(S_{\rm rep}, n=2)|) / \median(S_{\rm rep})
\end{equation}
If $n_{\rm chan} > 192$ and $r_{\rm signal} < 0.94$ and $d_{spec} < 0.65$, then the spectrum is 
surmised to be a channel-averaged FDM spectrum containing
a lot of line emission. In this case, in order to promote the detection of line emission,
$r_{\rm signal}$ is set to zero, and a linear slope is temporarily removed from the spectrum 
to compute the gross emission line SNR ($E_{\rm SNR}$), which
effectively uses all channels as the baseline:
\begin{equation}
E_{\rm SNR} = (\texttt{max}(S_{\rm rep}[1:-1])-\median(S_{\rm rep}))/\texttt{MAD}(S_{\rm rep}).
\end{equation}
where both edge channels of $S_{\rm rep}$ are ignored in the calculation of the peak.  
If $E_{\rm SNR}>20$, then the result for \texttt{MAD}($B_{\rm rep}$) is scaled by 0.33 
before proceeding with the calculation of Equation~\ref{medianTrue}.  Also, the \param{trimChannels} parameter is changed from `auto' to 6 for all groups.

The second modification occurs for spectra with weak (or no) line emission.  In this case, the \texttt{MAD}($B_{\rm rep}$) is raised in order to prevent small
changes in noise level (such as can occur between serial vs. parallel \ccmd{tclean}, or different versions of CASA) from causing drastically different
continuum range selections.  Specifically, if \texttt{max}($S_{\rm rep}$)/\texttt{MAD}($S_{\rm rep}$) $<$ 5 and $r_{\rm signal}>0.94$, then the ratio ($r_{\rm SNR}$) of the net emission line SNR ($E'_{\rm SNR}$) to the spectral dynamic range SNR ($DR_{\rm SNR}$) is computed via:
\begin{equation}
\begin{aligned}
E'_{\rm SNR} &= (\texttt{max}(S_{\rm rep})-{\rm Median}_{\rm true})/\texttt{MAD}(B_{\rm rep})\\
DR_{\rm SNR} &= (\texttt{max}(S_{\rm rep})-\texttt{min}(S_{\rm rep}))/\texttt{MAD}(S_{\rm rep})\\
r_{\rm SNR} &=  E'_{\rm SNR}/DR_{\rm SNR} 
\end{aligned}
\end{equation}
If $r_{\rm SNR} > 1$, then the \texttt{MAD}($B_{\rm rep}$) is scaled upward by this factor before proceeding with the calculation of Equation~\ref{medianTrue}.

\subsubsection{Candidate channel range identification}
\label{candidateRanges}
The channels with intensity $>T_+$ or $<T_-$ are identified, after which all of the other channels are
considered candidate continuum channels and are sorted into contiguous groups, where the width of 
the widest group is defined as $w_{widest}$.   If $w_{widest} < n_{\rm chan}/8$ and $n_{\rm chan} > 1000$
and $0 < r_{\rm signal} < 0.925$ and $d_{spec} < 1.3$, then the spectrum is surmised to be a very line-rich hot core, 
and the \param{trimChannels} parameter is changed from `auto' to 13 and the \param{narrow} parameter is changed from `auto' to 2 
to try to increase the recovery of continuum bandwidth, which might otherwise only return a few channels.
Otherwise, if \param{trimChannels} is still set to `auto', then it is modified to 0.1, which means that both
edges of each group of channels will eventually be trimmed by one tenth of the number in each group rounded 
up to the nearest integer, up to a limit of \param{maxTrim}=20, and subject to the rules in \S\ref{trimming}.

\subsubsection{Channel range trimming}
\label{trimming}
Trimming the edges of each candidate continuum channel range is meant to avoid low-level emission and absorption from the wings of lines
that were avoided in the selection of the channel ranges in \S\ref{candidateRanges}.
If more than one range is found, then each range is trimmed by the amount prescribed in \S\ref{candidateRanges}, except for the two ranges nearest the
spw edges, which are trimmed by only 2 or 3 channels, corresponding to $n_{\rm chan} < 75$ or $n_{\rm chan}\geq 75$.  Reducing the trimming
on the edge ranges intentionally biases their contribution higher as they are more important for establishing the slope of the baseline.
If after trimming a range it becomes narrower than \param{narrow} channels, it is eliminated, where the automatic setting of
\param{narrow} = 2, 3, or 4 depending on $n_{\rm chan} < 100$, $n_{\rm chan} < 1000$, or $n_{\rm chan} \geq 1000$.   If no ranges are found, then the trimming is reverted, \param{trimChannels} is reduced to 2, 
and trimming is tried again.  After this stage, if there are 3-15 ranges remaining, then 
any inner range that is narrower than both ranges nearest the spw edges will be removed, in order to avoid picking up line contamination from a feature near the
middle of the spw in return for little gain in continuum sensitivity.

\subsection{Moment difference analysis}
\label{momdiff}
There is a category of science cases where the joint mask used in \pcmd{findContinuum} misses some weak but extended line emission (e.g., Galactic $^{13}$CO) which can significantly impact the continuum subtraction and appear as obvious missed signal in a peak intensity image created from the line-free (findcont) channels, the so-called mom8fc image, even though it does not appear as a large spectral feature in the \pcmd{hif_findcont} spectrum (constructed from the joint mask).  After the initial set of continuum channel ranges have been found (\S\ref{trimming}), all FDM spws will follow the extended procedure described in the following subsections. TDM spws will follow it only if the max baseline ($D$) $\leq$ 400\,m, as inferred from the PSF beamsize ($\theta$) and the central wavelength ($\lambda$) of the spw, via the formula: $D=\lambda/\theta$. Note that manual usage of \pcmd{findContinuum} can still invoke this procedure for all TDM spws by setting \param{amendMaskIterations}=3 instead of the default value of `auto'.

\subsubsection{Generate moment difference image}

At the end of the first call of the \pcmd{runFindContinuum} function, termed iteration 0, the moment difference image (momDiff) is created to assess whether any signal exists outside the joint mask, and if so, take action to expand it and iterate to a achieve a better result.
The momDiff image is created by subtracting an appropriately scaled version of the moment 0 image (i.e. divided by the product of the \cubechanwidth\/ and number of channels) from the moment 8 image, where 0 and 8 correspond to the \param{moments} parameter of the CASA task \ccmd{immoments}, where 0 is the integrated intensity and 8 is the peak intensity. Both moment images are computed over the initial continuum selection of channels (fc), and masking pixels below the PB=0.23 level (to avoid the occasional ``hot pixel'' phenomenon in moment images near the PB cutoff). 
The momDiffSNR ($SNR_{\rm mD}$) is defined as the difference between the peak intensity of the momDiff ($mD$) image and its median, divided by its MAD:
\begin{equation}
SNR_{\rm mD} = (\texttt{max}(mD) - \texttt{Mdn}(mD)) / \texttt{MAD}(mD)
\end{equation}
The algorithm described in the following subsections has been developed primarily using a subset of over 200 spws of the 1768 pipeline benchmark cubes (\S\ref{benchmark}) that invoke these heuristics.  

\subsubsection{Iteration 1: ``Amend mask''}
If at least 9 pixels in the original momDiff image outside the joint mask have a signal (intensity$-$\median($mD$)) that exceeds either of two thresholds: $N_{\rm mD}\times$\texttt{MAD}($mD$) or $N_{\rm cube}\times$\texttt{MAD}($cube$), then the joint mask is expanded to include the pixels of that signal and  the
\pcmd{runFindContinuum} function is run again.  The coefficient $N_{\rm mD}$ is set to 8 if the variation in atmospheric transmission across the spw is small and 11.5 otherwise\footnote{$N_{\rm mD}$ is reduced by subtracting 0.5 if the ratio of momDiffSNR/mom8fcSNR $< 0.3$, which occurs only when strong maser lines are present.}, while $N_{\rm cube}$ = 7.5 regardless of the atmosphere.  The MAD is measured in two ways: (1) using pixels outside the joint mask, and (2) using pixels in the standard noise annulus accounting for image size mitigation and outside the joint mask (\S\ref{noiseAnnulus}), but applying a minimum of PB=0.21 for the annulus. The lower of these two values is used. If the MAD is zero, then the standard deviation is used instead. The median is measured from the annulus. If the joint mask expansion increases the number of pixels by less than a factor of 2, then \param{sigmaFindContinuumMode} is set to `autolower' instead of `auto', and $\sigma_{\rm fc}$ is set to the final floating point value from the first run (instead of `auto'), which will prevent the subsequent execution of \pcmd{runFindContinuum} to arrive at a higher threshold than the first value. Using the new continuum channels from this execution, a new momDiff image is constructed. If the signal in the new momDiff image (peakOutsideMask$-$medianOutsideMask)/MAD is reduced below  
$N_{\rm mD}$, then it will stop and return the new continuum channel ranges, and update the plot accordingly. Otherwise, the difference image of the momDiff.amendedMask minus momDiff.original will be created to see where in the field the excess lies.

\subsubsection{Iteration 2a: ``Extra mask''}

If there is significant positive signal in this momDiff difference image (threshold = $\sqrt{2}$ * momDiff critical threshold from iteration 1) in one or more pixels, then it will compute the intersection of the findcont ranges from the two runs and build a new momDiff image. If the signal in this momDiff image is now below the critical threshold, then it stops and returns the intersected ranges, updating the plot accordingly. Otherwise, a new spectrum is constructed from the pixels with excess emission in the momDiff image (a so-called ``Extra mask''), the intensities in the non-findcont channel ranges that have a peak SNR$>$15 are replaced with Gaussian noise of the same median and MAD (so that ranges with already-identified signal are ignored), and the resulting mean spectrum is passed to a third run of the \pcmd{runFindContinuum} function with \param{signalRatioTier1}=1.0 (instead of the default of 0.967) for a more sensitive look at the lines that remain. The random number generator uses a fixed seed value so that results are repeatable.  The continuum ranges from this final run are intersected with the ranges from the prior (amended mask) run, the plot is updated accordingly and the new ranges are returned. In the unlikely event that there is no intersection of ranges, it will revert to the prior range. If the intersection yields only a single channel, and the spectrum is wider than 500 channels, then the two adjacent channels will also be included, as long as they are not an edge channel of the spw.

\subsubsection{Iteration 2b: ``Only extra mask''}

If the signal in the original momDiff image outside the joint mask was not significant (i.e., ``Amend mask'' not needed), but the signal across the whole image is significant (according to the momDiff critical threshold from iteration 1 over at least 7 pixels), then it will check if there also appears to be line emission in the initial continuum channel selection by evaluating: 
\begin{equation}
SNR_{\rm cube} = (\texttt{max}(cube)-\median(cube))/\texttt{MAD}(cube)
\end{equation}
within those channels. If 
this ratio $>$ 7.5, then only the ``Extra mask'' procedure will be run.

\begin{figure*}
\centering
\hspace*{-4mm}\includegraphics[scale=0.145]{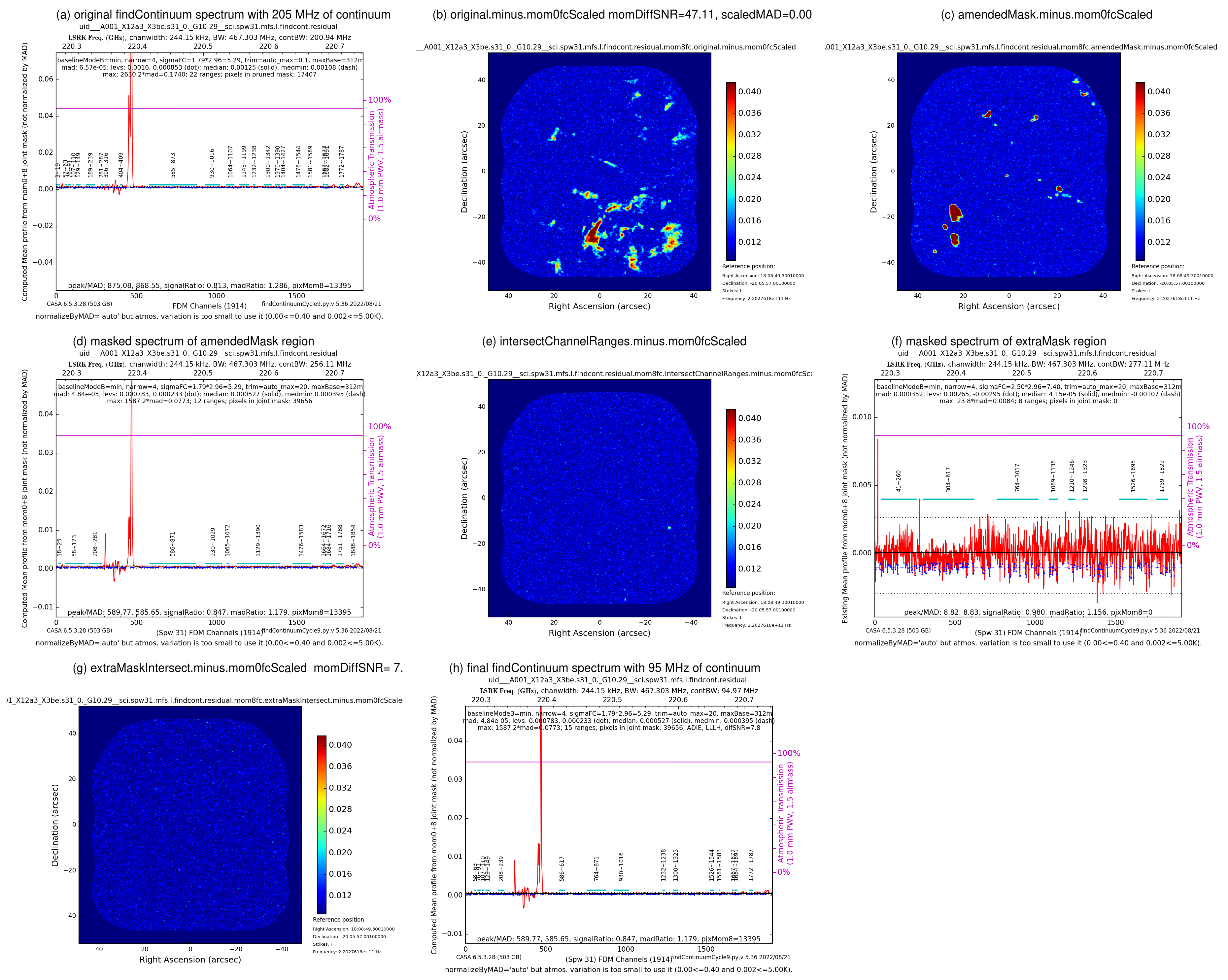}
\caption{The progression of improvements provided by the moment difference analysis (\S\ref{momdiff}). (a) The original \pcmd{findContinuum} result that selects 205\,MHz of candidate continuum channels;  (b) The moment difference image using the continuum channel selections from panel a showing the locations of the contamination; (c) Same as (b) but after expanding the mask and finding new continuum channel selections shown in panel d; (d) The spectrum of the amended mask;  (e) Same as (b) but using the intersection of the channel selections from panel a and d; (f) The spectrum of the extra mask; (g) Same as (b) but using the intersection of channel selections from panel a,  d, and f; (h) The final result which selects 95\,MHz of uncontaminated continuum channels. }
\label{fig:momdiffanalysis}
\end{figure*}

A graphical example illustrating how the result improves with iterations 1, 2a, and 2b of the moment difference analysis is shown in Figure~\ref{fig:momdiffanalysis}.  The original \pcmd{findContinuum }spectrum contains candidate continuum
selections which appear sensible based on viewing the spectrum alone; however, the 
subsequent moment difference
image shows a large amount of contamination.  The spectrum of the amended mask detects (and
thereby allows us to avoid) more velocity components of $^{13}$CO2-1, but other regions show contamination.  The intersection of these two sets of candidate continuum channels eliminates nearly all of the
contamination.  The extra mask step then eliminates the remaining faint contamination.

\subsubsection{Iteration 3: ``Autolower''}

If after the Extra Mask procedure completes there still exists excess signal that meets the same criteria for doing an Extra Mask, then another run of findContinuum will be performed, called AutoLower, using signalRatioTier1=1.0, \param{sigmaFindContinuumMode}='autolower', and $\sigma_{\rm fc}$ set to the final value from the previous run.  In this mode, the value of \param{signaFindContinuum} is prevented from being raised, it can only be lowered.

\subsubsection{Result codes}
\label{resultCodes}
After completion of the moment difference analysis,
the moment 8 images of the candidate continuum
channels are
assessed statistically to validate the procedure.
Four assessments are made: (1) the change in peak inside the mask; (2) the change in peak outside the mask; (3) the change in sum; and (4) the change in MAD.
If the change normalized by the initial value is greater than 1\%, then the result is deemed higher (H), if it is less than -1\%, then it is deemed lower (L), otherwise it is deemed the same (S). The four letters are assembled into a result code string. The continuum selection will be reverted to the value obtained prior to initiating the moment difference analysis
if the four-letter string is either HHHH or HHHS.  The rationale for reverting is to avoid producing a result that is clearly more highly contaminated with line emission that was found previously.

\subsection{Warnings for insufficient continuum ranges}
\label{warningRanges}
Warning messages are sent to the CASA log and written on the summary plot (\S\ref{findcontSummary}) if there is a small amount continuum bandwidth identified (less than 5\%, ``LowBW'') , and/or if the spread from the lowest to highest frequency continuum-identified channel is too small (less than 33\%, ``LowSpread''). In the event that both LowBW and LowSpread are invoked, then it will search for the widest list of contiguous baseline points (blue points on the plot) on the wider side of the spectrum devoid of ranges. If it finds a range, then it adds it to the list, creates a new mom0fc, mom8fc, and momDiff image, and reevaluates the statistics and warnings.   

\subsection{Identification of ``all continuum''}

If the amount of continuum bandwidth found is larger than 92.5\% of \spwbw\/ (91\% for TDM full polarization spws), and it is all in one contiguous group of channels, then the spectrum is termed ``AllCont''.  This label triggers the pipeline to forego cleaning of the cube in order to save processing time, as there is unlikely to be any line
emission present.

\subsection{WebLog display}
\label{findcontSummary}
The WebLog page for \pcmd{hif_findcont} contains a table with a row for each field and spw.  The columns contain the continuum channel ranges, each one translated to its corresponding LSRK frequency range, a summary plot of the final mean spectrum with an overlay of the atmospheric transmission in magenta and the continuum channel ranges marked as horizontal cyan lines and labeled, and an image of the final mask used to generate the mean spectrum.    If the moment difference analysis was invoked (\S\ref{resultCodes}), then the result codes are included in the top legend of the summary spectrum, as are any warnings for insufficient aggregate bandwidth or insufficient frequency spread of the channel ranges (\S\ref{warningRanges}).

\section{QA Scores}
\label{QAscores}
Each pipeline stage is assigned an overall numeric score in the range from 0.0---1.0 to indicate whether the pipeline tasks in the stage executed successfully, and if there were any identified problems or potential problems with the task results. This is referred to as a stage "Quality Assessment" (QA) score, with scores of 1.0 meaning that no problems were found, and progressively lower scores indicating progressively more worrisome results. For many tasks, this score is derived from a number of different checks or tests, each of which results in its own subscore in the range of 0.0--1.0. These subscores may result from different checks or heuristics, or the same check or heuristic run on different MS, or both. In these cases, the overall stage QA score is set to the minimum of all the QA subscores.

This section describes the overall philosophy for setting the QA scores, and the conditions or heuristics that are used to calculate them. The Pipeline Users' Guide\footnote{\url{https://almascience.org/documents-and-tools/cycle9/alma-science-pipeline-users-guide-casa-6-4.1}} (PLUG) for each Cycle gives a brief summary of the scores on a per-task basis.

\subsection{Philosophy}
The overall QA philosophy is to allow a quick assessment of whether a pipeline recipe completed with no problems, or if not which tasks are mostly likely to be responsible for the problems. For this, the pipeline has adopted a ``traffic light'' approach, where scores that indicate no problems are given range of scores that are mapped to a ``Green'' quality assessment; scores that indicate that something is potentially problematic with a stage so its results should be carefully reviewed are mapped to a ``Yellow'' assessment, and scores that indicate that something seriously went wrong with the stage are mapped to a ``Red'' assessment. To this we add a fourth ``Blue'' category of scores to indicate that the stage did not result in anything that necessarily should be investigated further, but if lower scores appear elsewhere in the recipe, there may be some clues found from the messages with the blue scores in this stage. This category is also used for assessments that by themselves do not mean that there is a problem with calibration or imaging, but might be worth knowing when using the products (e.g., a somewhat high level of flagging occurred, or properties of the \checksource\/ were less than ideal).

The score numerical ranges that map to each of these color categories are given in Table~\ref{QAtable}.
The conditions or heuristics used to set the stage QA scores are described in the remaining subsections of this section. One may notice that the terminology adopted in Table~\ref{QAtable} is subjective (e.g., {\it ``may mean there are issues''}). This is because ALMA uses a manual review process (currently for all WebLogs, but eventually only for WebLogs with no yellow or red QA scores), and can rely on trained data processing staff to investigate further and make a final QA determination. For this reason, the assignment of yellow QA scores is done conservatively. This is because delivering data with an uncaught issue results in much more work for the observatory than spending a bit more time manually reviewing a WebLog in a bit more detail. Therefore, one should not expect that all successful pipeline WebLogs will have all green or blue scores.

\begin{table}[ht]
\hspace{-1.2cm}
\begin{tabular}{p{1.6cm}p{1.1cm}p{4.5cm}}
\textbf{Score} & \textbf{Color} & \textbf{Meaning}\\
\hline
$>$0.90-1.00 & Green & No issues identified\\
$>$0.66-0.90 & Blue & No serious quality issues identified; there may be noteworthy QA messages\\
$>$0.33-0.66 & Yellow & QA warning triggered that may mean there are issues with the results of this stage; carefully review the yellow QA messages \\
0.00-0.33 & Red & Serious issue with the results of this stage; may not meet quality standards\\
\end{tabular}
\caption{QA Score Categories \label{QAtable}}
\end{table}

The overall QA assessment for the pipeline recipe is displayed on the ``By Task'' page of the WebLog, which lists all the pipeline stages in the recipe, each followed by a horizontal color bar indicating the stage overall QA score, and then the QA score itself. The bar is ``filled'' from left to right by the fractional amount corresponding to the numeric QA score and is colored according to the QA color category for the score given in Table~\ref{QAtable}. For example, a stage with a QA score of 1.0 will be filled 100\% to the right and have a green color, while a stage with a QA score of 0.5 will be filled half-way and have a yellow color. Stages with non-green QA scores will have a short message to the left of the color bar that gives a brief reason for the low QA score. Stages that had no quality assessment will have an unfilled color bar and ``N/A'' QA score.

This presentation allows a quick quality assessment of the pipeline results. Any stage with a less than optimal score will have additional QA messages of the appropriate color collected at the top of WebLog page for the stage, in an expandable table. By default, the table is collapsed and only displays a single QA message corresponding to the lowest QA subscore along with the number of subscores of the different color categories. The table can be expanded by clicking on the ``All QA Scores'' link.



As mentioned above, the WebLogs of all ALMA Pipeline runs that result in products in the ALMA archive will have been reviewed by a trained ALMA staff member who makes the decision on whether a data set meets the overall ALMA Quality Assurance standards at the product level \citep[QA2, see chapter 11 of][]{Remijan2019}. During this review, any stages with a yellow or red QA score will have been examined in detail and if some serious issue is found, corrective measures will be taken, for example by adding manually identified flagging commands to the \pcmd{hifa_flagdata} stage and re-running the pipeline, or sending the data for manual processing, or changing the QA0 classification for an execution (see \S\ref{InputData}) and sending the project back to the telescope for more observations.  Ideally, this workflow would ultimately result in a WebLog for which every pipeline stage has a green or blue QA score, but since, as mentioned above, the QA scores tend to be conservative, it is likely that some yellow scores still remain (and in rare cases some red scores). If the overall product evaluation is positive (``QA2 Pass''), one can safely assume that the stages with the lower QA scores have been reviewed and judged to pass ALMA quality standards.




\subsection{QA subscores based on simple checks}
\label{QAsimpleChecks}
The following QA subscores are based on simple deterministic checks for the listed stage, with no tunable parameters. The checks are performed for each measurement set being processed, and some are evaluated on a per-spw and/or a per-intent basis.

\begin{itemize}
\item \pcmd{hifa_importdata}: Check the following items for each dataset to be processed. If the checks are successful, the subscore is set to 1.0 (green), otherwise it is set to 0.0 (red).
\begin{itemize}
\item All required scan intents (\S\ref{requiredintents}) are present 
\item All calibrators have a valid RA \& DEC
\item The data were not previously processed by the pipeline
\item Give a blue subscore (0.90) if the data are high-frequency (ALMA bands 9 and 10), since the calibration of such data deserves more scrutiny (as finding bright calibrators at these frequencies can be challenging). 
\item If a polarization calibrator is present in the data but the polarization recipe 
is not specified in the context in the \pcmd{casa_pipescript.py} file (see \S\ref{Recipes}), the QA score will be set to 0.5 (yellow) with a message about unexpected polarization calibrations in the input data.
\end{itemize}

\item \pcmd{hifa_flagdata}: Check the mean transmission in each science spw.
\begin{itemize}
\item Set the subscore to 0.33 (red) if the representative spw has a mean transmission $<$5\%
\item Set the subscore to 0.9 (blue) for any other spw with a mean transmission $<$10\%
\item Otherwise the subscore is 1.0 (green)
\end{itemize}

\item \pcmd{hifa_fluxcalflag}: Check whether flagging in this stage results in any science spw being mapped to another spw. If so, set this subscore to 0.66 (yellow). Otherwise the subscore is 1.0 (green).

\item \pcmd{hif_refant} and (for polarization recipes) \pcmd{hifa_session_refant}: Check that a suitable reference antenna is found. If so, set this subscore to 1.0 (green), otherwise to 0.0 (red).

\item \pcmd{h_tsyscal}: Check that every science spw can be matched to a Tsys spw.  If so, set this subscore to 1.0 (green), otherwise to 0.0 (red).

\item \pcmd{hifa_antpos}: Check that antenna positional corrections were applied. If one or more antennas were corrected set the subscore to 0.9 (blue), otherwise to 1.0 (green, meaning no corrections were needed).

\item \pcmd{hif_lowgainflag}: Check that the flagging view is successfully created. If so, set this subscore to 1.0 (green), otherwise to 0.0 (red).

\item \pcmd{hif_setmodels}: Check that the model flux density of the Amplitude calibrator and the spectral index of the bandpass calibrator are successfully set for all science spws. If so, set this subscore to 1.0 (green), otherwise to 0.0 (red).

\item \pcmd{hif_makeimlist}: Assigns a score equal to the fraction of images in the imaging list compared to the total number expected. Set the score to 0.0 (red) if an error occurs. 

\item \pcmd{hif_makeimages}: All instances of this stage have a check based on convergence of the \ccmd{tclean} algorithm. If the task detects divergence of any of the cleaned images, this subscore is set to 0.0 (red), otherwise it is set to 1.0 (green).  For cubes, a reduction in QA score is applied if there are deviant point spread functions (PSF) in individual channels.  If the major axis of a channel's PSF differs by more than a factor of two from the median of all channels, the score is reduced by 0.11.  If more than 10 channels are deviant, the score is reduced by 0.34. 


\item \pcmd{hifa_imageprecheck}: Check that the dataset identifies a representative target \& frequency and has a defined ``Min / Max Acceptable Resolution'' (see \S\ref{weights}), hereafter referred to as the AR range. If not, set this subscore to 0.5 (yellow), otherwise evaluate the estimated resolution (major and minor axis of the synthesized beam) of the representative target \& spw for different values of the \param{robust} weighting parameter against the AR range, following the considerations given in \S\ref{weights}, with the following QA score assignments:

\begin{itemize}
\item Set the subscore to 1.0 (green) if both the major and minor axis with \param{robust}=0.5 are within the AR range
\item Set the subscore to 0.85 (blue) if both the major and minor axis are within the AR range with a \param{robust} value other than 0.5 but  between 0 -- 2
\item Set the subscore to 0.50 (yellow) if at least one axis is outside the AR range, but the beam area is within the range of areas corresponding to  the AR range 
\item Set the subscore to 0.25 (red) if the beam area falls outside the areas corresponding to the AR range for any value of \param{robust} between 0 -- 2, since this means that the imaging products are unlikely to meet the PI goals (so will not pass QA2)
\end{itemize}

\item \pcmd{hifa_checkproductsize}: Check whether target cubes need to be made with non-default imaging parameters or less data imaged in order to keep the cube or total product sizes within the \param{maxcubesize, maxcubelimit, maxproductsize} limits described in \S\ref{mitigation} (the values of these limits are specified in the PLUG for each pipeline release).
\begin{itemize}
    \item Set the subscore to 1.0 (green) if no mitigation was necessary
    \item Set the subscore to 0.50 (yellow) if the products had to be imaged with non-default parameters (larger cells, wider channels, smaller FOV, fewer targets) to fit within the limits
    \item Otherwise, set the subscore to 0.25 (red); such data will likely be sent to manual imaging where additional imaging mitigations can be implemented. 
\end{itemize}

\item \pcmd{hif_mstransform}: Check that target.ms files were successfully created. If so, set this subscore to 1.0 (green), otherwise to 0.0 (red).

\item \pcmd{hifa_flagtargets}: 
There is no check for this stage, and the score is always 1.0 (green). 

\item \pcmd{hif_findcont}: Check if the continuum identification algorithm (\S\ref{findcont}) identified suitable continuum ranges for all science spws. If so, set this subscore to 1.0 (green). Otherwise the numeric score is equal to the fraction of spws for which a continuum range was identified, unless size mitigation fails, in which case the score is set to 0.0 (red).  In operations, the latter situation can be 
detected earlier by the red score for the prior task \pcmd{hifa_imageprecheck}.

\item \pcmd{hif_uvcontfit}: Check if a continuum fit table was created for all science spws. If so, set this subscore to 1.0 (green), otherwise to 0.0 (red).

\item \pcmd{hif_uvcontsub}: Check if the continuum could be subtracted. If so, set this subscore to 1.0 (green).

\item \pcmd{hifa_exportdata}: Check that the standard pipeline products were copied to the \pcmd{/products} directory. If so, set this subscore to 1.0 (green), otherwise to 0.0 (red).
\end{itemize}

\subsection{Calibration QA subscores based on heuristics}

\subsubsection{Retrieving flux density from online catalog}
\label{QAsourcecat}

This subscore is calculated in the \pcmd{hifa_importdata} stage (\S\ref{importdata}) based on whether or not the online ALMA Calibrator Source Catalog (hereafter SC) returns a valid value. The value of the QA score depends on whether or not the pipeline was executed with the parameter \param{dbservice} set to True (meaning that the SC should be queried when the pipeline is run) or False (meaning that it should not be queried, in which case the calibrator fluxes will be read from the \file{flux.csv} file or the ASDM).

If \param{dbservice}=True (the pipeline default), and the SC can be reached and successfully returns a flux density measurement that is less than 14 days old, the subscore is set to 1.0 (green). If the SC can be reached but the measurement is more than 14 days old, the subscore is set to 0.5 (yellow). If the SC cannot be reached or returns a failure code, the subscore is set to 0.3 (red).

If \param{dbservice=False} and \param{Origin=Source.xml} in \file{flux.csv}, the subscore is set to 0.3 (red), since this means that the flux from the ASDM will be used for the FLUX calibrator. This is inadvisable, since such flux estimates are extrapolations from earlier measurements, and more accurate estimates result when the measurements bracket the observation (which can be done by running the pipeline with \param{dbservice}=True, or by setting the flux explicitly in the \file{flux.csv} file).  Otherwise, the subscore is set to 1.0 (green). 

\subsubsection{Percentage of incremental flagging}

All the pipeline stages that perform flagging have a QA subscore based on the additional percentage of data that are flagged. When more than one intent is evaluated, a subscore is calculated per intent, and an additional subscore is calculated from the products of the flagging subscores for each intent. 

The incremental flagging QA subscore is calculated by a linear function with a floor ($f_{\rm low}$) and a ceiling ($f_{\rm up}$), based on the percentage of data that is additionally flagged in the stage,  ($f_{\rm inc.flag}$), according to the following formula:

\begin{equation}
\begin{aligned}
QA_{\rm inc.flag} &= 0, \texttt{ if } f_{\rm inc.flag} > f_{\rm up}, \\
    &= \parbox[t]{5.5cm}{$1-(f_{\rm inc.flag}-f_{\rm low}) / (f_{\rm up}-f_{\rm low}),\\ \texttt{~if~} f_{\rm low}\leq f_{\rm inc.flag} \leq f_{\rm up}$,}\\
    &= 1, \texttt{ if } f_{\rm inc.flag} < f_{\rm low} 
\end{aligned} 
\end{equation}

The following tasks adopt $f_{\rm low}$=0.0 and $f_{\rm up}$=1.0 (for which the above equation simplifies to $1-f_{\rm inc.flag}$): \pcmd{hif_rawflagchans}, \pcmd{hifa_bandpassflag}, \pcmd{hifa_gfluxscaleflag} (calculated separately for each intent, with an additional subscore based on the product of all intents), \pcmd{hifa_polcalflag}, \pcmd{hifa_targetflag}. 

The following tasks adopt $f_{\rm low}$=0.05 and $f_{\rm up}$=0.50: \pcmd{hifa_fluxcalflag}, \pcmd{hifa_tsysflag}, \pcmd{hif_lowgainflag}, \pcmd{hif_applycal} (only for TARGET scan intents, if present; if not will use the score for any available intent).  

The task \pcmd{hifa_flagdata} uses $f_{\rm low}$=0.05 and $f_{\rm up}$=0.60, and only considers the sum of the following flag types: `Before Task', `QA0', `Online Flags', `Flagging Template', and `Shadowed Antennas'.


\subsubsection{Bandpass Solutions}

Because ALMA sets delays prior to running science EBs, the bandpass phase solutions in \pcmd{hifa_bandpass} should generally be flat
with some small instrumental features.  Delay errors or antenna position errors will induce slopes and possibly wraps.
The channel derivative ($dB/dc$) of the bandpass phase solution spectrum ($B_\phi$) is used to compute the fraction of outlier
channels ($f_{\rm outliers}$) per antenna, spw, and polarization via:
\begin{equation}
\begin{aligned}
dB/dc &= -\texttt{diff}(B_\phi)\\
f_{\rm outliers} &= \texttt{len}(\texttt{where}(dB/dc > 5\texttt{MAD}(dB/dc))) / \texttt{len}(B_\phi).
\end{aligned}
\end{equation}
A future improvement will define $dB/dc$ as an absolute value.

The QA score for outliers ($QA_{\rm DD}$) is computed via the derivative deviation ($DD$):
\begin{equation}
\begin{aligned}
DD &= \texttt{erf}(0.03/(\sqrt{2}f_{\rm outliers})),\\
QA_{\rm DD} &= 0.34 + (0.32)\frac{DD-0.03/\sqrt{2}}{0.2-0.03/\sqrt{2}}, \texttt{ if } DD\leq 0.2,  \\  
              &= 0.67 + (0.23)\frac{DD-0.2}{0.3-0.2}, \texttt{ if } 0.2 < DD < 0.3,\\
               &= 0.91 + (0.09)\frac{DD-0.3}{1-0.3}, \texttt{ if } DD \geq 0.3.
\end{aligned}
\end{equation}

This yields yellow scores when $f_{\rm outliers}>11.85\%$, blue scores when $7.78\% < f_{\rm outliers} < 11.85\%$, and green scores when $f_{\rm outliers} < 7.78\%$.



\subsubsection{WVR and phase RMS improvement}
The final QA score for the \pcmd{hifa_wvrgcalflag} stage is derived in a two step process primarily using the phase RMS improvement factor (how much the phases are improved with and without the WVR corrections applied), and then secondly assessing whether there were any flagged antennas, and whether the ``RMS'' or ``Disc'' (see \S\ref{wvrcorr}) values are elevated above 500\,$\mu$m, either per antenna, or globally. 

The QA score for the phase RMS improvement factor is calculated from an internal subscore that starts at 1.0 and is reduced as problems are identified. If the phase RMS improvement factor is $>$1, this secondary score maximum is set to 1.0. Provided that there are no issues, the secondary score propagates directly to the final stage score, QA score = 1.0 (green). Any issues that occur in the stage otherwise acts to reduce the secondary score. Any \textit{singular} antenna issue reduces the secondary score maximum value to 0.9 and thereafter any additionally flagged antenna or an antenna with an elevated ``RMS'' or ``Disc'' value further reduces the secondary score by 0.1. These single antenna issues are logically linked with a notification message as the data globally are good. 
The final QA score is calculated by a linear mapping of the secondary scores which range from 0.0 (clipped minimum) and 0.9 to a final QA score between 0.67 and 0.9 (blue). 

If during the phase improvement ratio assessment the phase RMS of the BANDPASS scan was found to exceed 85$^{\circ}$, then the secondary score is limited to a maximum of 0.66. The per antenna based issues then causes a further score reduction as outlined above. The logic to limit the secondary score to a maximum of 0.66 is in order to trigger a warning notification. By definition, the BANDPASS source should have a high SNR and thus the phase solutions themselves should not suffer from noise due to low signal. Any high phase RMS value therefore ties with underlying real variations of phase due to atmospheric fluctuations that may adversely affect the data quality (see Section \ref{ssf}). In this case the secondary score between 0.0 (clipped minimum) and 0.66 is mapped to a final QA subscore between 0.34 and 0.66 (yellow).

For any improvement factor $<$1, the secondary score is limited to 0.66 so as to issue a warning that the WVR correction is not improving the data. Single antenna based issues again reduce the score as described above and the same mapping to a final QA score between 0.34 and 0.66 (yellow) occurs. 

If there is a global elevation of the ``RMS'' and ``Disc'' values, the secondary score is otherwise limited to a maximum of 0.33, that relates to a severe warning message as to indicated that further action might be required. The final QA score mapping in this case is between 0.0 and 0.33    (red). 

The logic behind the separation of maximum scores is to distinguish between the very low water vapor conditions ($<$0.5\,mm), typical for higher frequency observing bands, where it is plausible that the WVR system solutions are not optimal but where the observations are otherwise good, from those where the solutions produced in the \pcmd{hifa_wvrgcalflag} stage are simply incorrect with large global ``RMS'' and ``Disc'' values that generally point to the presence of clouds and where \pcmd{remove_cloud} might be required (see \S\ref{wvrcorr}). When the improvement factors are $<$1 the WVR solutions are not applied to the data. 



\subsubsection{Gain solution SNR and phase RMS}
\label{spwphaseupQA}

The low SNR heuristics in the \pcmd{hifa_spwphaseup} stage establish the spw mapping scheme and the solution time interval as described in \S\ref{lowsnrheuristics}. After applying these heuristics, the gains are solved and the QA scores are generated as the final step (Figure~\ref{fig:hifa_spwphaseup}). Successful creation of the gain table produces a QA subscore of 1.0 (green).  Each science spw has another subscore ($QA_{\rm spwSNR}$) set by its median achieved SNR defined in terms of percentages (75\%, 50\%, and 30\%) of the target SNR of 32:
\begin{equation}
\begin{aligned}
QA_{\rm spwSNR} &= 1.0, \texttt{ if } {\rm SNR} >24,\\
&= 0.9, \texttt{ if } 16 < {\rm SNR} \leq 24,\\
&= 0.66, \texttt{ if } 9.6 < {\rm SNR} \leq 16,\\ 
&= 0.33, \texttt{ if } {\rm SNR} < 9.6.
\end{aligned}
\end{equation}


There is an additional QA subscore for this stage based on the phase RMS values calculated from the Phase RMS structure plots (\S\ref{ssf}). If the phase RMS is $<$30$^{\circ}$ then the subscore is 1.0 (green). When the phase RMS is between 30---50$^{\circ}$ the subscore is fixed to 0.9 (blue).  Otherwise, for phase RMS $>50^{\circ}$, the subscore is calculated via \texttt{max}(0, 1.0 - RMS/100), such that $50^{\circ}<$ RMS < 67$^{\circ}$ would result in QA subscores ranging from 0.5 to $>0.33$ (yellow), while RMS $\geq 67^{\circ}$ would result in QA subscores $\leq$ 0.33 (red), the latter indicative of very elevated phase RMS that can negatively impact the target data and perhaps should not have passed the input data QA0 assessment (\S\ref{InputData}).

\subsubsection{Fluxscale SNR and cross-spw consistency}

There are three subscores in the \pcmd{hifa_gfluxscale} stage to identify potential issues with the flux scale or with calibration transfer  by evaluating the derived fluxes for all calibrators except those with the FLUX intent.
The first subscore evaluates the completeness of the flux determination for each non-FLUX calibrator, and is simply the fraction of spws with derived fluxes (so, 1.0 if all spws have a flux determination, 0.50 if only half of them do, and so on). 


The second subscore evaluates the SNR of the flux determination for each non-FLUX calibrator, to give a blue score if the SNR falls below 20, and a yellow QA score if it falls below a SNR of 5. This is done by adopting a linear score, with a floor and a ceiling, according to the following formula:
\begin{equation}
\begin{aligned}
QA_{fluxSNR} &= 1, \texttt{ if } {\rm SNR}>26.25, \\
     &= \parbox[t]{5cm}{0.66 + (0.9$-$0.66)$\frac{({\rm SNR}-5)}{(20-5)}$,\\ \texttt{ if } 26.5$\geq$ SNR $\geq$5,}\\
     &= 0.66,  \texttt{ if } {\rm SNR} < 5.\\
\end{aligned} 
\end{equation}

The final subscore evaluates the consistency of the derived flux densities for each non-FLUX calibrator across all spws by comparing them to the expectation for the same calibrator extracted from the SC (\S\ref{QAsourcecat}). The comparison is made after scaling the SC flux values so they match for the highest SNR spw (since these sources often vary in time). So this is effectively a comparison of the inferred spectral index of the calibrator compared to the value in the SC. To calculate this QA subscore, the ratio $r_{\rm spw}$ of the derived flux for a given spw to the cataloged flux value at the central frequency of the same spw is calculated for each spw. If the flux in each spw exactly followed the behavior from the SC apart from a constant scale factor, then every spw would have the same value of $r_{\rm spw}$. To quantify deviations ($K_{\rm spw}$), each flux ratio is divided by the ratio for the spw with the highest SNR, $r_{\rm spw,maxSNR}$:

\begin{equation}
K_{\rm spw} = r_{\rm spw} / r_{\rm spw,maxSNR}.
\end{equation}

$K_{\rm spw}$ will be 1.0 when the derived flux in a spw is completely consistent with that from the highest SNR spw, otherwise it is the fractional amount the flux lies above or below the expectation from the SC. The overall QA subscore is based on the spw with the largest deviation of $K_{\rm spw}$ from 1.0, and mapped to a QA score as follows:

\begin{equation}
\begin{aligned}
QA_{\rm fluxSpw} &= 1.0, \texttt{ if max}( | 1 - K_{\rm spw} | ) < 0.1, \\
              &= 0.75, \texttt{ if } 0.1 \leq \texttt{max}( | 1 - K_{\rm spw} | ) \leq 0.2, \\
              &= 0.50, \texttt{ if max}( | 1 - K_{\rm spw} | ) > 0.2. \\
\end{aligned} 
\end{equation}

This gives the subscore a blue or green QA message if all deviations are less than 20\%, otherwise it is yellow. 

It is worth noting that this subscore can be low for reasons other than the spw fluxscale being incorrect, for instance if the spw has a low SNR (since excessive phase noise will lead to amplitudes biased low), or if there is an atmospheric absorption line in that spw. One should look at other pipeline stages (particularly the \pcmd{hifa_applycal} stage) to decide if the low QA actually represents a problem for the science target. 

\subsubsection{Consistency of cross-spw phase solutions}
\label{timegaincalQA}


In the \pcmd{hifa_timegaincal} stage (\S\ref{timegaincal}) there is a number of QA subscores to compare the phase calibrator's phase solution for an individual antenna/polarization/spw/scan against the phase solution that would be calculated if the signals from all spws were combined.  The difference between these solutions are plotted near the bottom of the WebLog page and should be zero (within the noise) if the variation in the phase solutions is due only to residual non-dispersive atmospheric path variations. Any systematic differences from zero indicate instrumental instabilities. These spw-to-spw offsets will correctly calibrate out unless spw mapping or combination is triggered due to the selection of a weak calibrator (\S\ref{spwmapping} \& Figure~\ref{fig:timegaincalLowSNR}).    Even with a strong calibrator, these instabilities can limit the SNR of observations with very high dynamic range.  In any case, it is always worth identifying these offsets since they generally indicate hardware problems.

The QA subscores are based on examining the mean offset (on a per-antenna/polarization/spw basis) for all scans, the maximum deviation for any given scan, and the scatter in the offsets for an antenna compared to the average scatter for all antennas. There is also a check for the scan-to-scan variation in the offsets considering all solutions, which indicates whether the solutions are too noisy to reliably calculate any score. 

The subscores are based on the following three metrics, where $\Delta \Phi(w)$ is the phase offset for the data selected by $w$ (a specific combination of spw, antenna, and polarization) and $N(w)$ is the number of solutions selected by $w$:


\begin{equation}
\begin{aligned}
Mean_{\rm Off}(spw_{\rm j},ant_{\rm k},pol_{\rm l}) & = \frac{\sum_{\rm scans} \Delta \Phi(j,k,l,scan)}{N(j,k,l)} \\
\end{aligned} 
\end{equation}

\begin{equation}
\begin{aligned}
Max_{\rm Off}(spw_{\rm j}, ant_{\rm k}) & = \texttt{max}(| \Delta \Phi(j,k,allpol,allscan) |) \\
\end{aligned} 
\end{equation}

\begin{equation}
\begin{aligned}
\parbox{8cm}{$RMS_{\rm Off1}(spw_{\rm j},ant_{\rm k},pol_{\rm l}) =$}\\ 
\sqrt{\frac{\sum_{\rm scans} (\Delta \Phi(j,k,l,scan) - Mean_{\rm Off}(j,k,l))^2} {N(j,k,l) - 1} } \\
\end{aligned} 
\end{equation}

\begin{equation}
\begin{aligned}
RMS_{\rm OffAll}(spw_{\rm j}) & = \texttt{MAD}(\Delta \Phi(j,all ants, pols, scans)) \\ 
\end{aligned} 
\end{equation}

The QA subscores are calculated on a per-spw, per-antenna basis, according to the following logic:
\begin{itemize}

\item If $RMS_{\rm OffAll}(spw_{\rm j})>15^{\rm o}$, then that spw is considered too noisy to reliably evaluate and given a subscore of 0.82 (blue) and no other tests are performed.

\item Otherwise, 
\begin{itemize}
\item if $Max_{\rm Off}(spw_{\rm j}, ant_{\rm k}) > 30^{\rm o}$ or $>6 \times RMS_{\rm OffAll}(spw_{\rm j})$, or 
\item if $Mean_{\rm Off}(spw_{\rm j},ant_{\rm k},pol_{\rm l}) >30^{\rm o}$ or $>6\times RMS_{\rm OffAll}(spw_{\rm j}) / \sqrt{N_{\rm ant}}$, or 
\item if the number of scan intervals is more than 3 and $RMS_{Off1(}spw_{\rm j},ant_{\rm k},pol_{\rm l})>30^{\rm o}$ or $>4\times RMS_{\rm OffAll}(spw_{\rm j})$, 
\end{itemize}
then that spw/antenna is given a subscore of 0.5 (yellow). 

\item Otherwise, 
\begin{itemize}
\item if $Max_{\rm Off}(spw_{\rm j}, ant_{\rm k})>15^{\rm o}$ or $>6\times RMS_{\rm OffAll}(spw_{\rm j})$, or 
\item if $Mean_{\rm Off}(spw_{\rm j},ant_{\rm k},pol_{\rm l})>5^{\rm o}$ or $6 \times RMS_{\rm OffAll}(spw_{\rm j}) / \sqrt{N_{\rm ant}}$, or 
\item if the number of scan intervals is more than 3 and $RMS_{\rm Off1}(spw_{\rm j},ant_{\rm k},pol_{\rm l})>15^{\rm o}$ or $>4\times RMS_{\rm OffAll}(spw_{\rm j})$, 
\end{itemize}
then that spw/antenna is given a subscore of 0.8 (blue). 

\end{itemize}

Finally, if any antenna has less than four scan intervals with phase solutions, there is an additional subscore of 0.85 (blue) and a message reporting that the standard deviation tests could not be evaluated (so only $Mean_{\rm Off}$ and $Max_{\rm Off}$ were evaluated).  As always, the overall stage QA score is the minimum of all these scores over all spw and antennas.

\subsubsection{Calibrator gain vs. frequency behavior}

The \pcmd{hif_applycal} stage presents various plots of the calibrated data. The efficacy of the per-antenna calibration solutions are evaluated by QA subscores that compare the results of each antenna against the median solution considering all antennas, and identifying antennas whose calibrated solutions are anomalous. 

For point sources, every antenna should have the same linear Amplitude vs. Frequency behavior, and should have Phase vs. Frequency behavior that is flat and centered at zero. Deviations from these behaviors could reflect bad data or calibration results which could corrupt the antenna-based gain solutions. Therefore, the QA subscores are calculated by fitting a linear function to the calibrated gains (Amplitude \& Phase) vs. frequency for all calibrators. 

The fits are done on a per-antenna, per-spw, per-polarization (considering only XX or YY correlations), per-scan basis, and again for the data combined over all scans (the all-scan fits). The fitting procedure is as follows.

The calibrated visibilities are loaded for each spectral window and scan, vector-averaged over the time dimension (per scan or all scans), and grouped per antenna. A standard error of the mean 
is calculated per channel as well. If the SNR of the data is less than 3, the data are considered too noisy to fit a slope reliably, so the slopes are set equal to zero and the midpoints are set equal to the median of the amplitudes and phases. Otherwise, the fits proceed as follows. 

First, the calibrated amplitudes are fit with the following linear function:

\begin{equation}
A_{\rm fit}(\nu') = m_{\rm amp}\nu' + b_{\rm amp}
\end{equation}
where $\nu' = (\nu-\nu_{\rm mid})/(\nu_{\rm max}-\nu_{\rm min})$,
$\nu_{\rm min}$ and $\nu_{\rm max}$ are the minimum and maximum frequencies in
the spectral window, $\nu_{\rm mid} = \frac{1}{2}(\nu_{\rm max}+\nu_{\rm min})$ is the central frequency, and $b_{\rm amp}$ and $m_{\rm amp}$ are the midpoint amplitude and amplitude slope that are being fitted for. The fits are performed with SciPy's module \pcmd{optimize.curve_fit}, which returns the best fitting values along with their fit errors ($\sigma_{b_{\rm amp}}, \sigma_{m_{\rm amp}}$).

For each antenna $k$, the fit parameters  are then compared to the median value of the fits for the same parameter for all $N_{\rm ant}$ antennas with fits, and a normalized absolute deviation ($\Delta AmpOffset_k$) is calculated by normalizing by the fit error and the standard error of the median ($\sqrt{\frac{\pi}{2}}\texttt{MAD}(b_{\rm amp})/\sqrt{N_{\rm ant}}$) added in quadrature:

\begin{equation}
    \Delta AmpOffset_k = \frac{| b_{\rm amp,k} - \texttt{Mdn}(b_{\rm amp}) |}{\sqrt{\sigma_{b_{\rm amp,k}}^2 + \pi {\texttt{MAD}}(b_{\rm amp})^2 /(2N_{\rm ant})}},
\end{equation}
and similarly for $\Delta AmpSlope_k$. 

The phases are fit a bit differently, since the fits can be confused if the phase offsets are large enough that some values wrap around $\pm\pi$. So first we do a 2-D minimization of the $\chi^{2}$ of the phases around the complex unit circle: 

\begin{multline}
\chi^{2}(m_{\rm phase1},b_{\rm phase1}) = \sum_{i=1}^{\mathrm{nchan}}\\
\frac{|\hat{V}(\nu'_i) - e^{i(m_{\rm phase1} \nu'_{\rm i} + b_{\rm phase1})}|^{2}}{|\hat{\sigma}(\nu'_i)|^{2}}
\end{multline}
where $\hat{V}(\nu'_i) = V(\nu'_i)/|V(\nu'_i)|$ and $\hat{\sigma}(\nu'_i) = \sigma(\nu'_i)/|V(\nu'_i)|$ are the per-channel calibrated visibilities and their errors normalized by the amplitude, and $b_{\rm phase1}, m_{\rm phase1}$ are the midpoint phase and phase slope that are being solved for. The minimization is performed with the SciPy module \pcmd{optimize.minimize}. 

After finding the best fit values $m_{\rm phase1}$ and $b_{\rm phase1}$, the channelized visibilities are multiplied by $e^{-i(m_{\rm phase1} \nu'_{\rm i} + b_{\rm phase1})}$ in order to center the phases around zero. The residual centered phases are then fit with the following linear function:

\begin{equation}
\Phi_{\rm fit}(\nu') = m_{\rm phase2}\nu' + b_{\rm phase2} 
\end{equation}
which returns the best fitting values for the midpoint phase ($b_{\rm phase2}$) and  phase slope ($m_{\rm phase2}$), along with their fit errors ($\sigma_{b_{\rm phase2}}, \sigma_{m_{\rm phase2}}$). The final phase slope and midpoint are $m_{\rm phase} = m_{\rm phase1} + m_{\rm phase2}$ and $b_{\rm phase} = b_{\rm phase1} + b_{\rm phase2}$, respectively. In order to have consistent statistical behavior, the errors used are $\sigma_{m_{\rm phase}} = \sigma_{m_{\rm phase2}}$ and $\sigma_{b_{\rm phase}} = \sigma_{b_{\rm phase2}}$.

For each antenna $k$, the normalized absolute deviation for the phase parameters ($\Delta PhaseOffset_k$) follows similarly as the amplitude, except instead of comparing to the median of the fit solutions for all antennas, both the midpoint phase and phase slope are compared to zero, which is the expectation for a properly calibrated point source at the phase center:

\begin{equation}
    \Delta PhaseOffset_k = \frac{| b_{\rm phase,k} |}{\sigma_{b_{\rm phase,k}}},
\end{equation}
and similarly for $\Delta PhaseSlope_k$. 

Finally, a QA score is calculated by comparing the normalized absolute deviations to the following empirically derived thresholds ($T$):

\begin{equation}
\label{applycalThresholds}
\begin{aligned}
T_{\Delta AmpOffset} &= 53.0,\\
T_{\Delta AmpSlope} &= 25.0,\\
T_{\Delta PhaseOffset} &= 60.5,\\
T_{\Delta PhaseSlope} &= 40.0.\\
\end{aligned}
\end{equation}

If any antenna has fit solutions (on a per-spw, per-pol, per-scan, or all-scan basis) that go above these thresholds, the QA score is set to 0.9 (blue), otherwise it is set to 1.0 (green). A text file is linked to from the {\tt hifa\_applycal} WebLog page that lists every outlier, its data selection, and its normalized absolute deviation.

The QA scores for this stage are only blue since we have not quantified how many outliers, or how large of an outlier, are likely to make a significant difference in the final products. We hope to rectify this in future versions of the pipeline. For now, these scores could help identify which antennas should be examined, should the WebLog plots or pipeline products suggest that something is amiss. 

Currently, these subscores are calculated for all calibrator intents, including CHECK. However, \checksources\/ are calibrated by phase transfer, and their solutions do not affect the calibration of the science target. Still, they can be diagnostic of other calibration errors, but one should not be concerned if many antennas appear to have phase offsets above the established thresholds (Equation~\ref{applycalThresholds}). 

\subsubsection{Renormalization scaling spectrum}

The QA subscores for the \pcmd{hifa_renorm} stage (\S\ref{renormalization}) are only calculated if there are FDM spectral windows, and are calculated on a per-spw basis. If all values of the calculated scaling spectrum are below 1.02 (so are not applied), then the score is set to 1.0 (green). Otherwise the values above 1.02 are checked to see if they align with absorption features in the atmospheric transmission curve, and the QA subscore depends on this result and the values of the parameters \param{apply} (default=True), \param{atm_auto_exclude} (default=True), and \param{excludechan} (default=unset). 

\begin{itemize}

\item The QA subscore is set to 0.90 (blue) for the following conditions:
\begin{itemize}
\item \param{apply}=True and the values above 1.02 do not align with atmospheric features,
\item \param{apply}=True and the values above 1.02 do align with an atmospheric feature but  \param{atm_auto_exclude}=True, with a message reporting the channels that were excluded,
\item \param{apply}=False and the values above 1.02 do align with atmospheric features, with a message reporting the recommended channels to exclude.
\end{itemize}

\item If \param{apply}=True and \param{atm_auto_exclude}=False and the values above 1.02 align with atmospheric features, then:
\begin{itemize}
\item if \param{excludechan} is unset, the QA score is set to 0.66 (yellow) with a message reporting the recommended channels to exclude.
\item if \param{excludechan} is set, the QA score is set to 0.85 (blue).
\end{itemize}

\end{itemize}

\subsection{Imaging QA subscores based on heuristics}

\subsubsection{Achieved RMS}
All stages of imaging (\pcmd{hif_makeimages} for calibrators or science targets, and for mfs continuum, aggregate continuum, or cubes) have a QA subscore aimed at identifying when significant imaging artefacts remain in the imaging product. The metric is based on a comparison of the measured noise ($\sigma_{\rm image}$, as measured in the image not corrected for primary beam response using the standard noise annulus which excludes pixels within the clean mask --- see \S\ref{noiseAnnulus}),
to the theoretical noise ($\sigma_{\rm theory}$, see \S\ref{sensitivity}) multiplied by the appropriate ``dynamic range correction factor'' ($f_{\rm DR}$, from Tables~\ref{calibratorDR} \& \ref{scienceDR}). The QA subscore is calculated by adopting a linear score, with a floor of 0 and a ceiling of 1, according to the following formula:

\begin{equation}
\begin{aligned}
QA_{\rm RMS} &= 1, \texttt{ if } \sigma_{\rm image} < f_{\rm DR}\sigma_{\rm theory} \\
            &= 1.25 - 0.25\sigma_{\rm image} / (f_{\rm DR}\sigma_{\rm theory}), \\
& ~~\texttt{ if } f_{\rm DR}\sigma_{\rm theory} \geq \sigma_{\rm image} \geq 5 f_{\rm DR}\sigma_{\rm theory} \\
     & = 0,  \texttt{ if } \sigma_{\rm image} > 5 f_{\rm DR} \sigma_{\rm theory}. \\
\end{aligned}
\end{equation}


\subsubsection{\Checksource\/ imaging}
\label{checksourceQA}
There are additional imaging QA subscores for the \checksource, since it is presumably a point source that should appear at the imaging phase center.  Therefore checking deviations from these assumptions is diagnostic of how well the phase transfer calibration worked for the science targets, since the \checksource\/ is observed in a similar manner.

The first subscore ($QA_{\rm pos}$) evaluates the astrometric quality of the phase transfer calibration by measuring the difference between the fitted position of the \checksource\/ (\S\ref{checksource}) and the position listed in the ALMA calibrator catalog, expressed in units of the synthesized beam width ($\Omega_{\rm check}$=| (catalog position -- fitted position) | / synthesized beam size). It is calculated as follows:

\begin{equation}
\begin{aligned}
QA_{\rm pos} & = 1.0, \texttt{ if } \Omega_{\rm check}  \leq 0.3, \\
              & = 1.0 - \Omega_{\rm check}, \texttt{ if } \Omega_{\rm check}  > 0.3. \\
\end{aligned}
\end{equation}

The second subscore ($QA_{\rm flux}$) evaluates the recovered flux for the \checksource\/ ($S_{\rm check}$) compared to the flux for it calculated from the \pcmd{hifa_gfluxscale} stage ($S_{\rm gflux}$), giving low scores if the \checksource\/ is fainter than expected:

\begin{equation}
\begin{aligned}
QA_{\rm flux} & = 1.0, \texttt{ if } S_{\rm check}/S_{\rm gflux}  \geq 0.8, \\
    & = S_{\rm check}/S_{\rm gflux}, \texttt{ if } S_{\rm check}/S_{\rm gflux} < 0.8. \\
\end{aligned}
\end{equation}

The third subscore ($QA_{\rm peak}$) evaluates the peak of the fitted flux ($P_{\rm check}$)
compared to its integrated flux ($S_{\rm check}$), as a measure of
atmospheric decorrelation and/or the presence of resolved emission:

\begin{equation}
\begin{aligned}
QA_{\rm peak} & = 1.0, \texttt{ if } P_{\rm check}/S_{\rm check}  \geq 0.7, \\
    & = P_{\rm check}/S_{\rm check}, \texttt{ if } P_{\rm check}/S_{\rm check} < 0.7. \\
\end{aligned}
\end{equation}

A per-image (i.e. per-spw, per-EB) aggregate QA score is constructed from these three subscores via $QA_{\rm check} = \sqrt{QA_{\rm pos}QA_{\rm flux}QA_{\rm peak}}$ if two or more of the subscores are less than 1, otherwise via $QA_{\rm check} = \texttt{min}([QA_{\rm pos}, QA_{\rm flux}, QA_{\rm peak}])$. 

As always, the aggregate score for the whole stage is the minimum of the individual EB/spw aggregate scores. These scores are quite diagnostic if the \checksource\/ SNR is high ($>20$) and it is located at a similar distance from the phase calibrator as the science target.  Otherwise, the results can be difficult to interpret. 

\subsubsection{Cube imaging}

There is a QA score for the cube imaging stage designed to identify if there is emission in the line-free channels identified by the \pcmd{hif_findcont} stage (\S\ref{findcont}). This is done by evaluating one of the diagnostic images that is created for each cube (and displayed in the WebLog ``Other QA Images'' pages of the \pcmd{hif_makeimages} stage for cube imaging), referred to as the ``line-free moment 8'' or ``mom8fc'' image. This is a two dimensional image in which each pixel contains the maximum intensity for that pixel along the third (frequency) axis, considering only the line-free channels identified in \pcmd{hif_findcont}. 

The QA subscore is based on three measures. The first is PeakSNR, representing the peak contrast seen in the mom8fc image, and calculated as the difference between the peak in the mom8fc image less the median of the mom8fc image calculated over the standard noise annulus (\S\ref{noiseAnnulus}) divided by the median absolute deviation in the cube measured over the findcont channels:

\begin{equation}
PeakSNR = \frac{\texttt{max}(mom8fc) - \median(mom8fc_{\rm annulus})}{\texttt{MAD}(cube_{\rm findcontChannels})}
\end{equation}

The second measure is the ``Histogram Asymmetry'' (HistAsym), which is calculated by comparing the histogram of intensity values from the mom8fc image with the absolute value of the histogram constructed from a similar two dimensional image which takes instead the minimum value at each pixel in the cube over the findcont channels (also called the ``line-free moment 10'' image). When the cube channels are purely emission free, these two histograms should be very close to each other. We quantify the difference by measuring the difference between the two histograms at intensities greater than the (median+2*MAD), divided by the area of the lower histogram above the same flux level. Values near zero mean that the positive and negative flux values are very similar, while large values mean that there is either a high or low intensity tail to one of the histograms. 

The last measure is the area of the largest ``segment'' of contiguous pixels with values above a set threshold in the mom8fc image, expressed in terms of the number of synthesized beams (MaxSeg = number of pixels in the largest segment divided by the number of pixels in the synthesized beam, which is given by $4\texttt{ln}(2) \theta_a \theta_b / \theta_{pixel}^2$, where $\theta_a\theta_b$ is the product of the synthesized beam major and minor axes). The larger this value, the more extensive any emission in the mom8fc image appears. From this last measure, an additional metric is calculated, which is the fractional area of the image covered by that largest segment ($f_{\rm segment}$ = the number of pixels in the largest segment divided by the number of pixels in the mom8fc image). This value will be used to calculate a QA score that becomes lower with the extent of any emission in the mom8fc image.

The set threshold for calculating segments is given by the following:
\begin{equation*}
\begin{aligned}
cut_1 &= \median(mom8fc)+3\texttt{MAD}(cube_{\rm findcontChannels})\\
cut_2 &= \median(mom8fc)+ \frac{\texttt{max}(mom8fc)-\median(mom8fc)}{2}\\
cut_3 &= \median(mom8fc)+2\texttt{MAD}(cube_{\rm findcontChannels})\\
Thresh &= \texttt{max}(\texttt{min}(cut_1,cut_2),cut_3)\\
\end{aligned}
\end{equation*}

The QA subscore is calculated based on the following thresholds:

\begin{center}
if ((PeakSNR > 5.0) and (HistAsym > 0.20)) \\
or \\
((PeakSNR > 3.5) and (HistAsym > 0.05) and (MaxSeg > 1.0)) \\
then $QA_{mom8fc}$ = \texttt{min}($QA_{\rm erf}$, 0.65)\\
else $QA_{mom8fc}$ = \texttt{max}($QA_{\rm erf}$, 0.67)\\
\end{center}

where $QA_{\rm erf}$=1.0, if $f_{\rm segment}=0$, otherwise:
\begin{multline}
QA_{\rm erf} = 0.33 + 0.5 * (1-0.33) * \\ [1 + \texttt{erf}(-\texttt{log10}(100 f_{\rm segment}) )].
\end{multline}

The first set of criteria (before the ``else'' condition) define the conditions for triggering QA scores between 0.33---0.65 (yellow), with lower scores for larger segments. If these criteria are not met, then the QA score will be greater than 0.67 (blue to green), with higher scores for smaller segments. 

These scores are defined to identify if there is noticeable emission in the mom8fc image, but that does not necessarily mean that something has gone wrong with continuum subtraction or cube imaging. It could be that there is spectrally unresolved line emission or that there are low SNR features that are missed by the \pcmd{findContinuum} algorithm. In the majority of cases, the cubes with poor $QA_{mom8fc}$ scores do not need to be regenerated, because a small change in the continuum ranges will have little to no effect on uv-based continuum subtraction. But in some cases one may want to create a new aggregate continuum image outside of the pipeline to avoid potential line contamination, or use different channel ranges when making moment images. 

\section{Testing and validation}
\label{validation}

\subsection{Testing of CASA tasks critical to the pipeline}

CASA provides the underlying general purpose data reduction and imaging tasks used by the pipeline to process data. Since these tasks are not specific to ALMA, other users and groups can request bug fixes and changes that may directly or indirectly impact the pipeline. The most complex and actively developed of the CASA tasks used by the Pipeline is \ccmd{tclean}. The ALMA imaging CASA Stakeholder test is designed to guard against unexpected changes in tclean affecting critical functionality needed for the ALMA pipeline \citep[e.g.,][]{2018nrao.reptE...2B}. It currently consists of 22 test cases whose inputs match those used by the ALMA pipeline and span the range of ALMA imaging use cases including combinations of the following usage modes: 
\begin{itemize}
\item single fields and mosaics, 
\item multi-frequency synthesis (mfs), multi-term multi-frequency synthesis (mtmfs), and cubes, 
\item per channel density weighting and briggs bandwidth taper weighting, and
\item ephemeris objects.
\end{itemize}
The tests are run on the development branch of a CASA ticket in the verification stage that occurs after a feature has been implemented. For each test case, the detailed properties of the resulting image products including the weights, point spread function, primary beam, mask, model, and residuals are checked and if any value varies by more than 1\%, the test fails for that use case. If any test fails, the results are inspected and either fixed, or, in cases where a change is expected, approval requested to change the fiducial numbers.

Additional ALMA CASA stakeholder tests may be implemented in the future. The CASA group also adds verification tests for all resolved CASA bugs, including those triggered by the pipeline. There are also a set of  CASA performance benchmark tests\footnote{\url{https://casangi.github.io/casabench/}} that include \ccmd{flagdata}, which is one of the most time-consuming tasks in the pipeline owing to the numerous flag summaries generated.

\subsection{Pipeline regression testing}
Regression tests ensure the consistency of pipeline results by quickly identifying unexpected changes to the pipeline output during pipeline software development. There are two automated regression tests dedicated to testing the ALMA interferometric pipeline. These regression tests run an entire pipeline recipe on an input dataset, extract specific quantities from the results, and then compare these new values to previously-saved reference values. At this time, 19 quantities (some per-field, spw, or scan) are used for this comparison from the \pcmd{hifa_gfluxscsaleflag}, \pcmd{hifa_gfluxscale}, and \pcmd{hif_applycal} pipeline stages. If the values do not match within a tolerance which is specified per-quantity, the test will fail. 
When a test fails, the reason for the failure is investigated and either the software will be updated to fix the issue or the reference data will be updated to reflect an expected change in output. 
Regression tests are run automatically each night a new change has been merged into the main branch of the pipeline software. The CASA version used is the most recent version of the branch intended for use by the next pipeline release. 

The regression tests use the pytest Python testing framework \citep{pytest}. Bamboo, a continuous integration tool from Atlassian, is used to run the automated tests. Both the number of regression tests and the number of quantities they compare from the pipeline results are currently being expanded to establish more comprehensive automated testing. 

\subsection{Pipeline benchmark testing}
\label{benchmark}
The pipeline is tested and validated using the pipeline benchmark suite, which consists of PI projects from Cycles 3 to 9 along with Observatory-led test projects which are used for end-to-end validation when preparing for a new Cycle. These 96 MOUS have been selected to cover a wide parameter space representing different bands, array configurations (both ACA and 12m), observing modes, spatial and spectral set-ups, calibration strategies, and target emission characteristics such as dynamic range, spatial and spectral distribution of emission, etc.  All of the distinguishing properties are tabulated on a shared spreadsheet with dozens of columns. 

The development cycle starts in October every year and ends in September of the next year when the final pipeline version for the new observing Cycle is released. During this period, these datasets are used to test the implementation of new features, new heuristics, and bug fixes in the pipeline, and to check the effect of any changes to underlying CASA tasks on the pipeline. The implementation of these features and bug fixes are tracked in individual pipeline tickets, with a separate branch of the pipeline created for each ticket. Subsets of projects from the benchmark are identified based on testing requirements for each ticket and run with the associated branch of the pipeline. The WebLogs and products are inspected to confirm that the results are scientifically valid, to identify any bugs in the implementation, and to ensure any changes in results compared to pipeline runs with previous releases are expected and understood. Once the ticket is validated, the changes are committed to the main branch.  

\subsection{Pipeline validation}
Towards the end of the development cycle when all of the scheduled pipeline tickets have been implemented and committed to the pipeline main branch, the entire benchmark suite is run, the WebLogs are reviewed and any bugs are reported and fixed, if possible. The latest end-to-end (E2E) datasets acquired for the upcoming Cycle are also run. Once all bug fixes are in and the final version of the pipeline is ready, the benchmark is run for a final time, to be saved and used for comparisons with future versions. At this stage, notes are made for future improvements/changes and any known issues that have not been resolved are documented (see \S\ref{bugfixes}). 

Once the pipeline has been validated by the international ALMA Pipeline Working Group (PLWG), a letter recommending its usage for ALMA regular operations is sent to the manager of the ALMA Data Management Group (DMG), and a shorter validation process starts at the Joint ALMA Observatory (JAO) in Santiago, Chile.
The PLWG provides a list of $\sim$15 datasets (representative of the entire benchmark) to process at JAO and compare the results with the pipeline results produced by PLWG. The goal here is to confirm that the pipeline results can be reproduced properly, even in different operating systems and with different processing node specifications, when keeping the required inputs fixed (the flux density measurements described in \S\ref{importdata}, and the antenna position corrections described in \S\ref{antposcorr}).
This validation is done in the same cluster used for ALMA science data processing. When required, a separate environment is created to use the updated versions of all ALMA systems that will be used in the upcoming observing Cycle, so the listed MOUSes are processed following the same workflow as ALMA science data.

Once pipeline executions have finished, the resulting WebLogs are compared to those produced by PLWG. Everything should be identical, from flagged data to imaging statistics, including flagging statistics and calibrated flux densities. When the results are not identical, an investigation is done to determine what causes the differences. This investigation ends with either a fix of the bug/problem and a new validation process, or with an understanding of the issue considered as acceptable (like rounding differences).
Once this validation at JAO is finished successfully, the new pipeline is ready to be used within the new ALMA Cycle. The shorter validation mechanism is employed when the need arises to  generate a patch release in the middle of an observing Cycle. Such releases only occur when an impactful bug is uncovered.

\section{Future plans and directions}
\label{future}


\subsection{Bug fixes and efficiency improvements}
\label{bugfixes}
The list of known issues in the pipeline (both current and past releases) is tracked on a publicly available website\footnote{\url{https://casaguides.nrao.edu/index.php?title=ALMA_Pipeline_Known_Issues}}.  Many of these issues, such as the one described in \S\ref{lowTransmission} for data older than Cycle 3, have already been resolved during the current development cycle, which is preparing for the next pipeline release (ALMA Cycle 10 using CASA 6.5.4).  In addition, the accuracy and precision of the QA scores are being improved to facilitate completely automated QA assessment.
Finally, some processing inefficiencies have been identified (see, e.g., \S\ref{fluxscale}), and will be addressed in future releases.

\subsection{Self-calibration}
\label{selfcal}
The use of antenna-based self-calibration on science targets allows one to model and remove residual calibration errors that cause image artifacts or limit image dynamic range  
\citep{Brogan2018}.
A significant feature has been added in the 2022 pipeline release to prepare for the future introduction of a new stage that implements iterative phase-only self-calibration of the science targets.  In previous releases, the calibrated data was written to the DATA column of the science target MS and the continuum subtracted version of the data was placed in the CORRECTED column.   In the 2022 release, as explained in \S\ref{uvcontsub}, the calibrated science target data is split into the DATA column of two separate MS, one containing the continuum+line data and one containing the continuum-subtracted data. This method frees the CORRECTED column of each MS to contain the future self-calibrated version of the data.  Initial deployment of an optional self-calibration stage is expected in the 2023 release.

\subsection{Polarization calibration}
\label{futurePolcal}
At present, the pipeline performs total intensity calibration in a ``polarization-friendly'' manner. A single optimal refant is chosen and used for all EBs (\S\ref{refantpolcal}), and bad data on the polarization calibrator is flagged in the extra stage \pcmd{hifa_polcalflag}.  A future release will include the additional stage needed to compute polarization calibration tables via the \ccmd{polcal} and \ccmd{polfromgain} tasks \citep{Moellenbrock2017}, which are currently run manually following pipeline calibration \citep{Petry2021}.

\subsection{Group OUS processing}

Group OUS (GOUS) processing encompasses several use cases: the combination of multiple 12m array observations from different antenna configurations, the combination of 12m with 7m array observations, and the combination of interferometric observations with total power observations \citep{Plunkett2023}. Such combinations are not yet supported in the pipeline, and the implementation of self-calibration and polarization calibration have a higher priority.

\subsection{Processing wider bandwidths and more channels}
\label{WSU}
As the first major initiative of the ALMA2030 road map \citep{Carpenter2020}, the ALMA Wideband Sensitivity Upgrade \citep[WSU,][]{WSU2022} will widen the instantaneous bandwidth of the receivers and replace all elements of the digital signal chain (including the digitizers, data transmission system, correlator, total power spectrometer, and associated physical infrastructure), which will improve the system efficiency \citep{Asayama2020}.   The first results producing double the current correlated bandwidth are expected toward the end of this decade.  Such wider bandwidths will allow the fitting of calibrator spectral indices in many of the lower frequency bands rather than relying on interpolation of the flux monitoring database.   The use of channelized weights will also become more important to obtain optimal sensitivity for these wider bandwidth observations.   Also, some of the current heuristics that make decisions based on spw bandwidth, channel width, or other characteristics, such as the channel response function and effective bandwidth, will need to be modified to also handle the new characteristics of WSU spws.

The WSU will also enable much higher spectral resolution at full correlated bandwidth, as fine as 13.5\,kHz, resulting in significantly higher data rates for some specific science use cases \citep{Brogan2023}.  The higher data rates can be partly mitigated by observing calibrators, such as the phase calibrator, at lower spectral resolution than the science target.  This technique represents a new ALMA observing mode -- ``bandwidth switching'' -- which is not yet supported by the pipeline, but is a near-term goal.    Such additions will need to be prioritized against other infrastructure-related work and QA scoring improvements, which are also aimed at helping the observatory process, assess, and deliver the larger data volumes enabled by the WSU.  

\subsection{Refactor for ngCASA}

On the longer term horizon, the pipeline code will require some changes to adapt to the architecture of next generation CASA \citep[ngCASA,][]{Raba2022}, which is being designed to enable further parallelization across cluster nodes \citep{CASA2022}. Such improvements will be crucial to process the much larger number of channels provided by the WSU as the data rate rollout plan unfolds, particularly once the full upgrade to four times the current correlated bandwidth is achieved and more front end receiver bands are upgraded.

\subsection{Impact on future observatories}

The successful heuristics and algorithms developed for the ALMA pipeline, along with those of the VLA pipeline \citep{Kent2020}, will form an advanced starting point for future interferometric pipelines, most notably for the next generation VLA \citep[ngVLA,][]{Selina22}.


\appendix
\section{Glossary of acronyms}
\label{appendixA}
\begin{longtable}[c]{ll}
\caption{Definition of acronyms used \label{glossary}}\\
\textbf{Acronym} & \textbf{Definition} \\
\hline
\endfirsthead
 \multicolumn{2}{c}{Continuation of Table \ref{glossary}}\\
 \hline
 \endhead
ACA & Atacama Compact Array \\
ACAC & Atacama Compact Array Correlator\\
ACD & ALMA Calibration Device\\
ADMT & ALMA Data Mining Toolkit\\
aggBW & aggregate bandwidth (of continuum)\\
ALMA & Atacama Large Millimeter/submillimeter Array \\
AQUA & ALMA Quality Assurance tool\\
AR & Acceptable Resolution\\
ASDM & ALMA Science Data Model \\
ASIAA & Academia Sinica Institute of Astronomy and Astrophysics\\
ATM & Atmospheric Transmission at Microwaves\\
AUI & Associated Universities Incorporated\\
B2B & Band To Band \\
BDF & Binary Data File\\
BLC & Baseline Correlator \\
CASA & Common Astronomy Software Applications package\\
DD & Derivative Deviation\\
DR & Dynamic Range\\
DRC & Dynamic Range Correction\\
E2E & End To End\\
EB & Execution Block \\
ESO & European Southern Observatory \\
FDM & Frequency Division Multiplexing \\
FITS & Flexible Image Transport System \\
FOV & Field Of View\\
FWHM & Full Width at Half Maximum\\
FX & Fourier transform then cross correlation\\
FXF & digital Filtering then cross correlation then Fourier transform\\
GOUS & Group Observation Unit Set \\
ICRS & International Coordinate Reference System\\
IDR & Interdecile Range\\
IERS & International Earth Rotation Service\\
IF & Intermediate Frequency\\
IQR & Interquartile Range\\
JAO & Joint ALMA Observatory\\
KASI & Korea Astronomy and Space Science Institute\\
LO & Local Oscillator\\
LSRK & kinematic Local Standard of Rest \\
MAD & Median Absolute Deviation from the median \\
MAPS & Molecules with ALMA at Planet-forming Scales\\
mfs & multi-frequency synthesis\\
MOST & Ministry of Science and Technology\\
MOUS & Member Observation Unit Set \\
MPI & Message Passing Interface\\
MS & MeasurementSet \\
mtmfs & multi-Taylor multi-frequency synthesis\\
NAOJ & National Astronomical Observatory of Japan\\
ngCASA & next generation Common Astronomy Software Applications\\
ngVLA & next generation Very Large Array\\
NINS & National Institute of Natural Sciences\\
NRAO & National Radio Astronomy Observatory\\
NRC & National Research Council\\
NRO & Nobeyama Radio Observatory \\
NSF & National Science Foundation \\
OT & Observing Tool \\
PB & Primary Beam\\
PDF & Portable Document Format\\
PHANGS & Physics at High Angular resolution in Nearby Galaxies \\
PI & Principle Investigator\\
PLUG & Pipeline User's Guide\\
PLWG & Pipeline Working Group\\
PPR & Pipeline Processing Request \\
PSF & Point Spread Function\\
QA & Quality Assessment \\
repBW & representative bandwidth \\
repSpw & representative spectral window\\
RF & Radio Frequency\\
RFI & Radio Frequency Interference\\
RMS & Root Mean Square\\
SB & Scheduling Block \\
SC & calibrator Source Catalog\\
SNR & Signal to Noise Ratio\\
spw & spectral window\\
SQLD & Square Law Detector\\
SSO & Solar System Object\\
TAI & International Atomic Time\\
TDM & Time Division Multiplexing \\
TFB & Tunable Filter Bank\\
TMCDB & Telescope Monitor and Control DataBase\\
TPA & Total Power Array \\
UTC & Coordinated Universal Time\\
VLA & Karl G. Jansky Very Large Array\\
VLBI & Very Long Baseline Interferometry\\
WSU & Wideband Sensitivty Upgrade\\
WVR & Water Vapor Radiometer\\
XF & cross correlation then Fourier transform\\
XML & Extensible Markup Language \\
XX & correlation product of the ``X'' linearly polarized feeds\\
YY & correlation product of the ``Y'' linearly polarized feeds\\
\end{longtable}

\section{Frequency ranges flagged by \titlelowercase{\pcmd{hifa_fluxcalflag}}}
\label{appendixB}
\label{ssolines}
Table~\ref{tab:ssolines} lists the molecular species and their corresponding spectral line frequencies that are flagged in the stage \pcmd{hifa_fluxcalflag} on specific solar system objects when they are observed as flux calibrators.

\begin{table}[ht]
\caption{Topocentric frequency ranges flagged by \pcmd{hifa_fluxcalflag} on selected solar system objects.}
\label{tab:ssolines}
\begin{flushleft}
\begin{supertabular}{lrl}
\textbf{Object} & \multicolumn{2}{l}{\textbf{~~~Species: ~Topocentric frequency ranges flagged (GHz)}} \\
\hline
\textbf{Mars} &
CO:& [115.204,115.338], [230.404,230.672], [345.595,345.997], 
[460.773,461.309], \\
& & [691.071,691.875], [806.184,807.120],
[921.265,922.335]\\
&$^{13}$CO:& [110.190,110.212], [220.377,220.421], [330.555,330.621],
[440.721,440.809],\\
& & [661.001,661.133], [771.108,771.260], 
[881.196,881.370]\\
\hline

\textbf{Venus} &
CO:& [115.206,115.337], [230.407,230.669], [345.600,345.992], 
[460.779,461.303], \\
& & [691.081,691.865], [806.194,807.110], 
[921.277,922.323]\\
&$^{13}$CO:& [110.192,110.210], [220.380,220.418], [330.560,330.616], 
[440.727,440.803],\\
& & [661.011,661.123], [771.118,771.250], 
[881.208,881.358]\\\hline

\textbf{Titan} &
CO:& [114.92,115.67], [229.49,231.74], [343.82,347.62], [458.29,463.80],\\
& & [687.75,694.66], [803.46,809.85]\\
&$^{13}$CO:& [110.18,110.22], [220.28,220.52], [330.36,330.82], [440.42,441.15],\\
& & [660.60,661.53], [770.65,771.72], [880.73,881.82]\\
&HCN: &[88.45,88.81], [176.73,177.80], [264.96,266.81], [353.29,355.72], \\
& & [441.74,444.52], [618.45,622.16], [707.01,710.74], [795.56,799.31], \\
& & [883.82,887.86]\\
&HC$^{15}$N:& [86.04,86.07], [172.05,172.16], [258.04,258.27], [430.02,430.45], \\
& & [601.95,602.60], [773.88,774.64], [859.84,860.62]\\
&H$^{13}$CN:& [86.33,86.35], [172.63,172.73], [258.91,259.11], [431.44,431.88],\\
& & [603.98,604.56], [776.48,777.16], [862.72,863.42]\\
&HCN v$_2$=1:& [177.192,177.286], [178.088,178.184], [265.782,265.924],
[267.128,267.270],\\
& & [354.365,354.555], [356.161,356.351], 
[442.942,443.178], [445.184,445.422],\\
& & [620.058,620.390], 
[623.197,623.529], [708.596,708.974], [712.182,712.562],\\
& & [797.117,797.543], [801.149,801.577], [885.619,886.091], 
[890.096,890.572]\\
&CH$_3$CN: & [91.938,92.008], [110.304,110.409], [128.659,128.809], 
[147.039,147.209],\\
& & [165.415,165.608], [183.792,184.006], 
[202.168,202.403], [220.543,220.798],\\
& & [238.916,239.194], 
[257.289,257.587], [275.660,275.980], [294.030,294.371],\\
& & [312.399,312.761], [330.766,331.149], [349.205,349.534],
[367.572,367.920], \\
& & [385.938,386.303], [404.301,404.683],
[422.735,423.062], [441.098,441.440],\\
& & [459.459,459.814], 
[477.881,478.186], [496.239,496.557]\\
\hline

\textbf{Neptune} &
CO:& [113.99,116.51], [226.98,234.52], [339.97,351.54], [454.95,467.55], \\
& & [685.93,696.57], [802.92,810.58]\\
\end{supertabular}
\end{flushleft}
\end{table}

\acknowledgments

We thank all of the programmers and scientists who have contributed to the pipeline development over the past two decades, with special mention
to Lindsey Davis, Ed Fomalont, Jeff Kern, and John Lightfoot.
Marcel Neeleman (NRAO) and Sergio Mart\'in (JAO) provided valuable comments and suggestions on this manuscript. We thank the anonymous referee for timely comments, corrections, and suggestions that improved this paper.
The National Radio Astronomy Observatory is a facility of the National Science Foundation operated under cooperative agreement by Associated Universities, Inc. ALMA is a partnership of ESO (representing its member states), NSF (USA) and NINS (Japan), together with NRC (Canada), MOST and ASIAA (Taiwan), and KASI (Republic of Korea), in cooperation with the Republic of Chile. The Joint ALMA Observatory is operated by ESO, AUI/NRAO and NAOJ.   This paper makes use of the following ALMA data: ADS/JAO.ALMA\#2016.1.00314.S and ADS/JAO.ALMA\#2017.1.00983.S.

\software{analysisUtils (\url{https://doi.org/10.5281/zenodo.7502160}), 
Astropy \citep{astropy2022}, 
ATM \citep{Pardo2001}, 
cachetools (\url{https://cachetools.readthedocs.io}),
CASA \citep{CASA2022}, 
Casacore \citep{CASACORE2019}, 
ImageMagick \citep{imagemagick}, 
intervaltree (\url{https://pypi.org/project/intervaltree}),
IPython \citep{IPython}, 
Javascript \citep{javascript},
mako (\url{https://www.makotemplates.org}),
Matplotlib \citep{hunter2007matplotlib}, 
NumPy \citep{2020NumPy-Array}, 
Poppler (\url{https://poppler.freedesktop.org}), 
pyparsing (\url{https://pyparsing-docs.readthedocs.io}),
pytest \citep{pytest}, 
Python \citep{python3}, 
SciPy \citep{2020SciPy-NMeth}
}

\facility{ALMA}

\bibliography{main}{}

\begin{thebibliography}{}
\expandafter\ifx\csname natexlab\endcsname\relax\def\natexlab#1{#1}\fi
\providecommand{\url}[1]{\href{#1}{#1}}
\providecommand{\dodoi}[1]{doi:~\href{http://doi.org/#1}{\nolinkurl{#1}}}
\providecommand{\doeprint}[1]{\href{http://ascl.net/#1}{\nolinkurl{http://ascl.net/#1}}}
\providecommand{\doarXiv}[1]{\href{https://arxiv.org/abs/#1}{\nolinkurl{https://arxiv.org/abs/#1}}}

\bibitem[{{Amestica} {et~al.}(2020){Amestica}, {Brandt}, {Marson}, {Rosen}, \&
  {Whiteis}}]{Amestica20}
{Amestica}, R., {Brandt}, P., {Marson}, R., {Rosen}, R., \& {Whiteis}, P. 2020,
  in Astronomical Society of the Pacific Conference Series, Vol. 522,
  Astronomical Data Analysis Software and Systems XXVII, ed. P.~{Ballester},
  J.~{Ibsen}, M.~{Solar}, \& K.~{Shortridge}, 663

\bibitem[{{Asayama} {et~al.}(2020){Asayama}, {Tan}, {Saini}, {Carpenter},
  {Hunter}, {Phillips}, {Nagai}, {Siringo}, \& {Whyborn}}]{Asayama2020}
{Asayama}, S., {Tan}, G.~H., {Saini}, K., {et~al.} 2020, in Society of
  Photo-Optical Instrumentation Engineers (SPIE) Conference Series, Vol. 11445,
  Society of Photo-Optical Instrumentation Engineers (SPIE) Conference Series,
  1144575, \dodoi{10.1117/12.2562272}

\bibitem[{{Astropy Collaboration} {et~al.}(2022){Astropy Collaboration},
  {Price-Whelan}, {Lim}, {Earl}, {Starkman}, {Bradley}, {Shupe}, {Patil},
  {Corrales}, {Brasseur}, {N{\"o}the}, {Donath}, {Tollerud}, {Morris},
  {Ginsburg}, {Vaher}, {Weaver}, {Tocknell}, {Jamieson}, {van Kerkwijk},
  {Robitaille}, {Merry}, {Bachetti}, {G{\"u}nther}, {Aldcroft},
  {Alvarado-Montes}, {Archibald}, {B{\'o}di}, {Bapat}, {Barentsen},
  {Baz{\'a}n}, {Biswas}, {Boquien}, {Burke}, {Cara}, {Cara}, {Conroy},
  {Conseil}, {Craig}, {Cross}, {Cruz}, {D'Eugenio}, {Dencheva}, {Devillepoix},
  {Dietrich}, {Eigenbrot}, {Erben}, {Ferreira}, {Foreman-Mackey}, {Fox},
  {Freij}, {Garg}, {Geda}, {Glattly}, {Gondhalekar}, {Gordon}, {Grant},
  {Greenfield}, {Groener}, {Guest}, {Gurovich}, {Handberg}, {Hart},
  {Hatfield-Dodds}, {Homeier}, {Hosseinzadeh}, {Jenness}, {Jones}, {Joseph},
  {Kalmbach}, {Karamehmetoglu}, {Ka{\l}uszy{\'n}ski}, {Kelley}, {Kern},
  {Kerzendorf}, {Koch}, {Kulumani}, {Lee}, {Ly}, {Ma}, {MacBride}, {Maljaars},
  {Muna}, {Murphy}, {Norman}, {O'Steen}, {Oman}, {Pacifici}, {Pascual},
  {Pascual-Granado}, {Patil}, {Perren}, {Pickering}, {Rastogi}, {Roulston},
  {Ryan}, {Rykoff}, {Sabater}, {Sakurikar}, {Salgado}, {Sanghi}, {Saunders},
  {Savchenko}, {Schwardt}, {Seifert-Eckert}, {Shih}, {Jain}, {Shukla}, {Sick},
  {Simpson}, {Singanamalla}, {Singer}, {Singhal}, {Sinha}, {Sip{\H{o}}cz},
  {Spitler}, {Stansby}, {Streicher}, {{\v{S}}umak}, {Swinbank}, {Taranu},
  {Tewary}, {Tremblay}, {Val-Borro}, {Van Kooten}, {Vasovi{\'c}}, {Verma}, {de
  Miranda Cardoso}, {Williams}, {Wilson}, {Winkel}, {Wood-Vasey}, {Xue},
  {Yoachim}, {Zhang}, {Zonca}, \& {Astropy Project Contributors}}]{astropy2022}
{Astropy Collaboration}, {Price-Whelan}, A.~M., {Lim}, P.~L., {et~al.} 2022,
  \apj, 935, 167, \dodoi{10.3847/1538-4357/ac7c74}

\bibitem[{{Bastian} {et~al.}(2022){Bastian}, {Shimojo}, {B{\'a}rta}, {White},
  \& {Iwai}}]{Bastian22}
{Bastian}, T.~S., {Shimojo}, M., {B{\'a}rta}, M., {White}, S.~M., \& {Iwai}, K.
  2022, Frontiers in Astronomy and Space Sciences, 9, 977368,
  \dodoi{10.3389/fspas.2022.977368}

\bibitem[{{Baudry} {et~al.}(2012){Baudry}, {Lacasse}, {Escoffier}, {Webber},
  {Greenberg}, {Platt}, {Treacy}, {Saez}, {Cais}, {Comoretto}, {Quertier},
  {Okumura}, {Kamazaki}, {Chikada}, {Watanabe}, {Okuda}, {Kurono}, \&
  {Iguchi}}]{Baudry2012}
{Baudry}, A., {Lacasse}, R., {Escoffier}, R., {et~al.} 2012, in Society of
  Photo-Optical Instrumentation Engineers (SPIE) Conference Series, Vol. 8452,
  Millimeter, Submillimeter, and Far-Infrared Detectors and Instrumentation for
  Astronomy VI, ed. W.~S. {Holland} \& J.~{Zmuidzinas}, 845217,
  \dodoi{10.1117/12.925700}

\bibitem[{{Blackburn} {et~al.}(2019){Blackburn}, {Chan}, {Crew}, {Fish},
  {Issaoun}, {Johnson}, {Wielgus}, {Akiyama}, {Barrett}, {Bouman}, {Cappallo},
  {Chael}, {Janssen}, {Lonsdale}, \& {Doeleman}}]{Blackburn19}
{Blackburn}, L., {Chan}, C.-k., {Crew}, G.~B., {et~al.} 2019, \apj, 882, 23,
  \dodoi{10.3847/1538-4357/ab328d}

\bibitem[{{Bonato} {et~al.}(2018){Bonato}, {Liuzzo}, {Giannetti}, {Massardi},
  {De Zotti}, {Burkutean}, {Galluzzi}, {Negrello}, {Baronchelli}, {Brand},
  {Zwaan}, {Rygl}, {Marchili}, {Klitsch}, \& {Oteo}}]{ALMACAL2018}
{Bonato}, M., {Liuzzo}, E., {Giannetti}, A., {et~al.} 2018, \mnras, 478, 1512,
  \dodoi{10.1093/mnras/sty1173}

\bibitem[{{Briggs} {et~al.}(1999){Briggs}, {Schwab}, \& {Sramek}}]{Briggs}
{Briggs}, D.~S., {Schwab}, F.~R., \& {Sramek}, R.~A. 1999, in Astronomical
  Society of the Pacific Conference Series, Vol. 180, Synthesis Imaging in
  Radio Astronomy II, ed. G.~B. {Taylor}, C.~L. {Carilli}, \& R.~A. {Perley},
  127

\bibitem[{{Brogan}(2023)}]{Brogan2023}
{Brogan}, C. 2023, in American Astronomical Society Meeting Abstracts, Vol.~55,
  American Astronomical Society Meeting Abstracts, 134.02

\bibitem[{{Brogan} {et~al.}(2018{\natexlab{a}}){Brogan}, {Hunter}, \&
  {Fomalont}}]{Brogan2018}
{Brogan}, C.~L., {Hunter}, T.~R., \& {Fomalont}, E.~B. 2018{\natexlab{a}},
  arXiv e-prints, arXiv:1805.05266, \dodoi{10.48550/arXiv.1805.05266}

\bibitem[{{Brogan} {et~al.}(2018{\natexlab{b}}){Brogan}, {Indebetouw},
  {Hunter}, {Meyer}, {Mason}, \& {CASA Team}}]{2018nrao.reptE...2B}
{Brogan}, C.~L., {Indebetouw}, R., {Hunter}, T.~R., {et~al.}
  2018{\natexlab{b}}, {ALMA Mosaic Imaging Issues Prior to Cycle 6 (CASA
  5.4.0)}, North American ALMA Science Center (NAASC) Memo 117, September 28,
  2018, 42 pages

\bibitem[{{Brogui{\`e}re} {et~al.}(2011){Brogui{\`e}re}, {Lucas}, {Pardo}, \&
  {Roche}}]{Telcal2011}
{Brogui{\`e}re}, D., {Lucas}, R., {Pardo}, J., \& {Roche}, J.~C. 2011, in
  Astronomical Society of the Pacific Conference Series, Vol. 442, Astronomical
  Data Analysis Software and Systems XX, ed. I.~N. {Evans}, A.~{Accomazzi},
  D.~J. {Mink}, \& A.~H. {Rots}, 277

\bibitem[{{Butler}(2012)}]{ButlerALMAMemo594}
{Butler}, B. 2012, Atacama Large Millimeter/Submillimeter Array Memo 594, 594,
  1

\bibitem[{{Carpenter} {et~al.}(2022){Carpenter}, {Brogan}, {Iono}, \&
  {Mroczkowski}}]{WSU2022}
{Carpenter}, J., {Brogan}, C., {Iono}, D., \& {Mroczkowski}, T. 2022, arXiv
  e-prints, arXiv:2211.00195.
\newblock \doarXiv{2211.00195}

\bibitem[{{Carpenter} {et~al.}(2020){Carpenter}, {Iono}, {Kemper}, \&
  {Wootten}}]{Carpenter2020}
{Carpenter}, J., {Iono}, D., {Kemper}, F., \& {Wootten}, A. 2020, Atacam Large
  Millimeter/Submillimeter Array Memo 594, arXiv:2001.11076,
  \dodoi{10.48550/arXiv.2001.11076}

\bibitem[{{CASA Team} {et~al.}(2022){CASA Team}, {Bean}, {Bhatnagar}, {Castro},
  {Donovan Meyer}, {Emonts}, {Garcia}, {Garwood}, {Golap}, {Gonzalez Villalba},
  {Harris}, {Hayashi}, {Hoskins}, {Hsieh}, {Jagannathan}, {Kawasaki},
  {Keimpema}, {Kettenis}, {Lopez}, {Marvil}, {Masters}, {McNichols},
  {Mehringer}, {Miel}, {Moellenbrock}, {Montesino}, {Nakazato}, {Ott}, {Petry},
  {Pokorny}, {Raba}, {Rau}, {Schiebel}, {Schweighart}, {Sekhar}, {Shimada},
  {Small}, {Steeb}, {Sugimoto}, {Suoranta}, {Tsutsumi}, {van Bemmel},
  {Verkouter}, {Wells}, {Xiong}, {Szomoru}, {Griffith}, {Glendenning}, \&
  {Kern}}]{CASA2022}
{CASA Team}, {Bean}, B., {Bhatnagar}, S., {et~al.} 2022, \pasp, 134, 114501,
  \dodoi{10.1088/1538-3873/ac9642}

\bibitem[{{Casacore Team}(2019)}]{CASACORE2019}
{Casacore Team}. 2019, {casacore: Suite of C++ libraries for radio astronomy
  data processing}, Astrophysics Source Code Library, record ascl:1912.002.
\newblock \doeprint{1912.002}

\bibitem[{{Castro} {et~al.}(2017){Castro}, {Gonzalez}, {Taylor}, {Bhatnagar},
  {Caillat}, {Ford}, {Golap}, {Jakobs}, {Kawasaki}, {Kern}, {Kuniyoshi},
  {Loveland}, {Mehringer}, {Miel}, {Moellenbrock}, {Nakazato}, {Petry},
  {Pokorny}, {Rao}, {Rawlings}, {Schiebel}, {Kugimoto}, {Suoranta}, \&
  {Tsutsumi}}]{Castro17}
{Castro}, S., {Gonzalez}, J., {Taylor}, J., {et~al.} 2017, in Astronomical
  Society of the Pacific Conference Series, Vol. 512, Astronomical Data
  Analysis Software and Systems XXV, ed. N.~P.~F. {Lorente}, K.~{Shortridge},
  \& R.~{Wayth}, 595

\bibitem[{Chacon \& Straub(2014)}]{git}
Chacon, S., \& Straub, B. 2014, Pro Git (Apress)

\bibitem[{{Chavan} {et~al.}(2016){Chavan}, {Tanne}, {Akiyama}, {Kurowski},
  {Randall}, {Vila Vilaro}, \& {Villard}}]{Chavan16}
{Chavan}, A.~M., {Tanne}, S.~L., {Akiyama}, E., {et~al.} 2016, in Society of
  Photo-Optical Instrumentation Engineers (SPIE) Conference Series, Vol. 9910,
  Observatory Operations: Strategies, Processes, and Systems VI, ed. A.~B.
  {Peck}, R.~L. {Seaman}, \& C.~R. {Benn}, 99101H, \dodoi{10.1117/12.2232426}

\bibitem[{{Cherednichenko} {et~al.}(2010){Cherednichenko}, {Emrich}, \&
  {Peacocke}}]{WVR2010}
{Cherednichenko}, S., {Emrich}, A., \& {Peacocke}, T. 2010, in Twenty-First
  International Symposium on Space Terahertz Technology, 389

\bibitem[{{Coud{\'e}} {et~al.}(2016){Coud{\'e}}, {Bastien}, {Kirk},
  {Johnstone}, {Drabek-Maunder}, {Graves}, {Hatchell}, {Chapin}, {Gibb},
  {Matthews}, \& {JCMT Gould Belt Survey Team}}]{Coude2016}
{Coud{\'e}}, S., {Bastien}, P., {Kirk}, H., {et~al.} 2016, \mnras, 457, 2139,
  \dodoi{10.1093/mnras/stv3009}

\bibitem[{{Crane} \& {Napier}(1989)}]{Crane89}
{Crane}, P.~C., \& {Napier}, P.~J. 1989, in Astronomical Society of the Pacific
  Conference Series, Vol.~6, Synthesis Imaging in Radio Astronomy, ed. R.~A.
  {Perley}, F.~R. {Schwab}, \& A.~H. {Bridle}, 139

\bibitem[{{Davis} {et~al.}(2012){Davis}, {Muders}, {Humphreys}, {Stoehr}, {Leon
  Tanne}, \& {Saito}}]{Davis2012}
{Davis}, L., {Muders}, D., {Humphreys}, E., {et~al.} 2012, in Astronomical
  Society of the Pacific Conference Series, Vol. 461, Astronomical Data
  Analysis Software and Systems XXI, ed. P.~{Ballester}, D.~{Egret}, \&
  N.~P.~F. {Lorente}, 185

\bibitem[{{Davis} {et~al.}(2015){Davis}, {Williams}, {Nakazato}, {Lightfoot},
  {Muders}, \& {Kent}}]{Davis15}
{Davis}, L., {Williams}, S., {Nakazato}, T., {et~al.} 2015, in Astronomical
  Society of the Pacific Conference Series, Vol. 495, Astronomical Data
  Analysis Software an Systems XXIV (ADASS XXIV), ed. A.~R. {Taylor} \&
  E.~{Rosolowsky}, 301

\bibitem[{{Davis}(2009)}]{Davis09}
{Davis}, L.~E. 2009, in Astronomical Society of the Pacific Conference Series,
  Vol. 411, Astronomical Data Analysis Software and Systems XVIII, ed. D.~A.
  {Bohlender}, D.~{Durand}, \& P.~{Dowler}, 306

\bibitem[{{Dick} \& {Richter}(2004)}]{IERS2004}
{Dick}, W.~R., \& {Richter}, B. 2004, in Astrophysics and Space Science
  Library, Vol. 310, Organizations and Strategies in Astronomy, Vol. 5, ed.
  A.~{Heck}, 159--168, \dodoi{10.1007/978-1-4020-2571-6_8}

\bibitem[{{Escoffier} {et~al.}(2007){Escoffier}, {Comoretto}, {Webber},
  {Baudry}, {Broadwell}, {Greenberg}, {Treacy}, {Cais}, {Quertier}, {Camino},
  {Bos}, \& {Gunst}}]{Escoffier07}
{Escoffier}, R.~P., {Comoretto}, G., {Webber}, J.~C., {et~al.} 2007, \aap, 462,
  801, \dodoi{10.1051/0004-6361:20054519}

\bibitem[{{Figueira} {et~al.}(2018){Figueira}, {Bronfman}, {Zavagno}, {Louvet},
  {Lo}, {Finger}, \& {Rod{\'o}n}}]{Figueira2018}
{Figueira}, M., {Bronfman}, L., {Zavagno}, A., {et~al.} 2018, \aap, 616, L10,
  \dodoi{10.1051/0004-6361/201832930}

\bibitem[{Flanagan(2006)}]{javascript}
Flanagan, D. 2006, JavaScript: the definitive guide, 5th edn. (" O'Reilly
  Media, Inc.")

\bibitem[{{Fomalont} {et~al.}(2014){Fomalont}, {van Kempen}, {Kneissl},
  {Marcelino}, {Barkats}, {Corder}, {Cortes}, {Hills}, {Lucas}, {Manning}, \&
  {Peck}}]{Fomalont2014}
{Fomalont}, E., {van Kempen}, T., {Kneissl}, R., {et~al.} 2014, The Messenger,
  155, 19

\bibitem[{{Geers} {et~al.}(2019){Geers}, {Davis}, {Hales}, {Kent}, {Kern},
  {Kosugi}, {Muders}, {Nakazato}, {Sugimoto}, {Williams}, \&
  {Wyrowski}}]{Geers19}
{Geers}, V.~C., {Davis}, L., {Hales}, C.~A., {et~al.} 2019, in Astronomical
  Society of the Pacific Conference Series, Vol. 521, Astronomical Data
  Analysis Software and Systems XXVI, ed. M.~{Molinaro}, K.~{Shortridge}, \&
  F.~{Pasian}, 366

\bibitem[{{Gibney}(2022)}]{leapSecond2022}
{Gibney}, E. 2022, \nat, 612, 18, \dodoi{10.1038/d41586-022-03783-5}

\bibitem[{{Goddi} {et~al.}(2019){Goddi}, {Mart{\'\i}-Vidal}, {Messias}, {Crew},
  {Herrero-Illana}, {Impellizzeri}, {Rottmann}, {Wagner}, {Fomalont},
  {Matthews}, {Petry}, {Phillips}, {Tilanus}, {Villard}, {Blackburn},
  {Janssen}, \& {Wielgus}}]{Goddi19}
{Goddi}, C., {Mart{\'\i}-Vidal}, I., {Messias}, H., {et~al.} 2019, \pasp, 131,
  075003, \dodoi{10.1088/1538-3873/ab136a}

\bibitem[{Gordon(1976)}]{Gordon76}
Gordon, M.~A. 1976, in Methods of Experimental Physics: Volume 12:
  Astrophysics, Part C: Radio Observations, ed. M.~L. Meeks (Oxford: Academic
  Press), 277--283

\bibitem[{{Groesbeck}(1995)}]{Groesbeck1995}
{Groesbeck}, T.~D. 1995, PhD thesis, California Institute of Technology

\bibitem[{{Hafok} {et~al.}(2006){Hafok}, {Caillat}, \& {McMullin}}]{Hafok2006}
{Hafok}, H., {Caillat}, M., \& {McMullin}, J. 2006, in Astronomical Society of
  the Pacific Conference Series, Vol. 351, Astronomical Data Analysis Software
  and Systems XV, ed. C.~{Gabriel}, C.~{Arviset}, D.~{Ponz}, \& S.~{Enrique},
  189

\bibitem[{Harris {et~al.}(2020)Harris, Millman, van~der Walt, Gommers,
  Virtanen, Cournapeau, Wieser, Taylor, Berg, Smith, Kern, Picus, Hoyer, van
  Kerkwijk, Brett, Haldane, Fernández~del Río, Wiebe, Peterson,
  Gérard-Marchant, Sheppard, Reddy, Weckesser, Abbasi, Gohlke, \&
  Oliphant}]{2020NumPy-Array}
Harris, C.~R., Millman, K.~J., van~der Walt, S.~J., {et~al.} 2020, Nature, 585,
  357–362, \dodoi{10.1038/s41586-020-2649-2}

\bibitem[{{Harris}(1978)}]{Harris78}
{Harris}, F.~J. 1978, IEEE Proceedings, 66, 51

\bibitem[{{Hills} \& {Beasley}(2008)}]{Hills2008}
{Hills}, R.~E., \& {Beasley}, A.~J. 2008, in Society of Photo-Optical
  Instrumentation Engineers (SPIE) Conference Series, Vol. 7012, Ground-based
  and Airborne Telescopes II, ed. L.~M. {Stepp} \& R.~{Gilmozzi}, 70120N,
  \dodoi{10.1117/12.787567}

\bibitem[{{H{\"o}gbom}(1974)}]{hogbom1974}
{H{\"o}gbom}, J.~A. 1974, \aaps, 15, 417

\bibitem[{Hunter(2007)}]{hunter2007matplotlib}
Hunter, J.~D. 2007, Computing in science \& engineering, 9, 90

\bibitem[{{Hunter}(2012)}]{Hunter2012}
{Hunter}, T.~R. 2012, {Integration Durations for ALMA FDM Observations}, North
  American ALMA Science Center (NAASC) Memo 111, August 21, 2012, 5 pages

\bibitem[{{Hunter} \& {Kimberk}(2015)}]{Hunter2015}
{Hunter}, T.~R., \& {Kimberk}, R. 2015, arXiv e-prints, arXiv:1507.04280,
  \dodoi{10.48550/arXiv.1507.04280}

\bibitem[{{Hunter} {et~al.}(2016){Hunter}, {Lucas}, {Brogui{\`e}re},
  {Fomalont}, {Dent}, {Phillips}, {Rabanus}, \& {Vlahakis}}]{Hunter2016}
{Hunter}, T.~R., {Lucas}, R., {Brogui{\`e}re}, D., {et~al.} 2016, in Society of
  Photo-Optical Instrumentation Engineers (SPIE) Conference Series, Vol. 9914,
  Millimeter, Submillimeter, and Far-Infrared Detectors and Instrumentation for
  Astronomy VIII, ed. W.~S. {Holland} \& J.~{Zmuidzinas}, 99142L,
  \dodoi{10.1117/12.2232585}

\bibitem[{{Hunter} {et~al.}(2023){Hunter}, {Petry}, {Barkats}, {Corder}, \&
  {Indebetouw}}]{aU}
{Hunter}, T.~R., {Petry}, D., {Barkats}, D., {Corder}, S., \& {Indebetouw}, R.
  2023, {analysisUtils}, v2.15, Zenodo,  Zenodo, \dodoi{10.5281/zenodo.7502160}

\bibitem[{{Iguchi} {et~al.}(2009){Iguchi}, {Morita}, {Sugimoto}, {Vilar{\'o}},
  {Saito}, {Hasegawa}, {Kawabe}, {Tatematsu}, {Sakamoto}, {Kiuchi}, {Okumura},
  {Kosugi}, {Inatani}, {Takakuwa}, {Iono}, {Kamazaki}, {Ogasawara}, \&
  {Ishiguro}}]{Iguchi2009}
{Iguchi}, S., {Morita}, K.-I., {Sugimoto}, M., {et~al.} 2009, \pasj, 61, 1,
  \dodoi{10.1093/pasj/61.1.1}

\bibitem[{{Johnson} {et~al.}(2021){Johnson}, {Paradies}, {Dembska}, {Lackeos},
  {Kl{\"o}ckner}, {Champion}, \& {Schindler}}]{Johnson2021}
{Johnson}, M. A.~C., {Paradies}, M., {Dembska}, M., {et~al.} 2021, arXiv
  e-prints, arXiv:2109.10759, \dodoi{10.48550/arXiv.2109.10759}

\bibitem[{{Kamazaki} {et~al.}(2012){Kamazaki}, {Okumura}, {Chikada}, {Okuda},
  {Kurono}, {Iguchi}, {Mitsuishi}, {Murakami}, {Nishimura}, {Mita}, \&
  {Sano}}]{Kamazaki12}
{Kamazaki}, T., {Okumura}, S.~K., {Chikada}, Y., {et~al.} 2012, \pasj, 64, 29,
  \dodoi{10.1093/pasj/64.2.29}

\bibitem[{{Kawamura} {et~al.}(2002){Kawamura}, {Hunter}, {Tong}, {Blundell},
  {Papa}, {Patt}, {Peters}, {Wilson}, {Henkel}, {Gol'tsman}, \&
  {Gershenzon}}]{Kawamura2002}
{Kawamura}, J., {Hunter}, T.~R., {Tong}, C. Y.~E., {et~al.} 2002, \aap, 394,
  271, \dodoi{10.1051/0004-6361:20021090}

\bibitem[{{Kent} {et~al.}(2020){Kent}, {Masters}, {Chandler}, {Tobin},
  {Marvil}, {Ott}, {Myers}, {Medlin}, {Kimball}, {Schinzel}, {Lacy}, {Kern},
  {Butler}, {Sugimoto}, {Muders}, {Williams}, \& {Geers}}]{Kent2020}
{Kent}, B.~R., {Masters}, J.~S., {Chandler}, C.~J., {et~al.} 2020, in
  Astronomical Society of the Pacific Conference Series, Vol. 527, Astronomical
  Data Analysis Software and Systems XXIX, ed. R.~{Pizzo}, E.~R. {Deul}, J.~D.
  {Mol}, J.~{de Plaa}, \& H.~{Verkouter}, 571

\bibitem[{{Kepley} {et~al.}(2023){Kepley}, {Lipnicky}, {Venkata}, \&
  {Indebetouw}}]{memo623}
{Kepley}, A.~A., {Lipnicky}, A., {Venkata}, U.~R., \& {Indebetouw}, R. 2023,
  Atacama Large Millimeter/Submillimeter Array Memo 623, 623, 11

\bibitem[{{Kepley} {et~al.}(2020){Kepley}, {Tsutsumi}, {Brogan}, {Indebetouw},
  {Yoon}, {Mason}, \& {Donovan Meyer}}]{Kepley2020}
{Kepley}, A.~A., {Tsutsumi}, T., {Brogan}, C.~L., {et~al.} 2020, \pasp, 132,
  024505, \dodoi{10.1088/1538-3873/ab5e14}

\bibitem[{Krekel {et~al.}(2004)Krekel, Oliveira, Pfannschmidt, Bruynooghe,
  Laugher, \& Bruhin}]{pytest}
Krekel, H., Oliveira, B., Pfannschmidt, R., {et~al.} 2004, pytest.
\newblock \url{https://github.com/pytest-dev/pytest}

\bibitem[{{Leroy} {et~al.}(2021){Leroy}, {Hughes}, {Liu}, {Pety}, {Rosolowsky},
  {Saito}, {Schinnerer}, {Schruba}, {Usero}, {Faesi}, {Herrera}, {Chevance},
  {Hygate}, {Kepley}, {Koch}, {Querejeta}, {Sliwa}, {Will}, {Wilson}, {Anand},
  {Barnes}, {Belfiore}, {Be{\v{s}}li{\'c}}, {Bigiel}, {Blanc}, {Bolatto},
  {Boquien}, {Cao}, {Chandar}, {Chastenet}, {Chiang}, {Congiu}, {Dale},
  {Deger}, {den Brok}, {Eibensteiner}, {Emsellem},
  {Garc{\'\i}a-Rodr{\'\i}guez}, {Glover}, {Grasha}, {Groves}, {Henshaw},
  {Jim{\'e}nez Donaire}, {Kim}, {Klessen}, {Kreckel}, {Kruijssen}, {Larson},
  {Lee}, {Mayker}, {McElroy}, {Meidt}, {Mok}, {Pan}, {Puschnig}, {Razza},
  {S{\'a}nchez-Bl'azquez}, {Sandstrom}, {Santoro}, {Sardone}, {Scheuermann},
  {Sun}, {Thilker}, {Turner}, {Ubeda}, {Utomo}, {Watkins}, \&
  {Williams}}]{phangs2021}
{Leroy}, A.~K., {Hughes}, A., {Liu}, D., {et~al.} 2021, \apjs, 255, 19,
  \dodoi{10.3847/1538-4365/abec80}

\bibitem[{{Lightfoot} {et~al.}(2006){Lightfoot}, {Wyrowski}, {Muders}, {Boone},
  {Davis}, {Shepherd}, \& {Wilson}}]{Lightfoot2006}
{Lightfoot}, J., {Wyrowski}, F., {Muders}, D., {et~al.} 2006, in Astronomical
  Society of the Pacific Conference Series, Vol. 351, Astronomical Data
  Analysis Software and Systems XV, ed. C.~{Gabriel}, C.~{Arviset}, D.~{Ponz},
  \& S.~{Enrique}, 315

\bibitem[{{Liszt}(2014)}]{Liszt14}
{Liszt}, H.~S. 2014, in Society of Photo-Optical Instrumentation Engineers
  (SPIE) Conference Series, Vol. 9149, Observatory Operations: Strategies,
  Processes, and Systems V, ed. A.~B. {Peck}, C.~R. {Benn}, \& R.~L. {Seaman},
  91490N, \dodoi{10.1117/12.2055781}

\bibitem[{{Lorente}(2012)}]{Lorente2012}
{Lorente}, N.~P.~F. 2012, in Astronomical Society of the Pacific Conference
  Series, Vol. 461, Astronomical Data Analysis Software and Systems XXI, ed.
  P.~{Ballester}, D.~{Egret}, \& N.~P.~F. {Lorente}, 165

\bibitem[{{Lucas} {et~al.}(2000){Lucas}, {Clark}, {Mangum}, {Schilke}, {Scott},
  {Viallefond}, \& {Wright}}]{Lucas2000}
{Lucas}, R., {Clark}, B., {Mangum}, J., {et~al.} 2000, ALMA Memo 293

\bibitem[{{Lucas} {et~al.}(2002){Lucas}, {Clark}, {Ezawa}, {Gueth}, {Handa},
  {Harris}, {Mangum}, {Muders}, {Myers}, {Nakanishi}, {Ohnishi}, {Richer},
  {Schilke}, {Schwarz}, {Scott}, {Tatematsu}, {Viallefond}, {Warmels}, \&
  {Wright}}]{Lucas2002}
{Lucas}, R., {Clark}, B., {Ezawa}, H., {et~al.} 2002, ALMA Computing Memo 11

\bibitem[{{Lundgren} {et~al.}(2012){Lundgren}, {Nyman}, {Saito}, {Vila Vilaro},
  {Mathys}, {Andreani}, {Hibbard}, {Okumura}, {Tatematsu}, {Dent}, {Rawlings},
  {Villard}, \& {Ball}}]{Lundgren2012}
{Lundgren}, A., {Nyman}, L.-A., {Saito}, M., {et~al.} 2012, in Society of
  Photo-Optical Instrumentation Engineers (SPIE) Conference Series, Vol. 8448,
  Observatory Operations: Strategies, Processes, and Systems IV, ed. A.~B.
  {Peck}, R.~L. {Seaman}, \& F.~{Comeron}, 844802, \dodoi{10.1117/12.926622}

\bibitem[{{Marson} \& {Hiriart}(2016)}]{Marson16}
{Marson}, R., \& {Hiriart}, R. 2016, in Society of Photo-Optical
  Instrumentation Engineers (SPIE) Conference Series, Vol. 9913, Software and
  Cyberinfrastructure for Astronomy IV, ed. G.~{Chiozzi} \& J.~C. {Guzman},
  991304, \dodoi{10.1117/12.2233584}

\bibitem[{{Masters} {et~al.}(2020){Masters}, {Kent}, {Sugimoto}, {Indebetouw},
  {Brogan}, {Hunter}, {Kepley}, {Hibbard}, {Nakazato}, {Kosugi}, {Ezawa},
  {Yoshino}, {Hayashi}, {Castro}, {Pouzols}, {Williams}, {Geers}, {Muders}, \&
  {Wyrowski}}]{Masters2020}
{Masters}, J.~S., {Kent}, B.~R., {Sugimoto}, K., {et~al.} 2020, in Astronomical
  Society of the Pacific Conference Series, Vol. 527, Astronomical Data
  Analysis Software and Systems XXIX, ed. R.~{Pizzo}, E.~R. {Deul}, J.~D.
  {Mol}, J.~{de Plaa}, \& H.~{Verkouter}, 639

\bibitem[{{Matsushita} {et~al.}(2017){Matsushita}, {Asaki}, {Fomalont},
  {Morita}, {Barkats}, {Hills}, {Kawabe}, {Maud}, {Nikolic}, {Tilanus},
  {Vlahakis}, \& {Whyborn}}]{Matsushita17}
{Matsushita}, S., {Asaki}, Y., {Fomalont}, E.~B., {et~al.} 2017, \pasp, 129,
  035004, \dodoi{10.1088/1538-3873/aa5787}

\bibitem[{{Maud} {et~al.}(2023){Maud}, {P{\'e}rez-S{\'a}nchez}, {Asaki},
  {Stoehr}, {Dent}, \& {D{\'\i}az Trigo}}]{Maud23}
{Maud}, L.~T., {P{\'e}rez-S{\'a}nchez}, A.~F., {Asaki}, Y., {et~al.} 2023, ALMA
  Memo 624, arXiv:2304.08318, \dodoi{10.48550/arXiv.2304.08318}

\bibitem[{{Maud} {et~al.}(2020){Maud}, {Asaki}, {Fomalont}, {Dent}, {Hirota},
  {Matsushita}, {Phillips}, {Carpenter}, {Takahashi}, {Villard}, {Sawada}, \&
  {Corder}}]{Maud20}
{Maud}, L.~T., {Asaki}, Y., {Fomalont}, E.~B., {et~al.} 2020, \apjs, 250, 18,
  \dodoi{10.3847/1538-4365/abab94}

\bibitem[{{Maud} {et~al.}(2022){Maud}, {Asaki}, {Dent}, {Hirota}, {Fomalont},
  {Takahashi}, {Matsushita}, {Phillips}, {Sawada}, {Corder}, \&
  {Carpenter}}]{Maud22}
{Maud}, L.~T., {Asaki}, Y., {Dent}, W. R.~F., {et~al.} 2022, \apjs, 259, 10,
  \dodoi{10.3847/1538-4365/ac3b57}

\bibitem[{{McCarthy} {et~al.}(2008){McCarthy}, {Hackman}, \&
  {Nelson}}]{McCarthy2008}
{McCarthy}, D.~D., {Hackman}, C., \& {Nelson}, R.~A. 2008, \aj, 136, 1906,
  \dodoi{10.1088/0004-6256/136/5/1906}

\bibitem[{{Moellenbrock}(2017)}]{Moellenbrock2017}
{Moellenbrock}, G. 2017, in Submm/mm/cm QUESO Workshop 2017 (QUESO2017), 19,
  \dodoi{10.5281/zenodo.1038085}

\bibitem[{{Morita} \& {Holdaway}(2008)}]{Morita2008}
{Morita}, K.-I., \& {Holdaway}, M.~A. 2008, in Society of Photo-Optical
  Instrumentation Engineers (SPIE) Conference Series, Vol. 7012, Ground-based
  and Airborne Telescopes II, ed. L.~M. {Stepp} \& R.~{Gilmozzi}, 70120O,
  \dodoi{10.1117/12.788346}

\bibitem[{{Muders} {et~al.}(2014){Muders}, {Wyrowski}, {Lightfoot}, {Williams},
  {Nakazato}, {Kosugi}, {Davis}, \& {Kern}}]{Muders2014}
{Muders}, D., {Wyrowski}, F., {Lightfoot}, J., {et~al.} 2014, in Astronomical
  Society of the Pacific Conference Series, Vol. 485, Astronomical Data
  Analysis Software and Systems XXIII, ed. N.~{Manset} \& P.~{Forshay}, 383

\bibitem[{{Nakazato} {et~al.}(2022){Nakazato}, {Sugimoto}, {Yoshino}, {Ezawa},
  {Hayashi}, {Kosugi}, {Maekawa}, {Takahashi}, \& {Tatematsu}}]{Nakazato2022}
{Nakazato}, T., {Sugimoto}, K., {Yoshino}, A., {et~al.} 2022, in Astronomical
  Society of the Pacific Conference Series, Vol. 532, Astronomical Society of
  the Pacific Conference Series, ed. J.~E. {Ruiz}, F.~{Pierfedereci}, \&
  P.~{Teuben}, 397

\bibitem[{{Nikolic} {et~al.}(2013){Nikolic}, {Bolton}, {Graves}, {Hills}, \&
  {Richer}}]{Nikolic2013}
{Nikolic}, B., {Bolton}, R.~C., {Graves}, S.~F., {Hills}, R.~E., \& {Richer},
  J.~S. 2013, \aap, 552, A104, \dodoi{10.1051/0004-6361/201220987}

\bibitem[{{Nikolic} {et~al.}(2012){Nikolic}, {Graves}, {Bolton}, \&
  {Richer}}]{Nikolic2012}
{Nikolic}, B., {Graves}, S.~F., {Bolton}, R.~C., \& {Richer}, J.~S. 2012, arXiv
  e-prints, arXiv:1207.6069.
\newblock \doarXiv{1207.6069}

\bibitem[{{{\"O}berg} {et~al.}(2021){{\"O}berg}, {Guzm{\'a}n}, {Walsh},
  {Aikawa}, {Bergin}, {Law}, {Loomis}, {Alarc{\'o}n}, {Andrews}, {Bae},
  {Bergner}, {Boehler}, {Booth}, {Bosman}, {Calahan}, {Cataldi}, {Cleeves},
  {Czekala}, {Furuya}, {Huang}, {Ilee}, {Kurtovic}, {Le Gal}, {Liu}, {Long},
  {M{\'e}nard}, {Nomura}, {P{\'e}rez}, {Qi}, {Schwarz}, {Sierra}, {Teague},
  {Tsukagoshi}, {Yamato}, {van't Hoff}, {Waggoner}, {Wilner}, \&
  {Zhang}}]{maps2021}
{{\"O}berg}, K.~I., {Guzm{\'a}n}, V.~V., {Walsh}, C., {et~al.} 2021, \apjs,
  257, 1, \dodoi{10.3847/1538-4365/ac1432}

\bibitem[{{Paine} {et~al.}(2000){Paine}, {Blundell}, {Papa}, {Barrett}, \&
  {Radford}}]{Paine2000}
{Paine}, S., {Blundell}, R., {Papa}, D.~C., {Barrett}, J.~W., \& {Radford}, S.
  J.~E. 2000, \pasp, 112, 108, \dodoi{10.1086/316497}

\bibitem[{{Pardo} {et~al.}(2001){Pardo}, {Cernicharo}, \&
  {Serabyn}}]{Pardo2001}
{Pardo}, J.~R., {Cernicharo}, J., \& {Serabyn}, E. 2001, IEEE Transactions on
  Antennas and Propagation, 49, 1683, \dodoi{10.1109/8.982447}

\bibitem[{{Pardo} {et~al.}(2022){Pardo}, {De Breuck}, {Muders}, {Gonz{\'a}lez},
  {Montenegro-Montes}, {P{\'e}rez-Beaupuits}, {Cernicharo}, {Prigent},
  {Serabyn}, {Mroczkowski}, \& {Phillips}}]{Pardo2022}
{Pardo}, J.~R., {De Breuck}, C., {Muders}, D., {et~al.} 2022, \aap, 664, A153,
  \dodoi{10.1051/0004-6361/202243409}

\bibitem[{P\'erez \& Granger(2007)}]{IPython}
P\'erez, F., \& Granger, B.~E. 2007, Computing in Science and Engineering, 9,
  21, \dodoi{10.1109/MCSE.2007.53}

\bibitem[{{Petry}(2021)}]{Petry2021}
{Petry}, D. 2021, arXiv e-prints, arXiv:2112.10050,
  \dodoi{10.48550/arXiv.2112.10050}

\bibitem[{{Plunkett} {et~al.}(2023){Plunkett}, {Hacar}, {Moser-Fischer},
  {Petry}, {Teuben}, {Pingel}, {Kunneriath}, {Takagi}, {Miyamoto}, {Moravec},
  {Suri}, {Hess}, {Hoffman}, \& {Mason}}]{Plunkett2023}
{Plunkett}, A., {Hacar}, A., {Moser-Fischer}, L., {et~al.} 2023, \pasp, 135,
  034501, \dodoi{10.1088/1538-3873/acb9bd}

\bibitem[{{Raba}(2022)}]{Raba2022}
{Raba}, R. 2022, in Astronomical Society of the Pacific Conference Series, Vol.
  532, Astronomical Society of the Pacific Conference Series, ed. J.~E. {Ruiz},
  F.~{Pierfedereci}, \& P.~{Teuben}, 67

\bibitem[{{Rau} \& {Cornwell}(2011)}]{Rau2011}
{Rau}, U., \& {Cornwell}, T.~J. 2011, \aap, 532, A71,
  \dodoi{10.1051/0004-6361/201117104}

\bibitem[{{Remijan} {et~al.}(2019){Remijan}, {Biggs}, {Cortes}, {Dent}, {Di
  Franceso}, {Fomalont}, {Hales}, {Kameno}, {Mason}, {Philips}, {Saini}, {Vila
  Vilaro}, \& {Villard}}]{TechHandRemijan19}
{Remijan}, A., {Biggs}, A., {Cortes}, P.~A., {et~al.} 2019, {ALMA Technical
  Handbook, ALMA Doc. 7.3, ver. 1.1}, 2019, ALMA Technical Handbook, ALMA Doc.
  7.3, ver. 1.1ISBN 978-3-923524-66-2, \dodoi{10.5281/zenodo.4511522}

\bibitem[{{Remijan} {et~al.}(2022){Remijan}, {Biggs}, {Cortes}, {Dent}, {Di
  Franceso}, {Fomalont}, {Hales}, {Kameno}, {Mason}, {Philips}, {Saini}, {Vila
  Vilaro}, \& {Villard}}]{Remijan2019}
---. 2022, {ALMA Technical Handbook,ALMA Doc. 9.3, ver. 1.0}, 2021, ALMA
  Technical Handbook, ALMA Doc. 9.3, ver. 1.0, ISBN 978-3-923524-66-2,
  \dodoi{10.5281/zenodo.7764458}

\bibitem[{{Santander-Vela} {et~al.}(2012){Santander-Vela}, {Bauhofer}, {Meuss},
  {Stoehr}, \& {Manning}}]{Santander2012}
{Santander-Vela}, J., {Bauhofer}, M., {Meuss}, H., {Stoehr}, F., \& {Manning},
  A. 2012, in Astronomical Society of the Pacific Conference Series, Vol. 461,
  Astronomical Data Analysis Software and Systems XXI, ed. P.~{Ballester},
  D.~{Egret}, \& N.~P.~F. {Lorente}, 435

\bibitem[{{Selina} {et~al.}(2022){Selina}, {Murphy}, \& {Beasley}}]{Selina22}
{Selina}, R., {Murphy}, E., \& {Beasley}, A. 2022, in Society of Photo-Optical
  Instrumentation Engineers (SPIE) Conference Series, Vol. 12182, Ground-based
  and Airborne Telescopes IX, ed. H.~K. {Marshall}, J.~{Spyromilio}, \&
  T.~{Usuda}, 121820O, \dodoi{10.1117/12.2627730}

\bibitem[{{Stoehr} {et~al.}(2014){Stoehr}, {Lacy}, {Leon}, {Muller}, {Manning},
  {Moins}, \& {Jenkins}}]{Felix2014}
{Stoehr}, F., {Lacy}, M., {Leon}, S., {et~al.} 2014, in Society of
  Photo-Optical Instrumentation Engineers (SPIE) Conference Series, Vol. 9149,
  Observatory Operations: Strategies, Processes, and Systems V, ed. A.~B.
  {Peck}, C.~R. {Benn}, \& R.~L. {Seaman}, 914902, \dodoi{10.1117/12.2055539}

\bibitem[{{Stoehr} {et~al.}(2012){Stoehr}, {Leon Tanne}, {Lacy}, {Saito}, \&
  {Santander-Vela}}]{Felix2012}
{Stoehr}, F., {Leon Tanne}, S., {Lacy}, M., {Saito}, M., \& {Santander-Vela},
  J. 2012, in Astronomical Society of the Pacific Conference Series, Vol. 461,
  Astronomical Data Analysis Software and Systems XXI, ed. P.~{Ballester},
  D.~{Egret}, \& N.~P.~F. {Lorente}, 697

\bibitem[{{Takahashi} {et~al.}(2021){Takahashi}, {Fomalont}, {Asaki}, {Crew},
  {Matthews}, {Cortes}, {Vila-Vilaro}, {Bastian}, {Shimojo}, {Biggs},
  {Messias}, {Hales}, {Villard}, \& {Humphreys}}]{Takahashi21}
{Takahashi}, S., {Fomalont}, E.~B., {Asaki}, Y., {et~al.} 2021, arXiv e-prints,
  arXiv:2104.12681.
\newblock \doarXiv{2104.12681}

\bibitem[{{Tarenghi}(2008)}]{Tarenghi2008}
{Tarenghi}, M. 2008, \apss, 313, 1, \dodoi{10.1007/s10509-007-9602-9}

\bibitem[{{Teuben} {et~al.}(2015){Teuben}, {Pound}, {Mundy}, {Rauch},
  {Friedel}, {Looney}, {Xu}, \& {Kern}}]{Teuben15}
{Teuben}, P., {Pound}, M., {Mundy}, L., {et~al.} 2015, in Astronomical Society
  of the Pacific Conference Series, Vol. 495, Astronomical Data Analysis
  Software an Systems XXIV (ADASS XXIV), ed. A.~R. {Taylor} \& E.~{Rosolowsky},
  305

\bibitem[{{The ImageMagick Development Team}(2021)}]{imagemagick}
{The ImageMagick Development Team}. 2021, ImageMagick, 7.0.10.
\newblock \url{https://imagemagick.org}

\bibitem[{{van Diepen}(2015)}]{MS2015}
{van Diepen}, G.~N.~J. 2015, Astronomy and Computing, 12, 174,
  \dodoi{10.1016/j.ascom.2015.06.002}

\bibitem[{Van~Rossum \& Drake(2009)}]{python3}
Van~Rossum, G., \& Drake, F.~L. 2009, Python 3 Reference Manual (Scotts Valley,
  CA: CreateSpace)

\bibitem[{{Viallefond}(2006)}]{Viallefond2006}
{Viallefond}, F. 2006, in Astronomical Society of the Pacific Conference
  Series, Vol. 351, Astronomical Data Analysis Software and Systems XV, ed.
  C.~{Gabriel}, C.~{Arviset}, D.~{Ponz}, \& S.~{Enrique}, 627

\bibitem[{Virtanen {et~al.}(2020)Virtanen, Gommers, Oliphant, Haberland, Reddy,
  Cournapeau, Burovski, Peterson, Weckesser, Bright, {van der Walt}, Brett,
  Wilson, Millman, Mayorov, Nelson, Jones, Kern, Larson, Carey, Polat, Feng,
  Moore, {VanderPlas}, Laxalde, Perktold, Cimrman, Henriksen, Quintero, Harris,
  Archibald, Ribeiro, Pedregosa, {van Mulbregt}, \& {SciPy 1.0
  Contributors}}]{2020SciPy-NMeth}
Virtanen, P., Gommers, R., Oliphant, T.~E., {et~al.} 2020, Nature Methods, 17,
  261, \dodoi{10.1038/s41592-019-0686-2}

\bibitem[{{Williams} \& {Bridger}(2013)}]{WilliamsBridger13}
{Williams}, S., \& {Bridger}, A. 2013, in Astronomical Society of the Pacific
  Conference Series, Vol. 475, Astronomical Data Analysis Software and Systems
  XXII, ed. D.~N. {Friedel}, 373

\bibitem[{{Wootten} \& {Emerson}(2005)}]{Wootten2005}
{Wootten}, A., \& {Emerson}, D. 2005, in IEEE International Conference on
  Acoustics, Vol.~5, 857--860, \dodoi{10.1109/ICASSP.2005.1416439}

\bibitem[{{Wootten} \& {Thompson}(2009)}]{Wootten2009}
{Wootten}, A., \& {Thompson}, A.~R. 2009, IEEE Proceedings, 97, 1463,
  \dodoi{10.1109/JPROC.2009.2020572}

\end{thebibliography}
\bibliographystyle{aasjournal}



\end{document}